\documentclass{aastex}
\usepackage{apjfonts}
\usepackage{emulateapj5}
\input{epsf}

\slugcomment{Accepted by the Astrophysical Journal, 18 April 2006}

\usepackage{graphicx}

\newcommand{\be}{\begin{equation}}
\newcommand{\ee}{\end{equation}}
\newcommand{\nn}{\mbox{} \nonumber \\ \mbox{} }
\newcommand{\ba}{\begin{eqnarray}}
\newcommand{\ea}{\end{eqnarray}}

\newcommand\lo{\mathrel{\raise.3ex\hbox{$<$}\mkern-14mu\lower0.6ex\hbox{$\sim$}}}
\newcommand\go{\mathrel{\raise.3ex\hbox{$>$}\mkern-14mu\lower0.6ex\hbox{$\sim$}}}
\begin{document}
%\date{}  
\title{DECELERATION OF A RELATIVISTIC, PHOTON-RICH SHELL: 
END OF PREACCELERATION, DAMPING OF MHD
TURBULENCE, AND THE EMISSION MECHANISM OF GAMMA-RAY BURSTS}
\author{Christopher Thompson}
\affil{CITA, 60 St. George St., Toronto, ON M5S 3H8, Canada}

\begin{abstract}
We consider the interaction of a relativistically-moving shell, composed of 
thermal photons, a reversing magnetic field and a small admixture of charged
particles, with a dense Wolf-Rayet wind. 
A thin outer layer of Wolf-Rayet material is entrained by the jet head;
it cools and becomes Rayleigh-Taylor unstable, thereby providing an
additional source of inertia and variability. 
The gamma-rays emitted by the shell
load the ambient medium with electron-positron pairs and, close to the
engine, force this medium to move at nearly the same speed as
the shell.  We show that this pre-acceleration of the wind material
defines a characteristic radiative
compactness at the point where the reverse shock has completed its passage
back through the shell. We argue that the prompt gamma-ray emission is
triggered by this external braking, at an optical depth $\sim 1$ to electron
scattering.  Torsional waves, excited by the forced reconnection of the 
reversing magnetic field, carry a fluctuating current, and are damped at 
high frequencies by the electrostatic acceleration of electrons and positrons.
We show that inverse Compton radiation by the accelerated charges is stronger
than their synchrotron emission, and is beamed along the magnetic field. 
Thermal radiation that is advected out from the base of the jet cools the 
particles. The observed relation between peak energy and isotropic 
luminosity -- both its amplitude and scaling -- is reproduced if the 
blackbody seeds are generated in a relativistic jet core that is subject 
to Kelvin-Helmholtz instabilities with the Wolf-Rayet envelope. This relation 
is predicted to soften to $E_{\rm peak} \sim L_{\rm iso}^{1/4}$ below an 
isotropic luminosity $L_{\rm iso} \sim 3\times 10^{50}$ ergs s$^{-1}$. 
Spectrally harder bursts will arise in outflows which encounter no dense
stellar envelope.  The
duration of spikes in the inverse-Compton emission is narrower at higher 
frequencies, in agreement with the observed relation. The transition from 
prompt gamma-ray emission to afterglow can be explained by the termination
of the thermal X-ray seed and the onset of synchrotron-self-Compton emission. 
GLAST will probe the mechanism of particle heating at the reverse shock by 
measuring the inverse-Compton scattering of the seed photons.
\end{abstract}
\keywords{gamma rays: bursts; supernovae: general; turbulence; 
radiation mechanisms: non-thermal}

\section{Introduction}

Models of gamma-ray burst (GRB) outflows have long been plagued by 
uncertainties in the mechanism by which gamma-rays are emitted, including
the underlying instability that triggers the rapid variability,
the mechanism by which particles are energized, and the location
of the gamma-ray emitting region.  These uncertainties are closely
interlocked.  Because there is no shortage of plausible mechanisms
which could, in principle, generate gamma-rays in a relativistic outflow,
the lack of convergence on one issue has tended to impede
progress on all the others.

The data offer some important clues which have not yet been fully
exploited by theoretical models.  For example, individual GRB light curves
show only weak changes in the pattern of variability from the beginning
to the end of the burst (Beloborodov, Stern, \& Svensson 1998;
Ramirez-Ruiz \& Fenimore 2000).  
This, in spite of the basic requirement --
based on simple considerations of gamma-ray opacity (Cavallo \& Rees
1978) -- that the non-thermal gamma-rays must be emitted in a region 
that is vastly larger than the central engine -- by 
some 6-8 orders of magnitude.  
One deduces that this dissipative zone must be concentrated within
a fairly narrow range of radius, in spite of the large distance
that the outflow must travel.   In addition, a well-defined correlation
has emerged between the peak energy of gamma-ray bursts and their
isotropic luminosity, in a sample of BeppoSAX and HETE-II bursts
with measured redshifts (Amati et al. 2002; Ghirlanda et al. 2004;
Lamb, Donaghy, \& Graziani 2005).  A third clue is that the 
the short gamma-ray bursts have systematically harder spectra than
the long bursts.  The recent burst GRB 050509b had and isotropic
luminosity of $\sim 3\times 10^{49}$ ergs s$^{-1}$ at a redshift of
$z = 0.2$, but a much harder spectrum than the Amati relation would imply
(Bloom et al. 2005).   This observation has interesting implications for 
the location of the dissipative zones in a GRB outflow.

Strong circumstantial evidence has emerged that long GRBs involve
the ejection of relativistically moving material from $\sim 2-3\,M_\odot$ 
black holes, which are formed in the collapse of massive stars
(Woosley 1993;  Paczy\'nski 1998).  The gamma-ray burst 
030329 coincided with the energetic Type Ic supernova
2003dh (Stanek et al. 2003).  The optical spectrum of this supernova
was similar to that of the 
energetic Type Ic SN 1998bw (Galama et al. 1998).  This previous
supernova coincided with a gamma-ray burst (GRB 980425) of a much
smaller isotropic luminosity,  which may therefore
have involved a buried jet.  The progenitors of the long GRBs are
apparently Wolf-Rayet stars surrounded by fast and dense
winds.  The origin of the short gamma-ray bursts is more uncertain,
and may involve a mix of sources including binary neutron
star systems (Eichler et al. 1989) and extragalactic Soft Gamma
Repeaters (Duncan 2001).

The Wolf-Rayet/jet model offers two specific locations where 
the physics of gamma-ray emission can be well constrained:
first, where the collimated relativistic outflow
breaks out of the Wolf-Rayet star;  and, second, where it begins to be
decelerated substantially by its interaction with the external medium
(including the outermost thin layer of stellar material that is
entrained during the breakout of the relativistic jet).
Our first goal in this paper 
is to define how the deceleration phase is modified by
pair creation in the Wolf-Rayet wind, and the associated radiative 
acceleration of the pair-loaded wind material.
Previous treatments of this process 
(Thompson \& Madau 2000; M\'esz\'aros, Ramirez-Ruiz, \& Rees 2001;
Beloborodov 2002) assume
a given pattern of gamma-ray emission, and do not examine its
feedback on the forward and reverse shocks at the very earliest stage where
the prompt burst is generated.   One finds that the relativistic ejecta
have a characteristic {\it compactness}
at the point where the reverse shock wave completes its passage through
the ejecta shell, because the process of pre-acceleration shuts off below
a critical value of the fluence of the gamma-rays flowing across the 
forward shock.  Rapid deceleration of the relativistic ejecta
follows once this threshold is reached.  The compactness is, 
in fact, regulated to a value somewhat larger than unity, and 
the optical depth through the swept-up $e^+$-$e^-$ pairs 
has a similar value.  One therefore obtains a direct feedback between
the physical conditions in the dissipative zone, and the radiative mechanism.

The gamma-ray emission depends on the presence of some irregularities
in the outflow.  These irregularities
could be caused by stochastic flips in the
sign of a dynamo-generated magnetic field that is swept into the 
outflow (Thompson 1994); or, alternatively, they could involve
fluctuations in momentum and
energy flux that manifest themselves as internal shock waves
in a relativistic particle outflow (Rees \& M\'esz\'aros 1994).  

The seed irregularities are, in both of these models, imprinted 
into the outflow near its
base.  Nonetheless, the models are distinguished by the {\it isotropy}
of the emission in the bulk frame.   The synchrotron or inverse-Compton
radiation of shock-accelerated particles is nearly isotropic
in the rest frame of the shocked fluid.  The observation of fast variability
in many gamma-ray bursts then requires that the gamma-rays be emitted
far inside the radius at which the bulk motion begins to be reduced
significantly (Sari \& Piran 1997).  The basic disadvantage of this
approach is that the location of the dissipative zone depends 
sensitively on the frequency and amplitude of the seed
irregularities.  Calculating the pattern of these irregularities from
first principals requires a detailed theoretical understanding of almost
every aspect of the central engine and its outflow.  
In spite of recent herculean efforts to simulate jet
outflows (McKinney \& Gammie 2004; de Villiers, Staff, \& Ouyed 2005; McKinney 2005a,b),
the model retains considerable freedom in practice.  

A hybrid model of jet dissipation is suggested by considering
the influence of jet breakout on the subsequent motion of the reverse shock
wave.  A thin outer layer of the Wolf-Rayet star accumulates energy
from the relativistic beam emerging from below, and is pushed
outward to higher speeds.  Existing calculations of this
`breakout shell' are limited to a one-dimensional approximation
(Waxman \& M\'esz\'aros 2003).  We calculate the column density
of this shell, and the growth of its Lorentz factor, by taking
into account the slippage of the swept-up matter to the side of
the beam.
Rayleigh-Taylor instabilities 
allow the relativistic fluid to emerge through the breakout shell
in isolated spots, which can radiate independently.

Our second goal in this paper 
is to establish a realistic mechanism by which the gamma-ray 
emission may be triggered directly by the deceleration off the
`breakout' shell and the external wind medium.  Calculations of 
prompt high-energy emission from the forward shock (M\'esz\'aros
\& Rees 1993; M\'esz\'aros, Rees, \& Papathanassiou 1994;
Chiang \& Dermer 1999; Sari \& Esin 2001) have neglected the
effects of pre-acceleration of the ambient medium, have assumed
that the radiating particles are shock-accelerated, and therefore
cannot easily 
account for fast variability in the emission.  This forces us to focus
on an alternative source of energy in the outflow -- in particular,
a strong non-radial magnetic field,
which is independently motivated by considerations of the launching
of the relativistic outflow.   The magnetic field provides a 
characteristic direction for particle acceleration, and provides
plausible instabilities which cause
impulsive and localized injections of energy. 

The relative proportions of energy carried by Poynting flux and by
ions are still very uncertain in observed extragalactic jet sources
(e.g. Sikora et al. 2005).  Basic conservation
laws limit the transfer of energy from the magnetic field to the
particles through internal instabilities, if the inner boundary
conditions of the outflow are time-invariant.  Magnetic flux, in particular, 
is not easily eliminated from a fast-moving 
astrophysical fluid with a very high magnetic Reynolds number.  
The ion luminosity can increase by a logarithmic factor
with respect to the Poynting luminosity in a smooth outflow (Begelman
\& Li 1994).  The magnetic field will, nonetheless,
have a strong softening effect on internal shock waves if
the ordered Poynting luminosity $L_P$ comprises 10 percent
of the outflow luminosity (Kennel \& Coroniti 1984; Zhang \& Kobayashi
2005);  and will
contribute substantially to the post-shock pressure
if the Poynting luminosity is somewhat smaller than this. 
Recent jet simulations suggest an approximate equipartition between
kinetic and magnetic energy at a distance of $\sim 10^{10}$ cm
from a stellar-mass black hole (McKinney 2005a,b).

The particle luminosity could also be increased by reconnection of
a reversing magnetic field.  This process is much more effective when the
field is being pushed into the external medium,  at a large distance 
from the engine where the sound-crossing time of the shell is smaller
than the flow time (Thompson 1994; see Lyubarsky \& Kirk 2001 for a
more detailed discussion in the context of pulsar wind termination shocks).
Tangling of the magnetic field allows a reverse shock wave to form
even when the fast mode speed is higher than the speed of the outflow
relative to the contact discontinuity.  

Radiation that is advected out from the base of the outflow has
a characteristic energy of $\sim 1$ MeV (Goodman 1986).  Seed black body
photons will be present in combination with a magnetic field in many
models of the central engine, and therefore can play an important role
as a source of inverse-Compton seeds at a large radius, and help to define the
spectral peak of the gamma-ray emission (Thompson 1994).
Our third goal in this paper is to investigate this effect in more
detail, in light of more recent ideas that the propagation of a
relativistic jet through the envelope of a Wolf-Rayet star
(Zhang, MacFadyen, \& Woosley 2003; Matzner 2003) can create a 
lower-energy thermal
component in the jet cocoon (Ramirez-Ruiz, MacFadyen, \& Lazzati 2002).
We find that the observed relation between peak energy and isotropic
luminosity in BeppoSAX and HETE-II bursts (Amati et al. 2002; 
Lamb et al. 2005) arises directly -- without free parameters --
from dissipation in a {\it relativistically-moving jet core} 
that is heated but not fully decelerated by Kelvin-Helmholtz 
instabilities with the jet cocoon.
(The cocoon acquires a similar temperature, but its specific
entropy is not realistically high enough to seed the gamma-ray
emission proper.)   Instead the portion of the jet material that
acquires a high entropy within the Wolf-Rayet star is 
preferentially accelerated to high Lorentz factors after breaking out
of the stellar photosphere. One can therefore establish a direct 
connection between the presence of a fireball component at the base
of the jet, and the observability of that portion of the jet
as a gamma-ray burst.  

It has also previously been suggested that the outflow can
dissipate over a much wider range of radii, thereby creating a low-energy
blackbody excess in gamma-ray burst spectra (M\'esz\'aros et al. 2002),
or possibly a harder blackbody component just inside
the photosphere which could help to regulate the peak energy in the 
observed manner (M\'esz\'aros \& Rees 2000; 
Rees \& M\'esz\'aros 2005).  If the jet is magnetically dominated
but very cold at its base, then continuing dissipation would be required
to push it to a large Lorentz factor (Drenkhahn \& Spruit 2002).
However, it is difficult to establish
clear predictions when the dissipation is driven by instabilities
in the freely expanding portion of the outflow, given the number 
of physical effects which influence the growth of these instabilities.

The energy density of a gamma-ray burst outflow is orders of magnitude
higher than that of the Solar corona, and the ambient radiation field has
a much stronger influence on the cooling of charged particles.
In fact, the conditions in the GRB outflow
are much closer to those expected in the magnetic corona of
an accretion disk around a stellar-mass black hole.
This suggests that, when drawing lessons from
Solar physics, we should focus on those physical processes which
appear to be {\it energetically dominant} in the Sun's atmosphere, and not
only those which are directly responsible for the hard X-ray emission
that emerges from that much more dilute environment.  In the largest
Solar flares, the output in thermal ($\sim 1$ keV) X-rays is smaller
than the kinetic energy of the ejected plasma by an order of magnitude;
and the output in hard X-rays and gamma-rays 
is smaller yet by six orders of magnitude (e.g. Somov 1992).
Processes very similar to those which heat the 
Sun's chromosphere and thermal corona will, in the more extreme 
environment of a gamma-ray burst, create a much harder  photon spectrum.

A non-potential magnetic field transfers energy to 
particles through various channels.  Energetically important channels include
the damping of large-scale magnetohydrodynamic motions (through the
formation of high-frequency spectrum of MHD waves);
and a sudden change in the topology of magnetic field lines through
reconnection.    In a GRB outflow,
charged particles are created at a sufficient rate 
by photon collisions ($\gamma + \gamma \rightarrow e^+ + e^-$)
to ensure than the MHD condition applies on large scales.
For this reason, we have previously suggested
that the main role of reconnection in GRB outflows is not to
cancel off the toroidal magnetic field (as in the Coroniti 1990 model
of the Crab pulsar wind), 
but instead to facilitate a tangling up of the field through a 
change in topology.

Damping of MHD turbulence is expected to be a ubiquitous process in
any relativistic outflow containing a magnetic field.  
In this paper, we examine, in more detail than previously
(Thompson \& Blaes 1998; Lyutikov \& Thompson 2005),
the regime in which the magnetic energy density is at least comparable
to the rest-energy density of the entrained charged particles.
The wave energy is transferred to the light
charges (electrons and positrons) primarily through electrostatic
acceleration along the magnetic field.  As a result, the radiation
is beamed along the local direction of the magnetic field.  
This means that inverse-Compton
cooling will dominate synchrotron cooling even if the magnetic energy
is larger than the energy density of the ambient radiation field.
Bulk relativistic motion of the magnetofluid 
provides another avenue for beamed emission (Lyutikov \& Blandford 2003),
but it is not required.

An overarching goal of this paper is to identify the key components
of the outflow which {\it must} be included for a proper description of
GRB emission.  We find that it is essential to include 
a thermal radiation field as well as a strong magnetic field,
these two components being in approximate equipartition
at the base of the outflow.  A small admixture of
baryons is probably present, which can carry a significant portion of
the luminosity in some parts of the jet; but the baryons actually play a less
essential role in the emission physics.   For the purposes of investigating
the prompt gamma-ray emission, the bulk of the outflow
can be assumed to expand freely in between the photosphere
of the Wolf-Rayet star and the dissipation zone.  Closer to the central
engine, the outflow probably is magnetically dominated and
is susceptible to global MHD instabilities
(Lyutikov \& Blandford 2003; Giannios \& Spruit 2006).  The field 
lines are then entrained 
in the heated outflow, and stretched back into a non-radial configuration 
outside the Wolf-Rayet envelope.  The fast motion of the jet implies strict
limits on the factor by which the field energy can be reduced this close
to the engine.

\subsection{Plan of the Paper}

The plan of this paper is as follows.  We begin in \S \ref{param}
by defining the inner boundary conditions of the relativistic outflow,
and how they are influenced by the interaction between jet and 
Wolf-Rayet envelope.  Basic constraints on the tangling of the jet
magnetic field interior to the stellar photosphere are outlined
in \S \ref{tangle}.  The fundamental relation between black-body
temperature an isotropic luminosity is worked out in \S \ref{seed}.
The baryons and the magnetic field are tightly coupled, and are
accelerated together by the flux of thermal photons outside the
baryonic photosphere (\S \ref{contam}).  Next we examine the 
interaction between the relativistic shell and the Wolf-Rayet
wind (\S \ref{shockdec}), with a particular focus on the
radiative compactness in the deceleration zone
(\S \ref{comp}) and optical depth in electron-positron
behind the forward shock (\S \ref{taup}).  These calculations are
repeated in \S \ref{sheath} in the case where the fragmenting
`breakout' shell is the dominant source of inertia.  

The evolution of the magnetic
field near the contact discontinuity is the subject of \S \ref{fieldyn}.
We consider whether a reverse shock will form, and how its structure
is modified by small-scale structure in the magnetic field (\S \ref{shockrel}).
We note that magnetic reconnection
across a neutral sheet can be facilitated by rapid synchrotron cooling
in some circumstances (\S \ref{recon}).
The Rayleigh-Taylor instability of the contact discontinuity is
analyzed in \S \ref{rtcon}.  The instability
is weak except when a dense breakout shell is present.  

We argue in \S
\ref{aldamp} that the damping of torsional MHD waves in a background
radiation field provides a
promising approach to the non-thermal and strongly 
variable emission of gamma-ray bursts.  The section begins 
(\S \ref{electro}) by reviewing the fact that electrons and positrons will be
electrostatically accelerated along the background magnetic field
by high-frequency waves near the inner scale of the turbulent spectrum
(Thompson \& Blaes 1998; Lyutikov \& Thompson 2005).
It is argued that electrostatic heating occurs primarily through 
many small impulses, at a scale somewhat larger than
the scale at which the torsional waves become charge-starved.  The
beaming of the inverse-Compton radiation of the heated charges is
calculated in \S \ref{compcool}.  When the Kolmogorov energy flux
is above a critical value, the electrostatic heating cannot be balanced
instantaneously by Compton cooling  (\S \ref{flasheat}).  

The beaming of the inverse-Compton emission provides a nice explanation for
the rapid variability of the gamma-ray flux at higher energies.
The correlation between variability timescale and photon energy is
analyzed in \S \ref{evar}.  
The effect of inverse-Compton scattering on pair creation in the 
relativistic ejecta is considered in \S \ref{gamgam}, and it is noted
that strong anisotropy in the gamma-ray emission can cause a significant
increase in the pair creation rate.  

The paper concludes in \S \ref{summar}
with a critical summary
of how the model fares in explaining the non-thermal spectra
of gamma-ray bursts.  We show in \S \ref{ic} that an inhomogeneous
distribution of particle energies will naturally arise behind the
reverse shock wave, and can explain the presence of a high-energy
continuum above the peak energy in the spectrum of a typical GRB.
The pairs swept up by 
the foward shock during the deceleration phase leave a radiative
signature in the form of inverse-Compton photons peaking at an
energy $\sim 10^2-10^3$ times the peak burst energy (\S \ref{hetail}).
High-energy spectral measurements will therefore provide a direct
diagnostic of the process of pair-loading and pre-acceleration.
The implications of weak optical emission
in the prompt GRB for the mechanism of particle heating are drawn
in \S \ref{opsync},
the transition from prompt GRB emission to afterglow is addressed
in \S \ref{spectrans} and, finally, the `peculiar' burst 941017
is discussed in light of these results in \S \ref{941017}.

A summary list of our results is provided in \S \ref{conc}.

\section{Geometry and Inner Boundary Conditions of the Relativistic Outflow}
\label{param}

Our approach to the gamma-ray emission problem starts with a
basic simplification:  the relativistic outflow that powers the
GRB suffers only a negligible dissipation
between the central engine and the non-thermal emission
zone at $\sim 10^{14}-10^{15}$ 
cm.  The base of this freely expanding outflow coincides with the
photosphere of the progenitor star.  The temperature in the outflow
is already small enough that thermal pairs have annihilated
(Goodman 1986; Shemi \& Piran 1990), and the entrained baryons and
electrons provide the main scattering opacity.  We now discuss
the relative partitioning of energy between the thermal photons,
magnetic field, and baryons.

In this paper we focus on a Wolf-Rayet progenitor,
with a radius of $\sim 2\times 10^{10}$ cm.  The star is a source of a strong
stellar wind, $\dot M_w \sim 0.5\times 10^{-5}-10^{-4}\,M_\odot$ yr$^{-1}$ 
(Nugis \& Lamers 2000) that
provides an effective stopping agent for the relativistic ejecta.
The central engine is a $\sim 2-3\, M_\odot$ 
black hole that is surrounded by a neutronized torus (Woosley 1993).
This torus is fed by the collapse of the inner core of the Wolf-Rayet
star.  Accretion and the emission of a Poynting-dominated jet continues
until the star explodes as the result of the energy deposited in it
by the jet (Paczy\'nski 1998).  The duration of the 
outflow is not lengthened compared with the period of activity
$\Delta t$ of the central engine, and so we leave $\Delta t$ as
a free parameter, normalized to the observed durations of long
GRBs, $\Delta t \sim 10$ s in the cosmological rest frame. 

It should be noted that careful studies of the X-ray/optical/radio
afterglow radiation of GRBs have called into question the
presence of a dense Wolf-Rayet wind environment in several burst events.
A lower mass-loss rate is, indeed, expected from Wolf-Rayet stars of
lower metallicity (Chevalier, Li, \& Fransson 2004); but the mass-loss
rates from luminous WC stars are also often inferred to be higher than 
$\sim 10^{-5} M_\odot$ yr$^{-1}$. 
Observations of strong radio emission at $\sim 1$ month post-burst
can place interesting constraints on the wind density (e.g.
Chevalier et al. 2004), but under some assumptions, e.g.
that the magnetic field energy fraction $\varepsilon_B$ is constant
between the early and later synchrotron-emitting phases.  (The
wind density derived for GRB 020405 scales as 
$\dot M \sim (\varepsilon_{B,\rm 1~month}/\varepsilon_{B,\rm 1~day})^2$:  
see eqs.[1] and [2] of Chevalier et al. 2004.)

The sharpness of the break in the optical
light curve provides a valuable probe of the density profile surrounding
the burst source (Panaitescu \& Kumar 2000, 2001, 2002).
In the case of a jet with non-radial structure, it is difficult
to reproduce the sharpness of the break in about 1/2 of events, if
the energy flux but not the Lorentz factor is allowed to vary as
a power-law about the jet axis (Panaitescu 2005).  On the other hand,
the Lorentz factor $\Gamma_{\rm rel}$ of the jet material is expected also to 
vary about the jet axis.  Some calculations have included a strong 
angular gradient
in $\Gamma_{\rm rel}$ (Kumar \& Granot 2003), but more general
distributions of $\Gamma_{\rm rel}$ have not been explored.
Recently Swift has made available simultaneous X-ray and optical light
curves which have revealed different break patterns in these bands.
GRB 050525a provides an example of an afterglow showing a much sharper
break in the optical band than in the X-ray band $\sim 10^4$ s post-burst
(Blustin et al. 2006). 
The implications for the density profile from jet break phenomenology
therefore appear to us to be somewhat ambiguous (see also Yost et al. 2003).
Given these caveats, we will focus solely on 
a $r^{-2}$ density profile in this paper and generally normalize $\dot M$ to 
a value $\sim 10^{-5}\,M_\odot$ yr$^{-1}$.  The implications of
more dilute winds for the dissipation of the magnetized fireball
are examined briefly in \S \ref{spectrans}.

The simplest relativistic jet expands ballistically into a fixed
solid angle.  A linear increase of Lorentz factor with
radius can be a good approximation to the actual 
situation even if the magnetic field is dynamically dominant close 
to the central engine.  The transfer
one half of the magnetic energy to thermal radiation allows the outflow
to continue to accelerate by pressure gradient forces inside its photosphere
(Drenkhahn \& Spruit 2002). We argue here that this process can be
effective even when the thermalization is localized at some small radius,
and that the acceleration can continue outside the outflow photosphere 
if the matter loading is very small.  One then has
\be\label{gamvsr}
\Gamma(r) = \Gamma(R_\star)\,\left({r\over R_\star}\right),
\ee
up to some maximum Lorentz factor $\Gamma_{\rm max}$ that depends
on the thermalization radius.    
It is $\Gamma_{\rm max} \sim 10^2$ for a jet emerging from a 
Wolf-Rayet envelope.  The remaining irregularities in the magnetic field
are then frozen in, and fall out of causal contact either because $\Gamma$
saturates, or because the fluid passes through the reverse shock 
wave (Thompson 1994; Lyubarsky \& Kirk 2001).  

\subsection{Tangling of a Dynamo-Generated Magnetic Field}\label{tangle}

The most plausible mechanism for generating a baryon-poor outflow invokes a 
large-scale ordered magnetic field threading the black hole
(Blandford \& Znajek 1977), but such a Poynting-dominated outflow
can undergo global current-driven instabilities as it propagates through
the envelope of the Wolf-Rayet star.  In our approach to the burst
emission problem, these instabilities are not the direct
source of the gamma-ray emission (as suggested by Lyutikov \& Blandford
2003), but they will facilitate a tangling up of the magnetic field
inside the Wolf-Rayet photosphere.   The limitations
of causality prevent the ordered Poynting flux from decreasing to
a small fraction of the total energy flux during the propagation
of the jet through the Wolf-Rayet envelope.
The field can be smoothed over a lengthscale $\Delta r_B'$ (in the bulk
frame) only if the Alfv\'en crossing time $\Delta r_B'/V_A'$ is
shorter than the flow time $t'$.  In a cold medium, the 
Alfv\'en speed can be normalized to the speed of light through the
fraction $\varepsilon_B'$ of the energy density carried by the magnetic 
field:  $V_A'/c \simeq (2\varepsilon_B')^{1/2}/(1+2\varepsilon_B')^{1/2}$.
The duration of a flow that expands with Lorentz factor
$\Gamma$ is $t' \sim r/c\Gamma$, and so one obtains  a lower bound
$\varepsilon_B' \ga (\Delta r_B'/ct')^2 \sim (\Gamma\Delta r_B'/r)^2$ 
on the magnetic energy density.  If the field starts out with a reversal
scale $\Delta r_{B0}'$ and $\varepsilon_{B0} \sim 1$, and if its
reversals are random, then conservation of 
magnetic flux implies that $\varepsilon_B' \sim 
\Delta r_{B0}'/\Delta r_B'$.  Combining these expressions gives
\be\label{epsb}
\varepsilon_B' \ga \left({\Delta r_{B0}'\over r/\Gamma}\right)^{2/3}.
\ee

The magnetic field that is carried outward by the relativistic outflow
is ultimately generated by
a dynamo process within the central engine.  The fluid shear
in the neutron torus surrounding a $2-3\,M_\odot$ black hole 
with a modest spin has a
characteristic timescale of $t_{\rm dyn} \sim 10^{-3}$ s.  The
convection motions inside a nascent neutron star have a similar
duration.  The sign of the magnetic field that is advected
into the outflow will vary with the sign of the dynamo generated
field (Thompson 1994).  The timescale for field reversals must be 
longer than the flow time $r/c \sim 10^{-4}$ s at the base of 
the outflow; but simulations of the MRI instability also suggest
that it will be much shorter than the total duration
$\Delta t \sim 10$ s of the inflow to the black hole from the
collapsing core of the progenitor star
(Woosley 1993; Paczy\'nski 1998).  This means that the
characteristic radial scale $\Delta r_B$ for toroidal field
reversals in the outflow will not be much smaller than the radius $R_\star$ 
of the Wolf-Rayet star.  Taking $\Delta r_{B0}/c \sim 0.1$ s
(before untangling) and $R_\star = 2\times 10^{10}$ cm gives 
$\Delta r_{B0}/R_\star \sim 0.15$ and
\be
\varepsilon_B' \ga 0.3\,\Gamma^{4/3}\,
\left({\Delta r_{B0}/c\over 0.1~{\rm s}}\right)^{2/3}
\ee
from eq. (\ref{epsb}).
This lower bound is large enough that the field will have a significant
effect on internal shocks farther out (\S \ref{shockrel}).

\subsection{Seed Blackbody Photons and the $E_{\rm peak}-L_{\rm iso}\,
\Delta t$ Relation}\label{seed}

A nearly black body photon gas is generated by the damping of turbulent
motions in the jet.  This seed radiation field can carry
a significant fraction $\varepsilon_{\rm bb}$ of the outflow luminosity $L$.  
At the stellar photosphere, $r = R_\star$, 
the rest-frame temperature $T_{\rm bb}'$ of the photons is given by
\ba\label{bbrel}
{4\over 3}\Gamma^2\, a {T'_{\rm bb}}^4 &=& \varepsilon_{\rm bb}\, 
{L\over \Delta\Omega R_\star^2 c} =
\varepsilon_{\rm bb}\,{L_{\rm iso}\over 4\pi R_\star^2 c};\nn
k_{\rm B} T'_{\rm bb} &=& 46\,{(\varepsilon_{\rm bb}L_{50})^{1/4}\over
\Gamma(R_\star)^{1/2}}\,\left({\Delta\Omega\over 10^{-2}}\right)^{-1/4}
\;\;\;\;{\rm keV}.
\ea
Here $L_{\rm iso} = L\,(4\pi/\Delta\Omega)$ is the isotropic luminosity
of a two-sided jet flowing through a solid angle $\Delta\Omega = 
2\times\pi\theta^2$.
Note that when the Lorentz boost of the jet material
is taken into account, the temperature of the blackbody component
could be as large as 
${4\over 3}\Gamma(R_\star)T'_{\rm bb} \sim 100$ keV.
It therefore
provides a useful seed for the prompt gamma-ray emission.

As we now argue, the jet material that develops a high 
entropy inside the Wolf-Rayet star is favored as an observable
source of rapidly variable GRB emission for two reasons.
First, one obtains a simple relation between isotropic
luminosity $L_{\rm iso}$ and the mean energy per photon 
carried by the thermal component of the outflow.   This agrees 
closely with the relation between $E_{\rm peak}$ and isotropic 
burst energy $L_{\rm iso}\Delta t$ that was obtained by Amati et al. (2002)
from a sample of BeppoSAX bursts with measured redshifts, and 
confirmed by HETE-II burst localizations (Lamb et al. 2005).  
Second, the strong momentum flux of the black body photons allows this
material to be accelerated to a higher Lorentz factor outside
the Wolf-Rayet photosphere.  The limiting Lorentz factor and the
maximum tolerable matter loading are considered in the following
section \ref{contam}.

Jet material
moving at a Lorentz factor $\Gamma \gg \theta^{-1}$, where $\theta$
is the angle from the axis of the jet, will sustain only a small
velocity shear in the fluid rest frame, and will remain cold.  
Material that is decelerated to $\Gamma \ll \theta^{-1}$ mixes easily
with the much heavier cocoon material and decelerates to $\Gamma \sim 1$.
At an angle $\theta$, there is therefore a characteristic Lorentz
factor $\Gamma \sim 1/\sqrt{3}\theta$ above which a Kelvin-Helmholtz mode will
not have time to grow on a lengthscale $\sim \theta R_\star$
(in the bulk frame).  (Here we have equated the sound-crossing time
$\theta R_\star/c_s = \sqrt{3}\theta R_\star/c$ with the flow time 
$R_\star/\Gamma c$ in the bulk frame.)   

If the entire jet fluid were to decelerate to $\Gamma \sim 1$ within
the stellar envelope, then the relation
between temperature and (isotropic) luminosity would follow the usual
blackbody relation, $T_{\rm bb} \sim L_{\rm iso}^{1/4}$.  The 
temperature-luminosity relation becomes harder with $\Gamma$ at the 
above critical value.   Observations of GRB afterglows suggest
that the total energy $E$ carried by the jet is roughly independent of
opening angle (Frail et al. 2001; Bloom et al. 2003).  In that case, one has
\be\label{tliso}
  L_{\rm iso}\Delta t\;\left({\Delta\Omega\over 4\pi}\right)
  = L_{\rm iso}\Delta t\,\left({\theta^2\over 2}\right) 
  = 5\times 10^{50}~{\rm ergs}.
\ee
The corresponding Lorentz factor is then
\be\label{gamvsl}
\Gamma(R_\star) = {1\over \sqrt{3}\theta} = 
1.8\,L_{\rm iso\,51}^{1/2}\,\Delta t_1^{1/2}.
\ee
Combining eqs. (\ref{bbrel}) and (\ref{tliso}) gives
\be\label{trest}
k_{\rm B}T_{\rm bb}' = 10\,\varepsilon_{\rm bb}^{1/4}\,
        \Delta t_1^{-1/4}\, \left({R_\star\over 
      2\times 10^{10}~{\rm cm}}\right)^{-1/2}\;\;\;\;{\rm keV}.
\ee
This temperature is low enough that the thermal creation of electron-positron
pairs can be neglected.  
The relation between Lorentz-boosted temperature and isotropic
luminosity can then be expressed in terms of the mean energy per
black-body photon, 
\ba\label{amati}
2.7\,k_{\rm B} T_{\rm bb} &=& 2.7\left[{4\over 3}\,\Gamma(R_\star)\right]
 k_{\rm B} T'_{\rm bb}\nn
  & =& 70\,{\varepsilon_{\rm bb}^{1/4}\over \Delta t_1^{1/4}}\,
      \left({L_{\rm iso}\Delta t\over 10^{52}~{\rm ergs}}\right)^{1/2}\,
      \left({R_\star\over 
      2\times 10^{10}~{\rm cm}}\right)^{-1/2}\;\;\;\;{\rm keV}.\nn
\ea
If one substitutes
$\varepsilon_{\rm bb} \sim {1\over 2}$ and takes into account
that the isotropic luminosities of long GRBs
appear to have a much broader distribution than do their
durations\footnote{The observed $T_{90}$ distribution covers the range
$4-40$ s at half maximum (Paciesas et al. 1999), which 
of couse includes the broadening
effects of a distribution of source redshifts.}, then 
one very nearly reproduces the observed relation (Amati et al. 2002;
Lamb et al. 2005).  
The normalization is $\sim 60\,\%$ of the measured value, but
an additional $\sim {1\over 2}$ of the outflow energy must be carried
by a separate component, whose dissipation generates the
high-energy, non-thermal tail to the gamma-ray burst spectrum.
We emphasize that this relation depends on only one free parameter, 
the radius of the Wolf-Rayet star, which is of course
contrained independently.  

The inferred width of the burst energy distribution depends on
the details of the beaming correction that is employed.  The most
detailed effort, by Ghirlanda et al. (2004), indicates a spread of
$\sim 1$ order of magnitude in the case of a $r^{-2}$ circumburst
density profile.  
However, Swift has observed additional time-structure in the
afterglow light curves at times intermediate between the burst
duration $\Delta t \sim 10-10^2$ s, and the later achromatic breaks
that were used by Ghirlanda et al. (2004) to normalize the jet opening
angle (Tagliaferri et al. 2005).  
There remains, therefore, some ambiguity in the mapping between
the prompt gamma-ray emission and the later phases of the afterglow.

The distribution of burst energies also depends on the mechanism
by which the jet is collimated.  The beaming-corrected burst energies
are remarkably similar to the typical kinetic energy of a core-collapse
supernova (Frail et al. 2001).  They are also similar to the gravitational
binding energies of the cores of 20-30 $M_\odot$ evolved stars
(typically $E_B \sim 1-2
\times 10^{51}$ ergs; see Figs. 5 and 6 of Woosley \& Weaver 1995).  
Current modelling of Blandford-Znajek jets from black holes suggests
that strong collimation requires the presence of a geometrically thick torus
around the hole (e.g. McKinney \& Gammie 2004).  The extent to which
this condition is satisifed in hyper-accreting tori
(accretion rates $\dot M \sim 2\,M_\odot/10~{\rm s} = 0.2\, M_\odot$ s$^{-1}$)
depends on the density profile in the outer atmosphere of the torus, which
in turn is sensitive to the dissipative processes operating there.
A geometrically thin, neutrino-cooled disk can form self-consistently
at this accretion rate (e.g. Narayan, Piran, \& Kumar 2001).  When 
the outflow is broad close to the engine, strong collimation 
is possible only if the outflow energy is comparable to $E_B$:
if the outflow energy is smaller, it will be swept back by the
accretion flow, and if it is substantially
larger it will trigger a quasi-spherical blast wave.  These considerations
provide a physical motivation for our assumption of a relatively narrow
distribution of burst energies, compared with current
beaming corrections (Ghirlanda et al. 2004).

Rees \& M\'esz\'aros (2005) have outlined a related approach to the
$E_{\rm peak}-L_{\rm iso}\,\Delta t$ 
relation, starting with the idea that the peak 
energy is regulated by dissipation and thermalization just inside the 
photosphere of the outflow.  The radius of the photosphere 
is determined self-consistently by the luminosity and composition
of the outflow, the scaling of
$\Gamma$ with radius, and the particular radiative processes occurring
in the outflow.  As a result, definite predictions for the slope
of the relation, or its normalization, are not easily obtained.
Note also that the 
causal relation $\Gamma \sim 1/\sqrt{3}\theta$ is more difficult to satisfy 
this far out in the outflow.  
For all of these reasons, we conclude that
the Amati et al. (2002) relation arises more naturally at a {\it fixed}
radius, during jet breakout from the Wolf-Rayet star.  The prompt gamma-ray
emission is, in this picture, reprocessed `fireball' radiation, with the
position of the fireball displaced outward from the radius of the 
black hole accelerator ($\sim 10^7$ cm) to its surrounding stellar envelope
($\sim 10^{10}$ cm).

We expect that the
relation (\ref{amati}) between $T_{\rm bb}$ and $L_{\rm iso}\Delta t$
will be modified at low luminosities, because
its derivation depends on $\Gamma$ being larger than unity.
One sees from eq. (\ref{gamvsl}) that this assumption must break down
below a luminosity $L_{\rm iso} \sim 3\times 10^{50}\,\Delta t_1^{-1}$ 
ergs s$^{-1}$.  At lower luminosities,  the relation 
is predicted to soften
to $T_{\rm bb} \sim L_{\rm iso}^{1/4}$ when the jet material
moves transrelativistically [$\Gamma(R_\star) \simeq 1$].

An apparent inconsistency between the Amati et al. (2002) relation and
the BATSE burst sample has been uncovered by Nakar \& Piran (2005).
These authors find that $\sim {1\over 2}$ of the BATSE
bursts are harder spectrally than the $E_{\rm peak}-L_{\rm iso}\,
\Delta t$ relation
would imply, no matter where each burst source is placed in redshift space.
Note that the BATSE sample has a lower flux threshold
than the BeppoSAX sample, so that the bulk of the BATSE bursts need not
sit in the $E_{\rm peak} \propto L_{\rm iso}^{1/2}$ portion of the
color-luminosity relation.  Our predicted modification of the Amati 
et al. (2002) relation, at energies lower than $E_{\rm iso} \sim 
3\times 10^{51}$ ergs, could therefore resolve this apparent contradiction
without recourse to beaming or orientation effects.

The $T_{\rm bb}-L_{\rm iso}$ relation that we have derived is also
consistent with a `structured' jet, but we note that 
a range of Lorentz factors within the jet core is required.  The published
structured-jet models generally assume a distribution of
energy flux across the jet, but a constant Lorentz factor.
In practice, there is no reason to expect that $\Gamma$ varies
much more weakly than $L_{\rm iso}$.  The kinematic
model considered here 
implies that $\Gamma \propto L_{\rm iso}^{1/2}\Delta t^{1/2}$ at the boundary
of the Wolf-Rayet star.  The asymptotic Lorentz factor of the 
freely-expanding jet material 
scales as $\Gamma \propto L_{\rm iso}^{3/8}\Delta t^{1/8}$
(see eq. [\ref{gambcrit}] below).

\subsection{Baryon Contamination and Terminal Lorentz Factor}\label{contam}

Effective thermalization of the radiation field at a temperature
(\ref{trest}) requires the presence of a small baryon component,
since the temperature (\ref{trest}) is too small to
excite pairs thermally.  The mass in
baryons depends, in turn, on the details of the launching of the jet,
and can be expected to vary from burst to
burst (depending especially on the viewing angle if additional
material is mixed from the envelope).   The baryon loading
of the outflow is represented by the parameter 
\be\label{gamb}
\Gamma_b = {L\over\dot M_bc^2}.
\ee
The associated electrons provide a scattering depth
\be\label{taub}
\tau_{{\rm T},b} = {\sigma_{\rm T}Y_e L_{\rm iso}\over 
8\pi \Gamma_b \Gamma^2(R_\star) m_p R_\star c^3}
\left({r\over R_\star}\right)^{-3}.
\ee
Here $Y_e \sim {1\over 2}$ is the ratio of the number of protons
to total nucleons that are bound up in charged ions.  (The role of
free neutrons is briefly addressed below.)

A substantial optical depth at radius $R_\star$ is needed if
the photon distribution function is to relax to a black body.
The jet material is still optically thin to free-free absorption
at the temperature (\ref{trest}),  and so
effective thermalization requires a
Compton parameter $y_{\rm C} =  4(k_{\rm B}T_{\rm bb}'/m_ec^2)\tau_{{\rm T},b}
\ga 5$.  Thermalization is therefore possible only if the baryon loading
is larger than
\be
\Gamma_b \la 7\times 10^4\,\varepsilon_{\rm bb}^{1/4}\,\Delta t_1^{-5/4}
\ee
at $r = R_\star = 2\times 10^{10}$ cm.  
The large scattering depth for thermalization means that
the blackbody photons must be created within the jet.
More generally, unless $\Gamma_b$ is extremely large and the jet is
extremely clean, the jet core will not see the soft thermal photons
that are created in the surrounding cocoon (cf. Ghisellini et al. 2000).

Above a critical baryon density, the Lorentz-boosted photon temperature
${4\over 3}\Gamma T_{\rm bb}'$ will be cooled by adiabatic 
expansion before the photons and matter decouple.
To avoid this, one requires that the scattering depth (\ref{taub})
drop below unity before $\Gamma$ reaches $\Gamma_b$.  
Combining eqs. (\ref{gamvsr}), (\ref{gamvsl}), and (\ref{taub}) 
gives the required lower bound on $\Gamma_b$,
\ba\label{gambcrit}
\Gamma_b > \Gamma_{b\,\rm crit} &=&
\left[{\sigma_{\rm T} Y_e \varepsilon_{\rm bb} L_{\rm iso} \Gamma(R_\star)
\over 8\pi m_pc^3 R_\star}\right]^{1/4}\nn
&=& 85\,L_{\rm iso\,51}^{3/8}\,\Delta t_1^{1/8}\,
(Y_e\varepsilon_{\rm bb})^{1/4}
\left({R_\star\over 2\times 10^{10}~{\rm cm}}\right)^{-1/4}.\nn
\ea
The key point here is that $\Gamma_{b\,\rm crit}$ depends
on the launching radius of the freely expanding jet.
Increasing $R_\star$ from $\sim 10^6-10^7$ cm (the physical size of the
central engine) to $R_\star \sim 10^{10}$ cm has the effect of
reducing $\Gamma_{b\,\rm crit}$ by an order of magnitude.

The magnetic field carries inertia, which can limit the
bulk motion of the outflow if it is cold at its
base.  The role of thermal pressure in accelerating the outflow 
has been examined by Drenkhahn \& Spruit (2002).  At a large
optical depth, the radiation field is advected with the other components
of the flow, and a pressure-gradient force is responsible for pushing 
the flow to a higher Lorentz factor.  Although some acceleration is 
possible by this mechanism in the optically thin region of the flow,
we focus here on the force imparted by the radiation field
to the magnetofluid.  This effect is already familiar from studies
of particle-dominated fireballs, and is especially important if
the outflow undergoes strong heating near its base.
It is straightforward to see that the limiting Lorentz factor 
(\ref{gambcrit}) is not modified significantly by the magnetic field
as long as the particle Lorentz factor $\Gamma$
is much larger than\footnote{Here $\rho$ is the rest energy
density of the entrained charges in the frame of the engine.} 
$(B^2/8\pi \rho c^2)^{1/3}$.  In a steady
state, both the Poynting flux $S = (B^2/8\pi)v_r$ and the momentum flux 
$P = (B^2/8\pi)(1+v_r^2/c^2)$
carried by the non-radial magnetic field $B$ receive (negative) corrections
$\Delta S/S \simeq \Delta P/P \simeq -\Delta\Gamma/\Gamma^3$
in response to a change $\Delta\Gamma$ that is driven by Compton scattering.
When $\Gamma$ satisfies the above inequality, both $\Delta S$ and $\Delta P$ 
are negligible compared with the increase $\rho c^3\Delta\Gamma$ 
in the particle kinetic energy flux.   Further out in the flow,
the inertia of electron-positron pairs can dominate the inertia
of the baryons,
and some of the energy carried by the magnetic field is
converted to non-thermal photons.

It is possible to set an upper bound to the baryon kinetic luminosity,
and to the terminal Lorentz factor of outflow.
Outside the photosphere of the Wolf-Rayet star,
the baryons will continue to accelerate off the flux of thermal photons.
The collimation of the photons grows with radius,
$\Gamma_\gamma(r) \simeq \Gamma(R_\star)\,(r/R_\star)$.  The baryons
maintain a Lorentz factor $\Gamma \simeq \Gamma_\gamma$ as long as
\be
\Gamma_\gamma \la {1\over 2}\ell_\gamma = {Y_e\,\sigma_{\rm T}\,
     (\varepsilon_{\rm bb} L_{\rm iso})\over 8\pi r^2 m_p c^2}\,
      \left({r\over 2\Gamma_\gamma^2c}\right).
\ee
This leads to a maximum Lorentz factor essentially equal to eq. 
(\ref{gambcrit}) (e.g. Shemi \& Piran 1990).  When 
$\Gamma_b > \Gamma_{b\,\rm crit}$, the Lorentz factor saturates at a value 
\be
\Gamma \simeq \Gamma_{b\,\rm crit}\;\;\;\;\;\;\;\;\;\;\;\;
        (\Gamma_b \gg \Gamma_{b,\rm crit}),
\ee
and the baryons carry only a fraction
$\varepsilon_b \simeq \Gamma_{b\,\rm crit}/\Gamma_b$
of the total energy of the outflow.   
The proportions of the energy flux carried by thermal photons, baryons, 
and magnetic field must of course sum to unity,
\be
\varepsilon_{\rm bb} + \varepsilon_b + \varepsilon_B = 1.
\ee
In a cold flow, $\varepsilon_B$ is related to the rest-frame magnetic
energy density $\varepsilon_B'$ by $\varepsilon_B \simeq 2\varepsilon_B'$.  

An important feature of this acceleration mechanism is the 
correlation between the presence of a strong thermal photon flux at the
base of the jet, and the development of a large terminal Lorentz factor
outside the Wolf-Rayet envelope. Portions of the jet which dissipate less 
inside the Wolf-Rayet star may attain a lower terminal Lorentz factor,
and their deceleration will be pushed to a larger radius.

Neutrons play a limited role in this jet model.  Advected outward
from the central engine, they are entrained by the protons and
alpha particles at the Wolf-Rayet photosphere, and decouple from
them somewhat inside the baryonic scattering photosphere.  
The neutron mass flux $\dot M_n$ can be much larger
than the proton mass flux, due to the effects of neutronization
in the central engine (e.g. Beloborodov 2003).  
We take a (representative) ratio $\dot M_n/\dot M_p = 8$, which 
corresponds to $\dot M_n/\dot M_\alpha = 4.5$ if all
of the protons have combined into alpha particles.  The neutrons then
decouple from the alpha particles just after the combined 
neutron-ion fluid (temporarily) reaches the limiting Lorentz
factor 
\ba
\Gamma_n &=&  {L_{\rm iso}\over (\dot M_n+\dot M_\alpha)c^2}\nn
   &=& 13\,L_{\rm iso\,51}^{3/8}\,\Delta t_1^{1/8}\,
     \varepsilon_{\rm bb}^{1/4}
\left({R_\star\over 2\times 10^{10}~{\rm cm}}\right)^{-1/4}\,
\left({\Gamma_b\over \Gamma_{b\,\rm crit}}\right).\nn
\ea
The neutrons fall behind the outer half of the ejecta shell at a radius 
$\Gamma_n^2c\Delta t = 5\times 10^{13}\,L_{\rm iso\,51}^{3/4}$
$\Delta t_1^{1/4}\,\varepsilon_{\rm bb}^{1/2}\,
   (R_\star/2\times 10^{10}~{\rm cm})^{-1/2}\,
(\Gamma_b/\Gamma_{b\,\rm crit})^2$ cm.
At this radius, a fraction $\ln 2\Gamma_n \Delta t/t_{1/2}
= 0.1\,L_{\rm iso\,51}^{3/8}$
$\Delta t_1^{9/8}(R_\star/2\times 10^{10}~{\rm cm})^{-1/4}$
of the neutrons have decayed.  The effective inertia of the decayed
neutrons (mass $\Delta M_n$) is larger than the
inertia of the alpha particles, $(\Gamma_{b\,\rm crit}/\Gamma_n)\Delta M_n
> M_\alpha$, when $L_{\rm iso} \ga 10^{51}$ erg/s.  Even in this case,
the neutron decay will have a modest effect on the dynamics of
the contact discontinuity, which at this point has a significantly
smaller Lorentz factor than that of the relativistic outflow
(\S \ref{sheath}).  The effect of the neutrons is also limited
when $\Gamma_b$ is larger than $\Gamma_{b\,\rm crit}$ and the
inertia of the outflow is dominated by an advected magnetic field.

\section{Deceleration of the Relativistic Outflow}\label{outdecel}

We now consider the interaction between the relativistic 
ejecta and the Wolf-Rayet wind.  The ejecta form a flattened shell outside
a distance $c\Delta t$ from the star.  A thin layer of shocked
wind material collects in front of the ejecta shell, and is preceded
by a forward shock wave.   We focus first on the idealized,
one-dimensional treatment of this problem, and defer consideration of
Rayleigh-Taylor instabilities to \S \ref{sheath} and \S \ref{rtcon}.

There is an extensive treatment of the dynamics of the reverse shock wave
in the literature, assuming ideal shock jump conditions (Sari \& Piran
1995; M\'esz\'aros \& Rees 1997; Sari \& Piran 1999; Nakar \& Piran 2004),
and taking into account the effects of magnetic pressure
in that approximation (Zhang \& Kobayashi 2005).  Our goal here is to
include i) the effects of pair loading and preacceleration of the medium
ahead of the forward
shock; and ii) the possibility of strong dissipation of the non-radial magnetic
field behind the reverse shock.  The first effect is an inevitable consequence
of the observed gamma-ray emission if it occurs inside $\sim 10^{16}$ cm; 
and the second is necessary if the emission is triggered by the 
interaction of the ejecta with the external medium (cf. \S \ref{spectrans}). 
The relativistic outflow will also interact with the thin shell of material 
that is collected by the jet head following breakout.  We defer a discussion
of the associated effects to \S \ref{sheath}, after we have analyzed
the interaction with the expanding Wolf-Rayet wind material.

We distinguish between the luminosity $L_{\rm rel}$ carried by the 
particles and magnetic field; and the total isotropic luminosity
$L_{\rm iso}$, which also has a contribution from the seed thermal photons.  
They are related by
\be
L_{\rm rel} = (1-\varepsilon_{\rm bb})L_{\rm iso}.
\ee
The mass profile in the Wolf-Rayet wind is 
\be
\rho_w(r) = {\dot M_w\over 4\pi r^2 V_w}
\ee
when the mass loss rate $\dot M_w = \dot M_{w\,-5}\times 10^{-5}\,M_\odot$ 
yr$^{-1}$ and wind speed $V_w = V_{w8}\times 10^8$ cm s$^{-1}$ are 
constant over the emission region.  The mass swept up from the wind
inside a radius $r$ is
\be\label{mwr}
M_w(r) = {\dot M_w r\over V_w}.
\ee
A relativistic jet emerging from the photosphere of a Wolf-Rayet
star will also collect a thin shell of matter from the stellar
envelope, which can exceed $M_w(r)$ for sufficiently small $r$
(\S \ref{sheath}).

\subsection{Dynamics of the Contact Discontinuity}\label{shockdec}

Pair creation by gamma-rays streaming into the 
external medium has a strong influence on the
propagation of the forward shock (Thompson \& Madau 2000;
M\'esz\'aros, Ramirez-Ruiz, \& Rees 2001; Beloborodov 2002).
What has not been examined is the feedback of the pre-acceleration
of the pair-loaded Wolf-Rayet wind on the dissipative processes
occurring in the relativistic outflow.
We show here that the the bulk of the dissipation is delayed to a radius 
where the relativistic motion of the ambient medium has nearly, but
not completely, decayed away.  As
a result, the compactness of the radiation emitted behind the
forward shock is regulated to a characteristic value that depends
weakly on $L_{\rm rel}$.  

The structure of the flow near the contact is displayed in Fig. 1.
If the wind material
were held static before passing through the forward shock, then
$\Gamma_c$ would be obtained by moving to the rest frame of the contact
and balancing the momentum flux on both sides,\footnote{This expression
holds if the relativistic outflow and the ambient medium are
both cold and particle-dominated 
in their respective rest frames.  This condition will be
approximately satisfied in the ambient medium even after it is loaded with 
electron-positron pairs and pre-accelerated (Beloborodov 2002).  
The momentum balance
at the reverse shock also remains similar when the relativistic outflow is
magnetically dominated, as long as the Poynting flux is converted
efficiently to radiation.  This is plausible if the magnetic field
reverses sign on a small scale (e.g. Thompson 1994; Drenkhahn \&
Spruit 2002).}
\be\label{mombal}
\Gamma_c^2 \rho_wc^2 = \Gamma_c^2(1-\beta_{\rm rel}\beta_c)^2
\left({L_{\rm rel}\over 4\pi r^2}\right) \simeq
{1\over \Gamma_c^2(1+\beta_c)^2}
\left({L_{\rm rel}\over 4\pi r^2}\right).
\ee

We will separate our treatment of the deceleration of the contact
from the acceleration of the relativistic outflow closer to the engine,
and focus on the
regime $\Gamma_{\rm rel} = (1-\beta_{\rm rel}^2)^{-1/2}
\gg \Gamma_c$.  It should be kept in mind that,
in the specific case of a breakout jet from a Wolf-Rayet star,
the seed thermal photons are able to push the baryons and entrained
magnetic field to a limiting Lorentz factor (\ref{gambcrit}) 
that is close to the  values of $\Gamma_c$ derived below. 
Continuing dissipation of the magnetic field combined with pair
creation could push the particles
to yet higher speeds (e.g. Drenkhahn \& Spruit 2002); but the observed
correlation between peak energy and isotropic gamma-ray burst energy
(Amati et al. 2002; Lamb et al. 2005) suggests that this dissipation
is not energetically important in between the photosphere of the Wolf-Rayet
star and the gamma-ray photosphere at $\sim 10^{14}$ cm (eq. [\ref{rtaupm}]).
We therefore approximate $\Gamma_{\rm rel} \gg \Gamma_c \gg 1$ in 
eq. (\ref{mombal}), which gives the equilibrium solution
\be\label{gameq}
\Gamma_c = 
\Gamma_{\rm eq} = \left({L_{\rm rel} V_w\over 4\dot M_w c^3}\right)^{1/4}
= 35\,L_{\rm rel\,51}^{1/4}\,V_{w\,8}^{1/4}\,\dot M_{w\,-5}^{-1/4}.
\ee

\vskip .2in
\centerline{{
\vbox{\epsfxsize=7.5cm\epsfbox{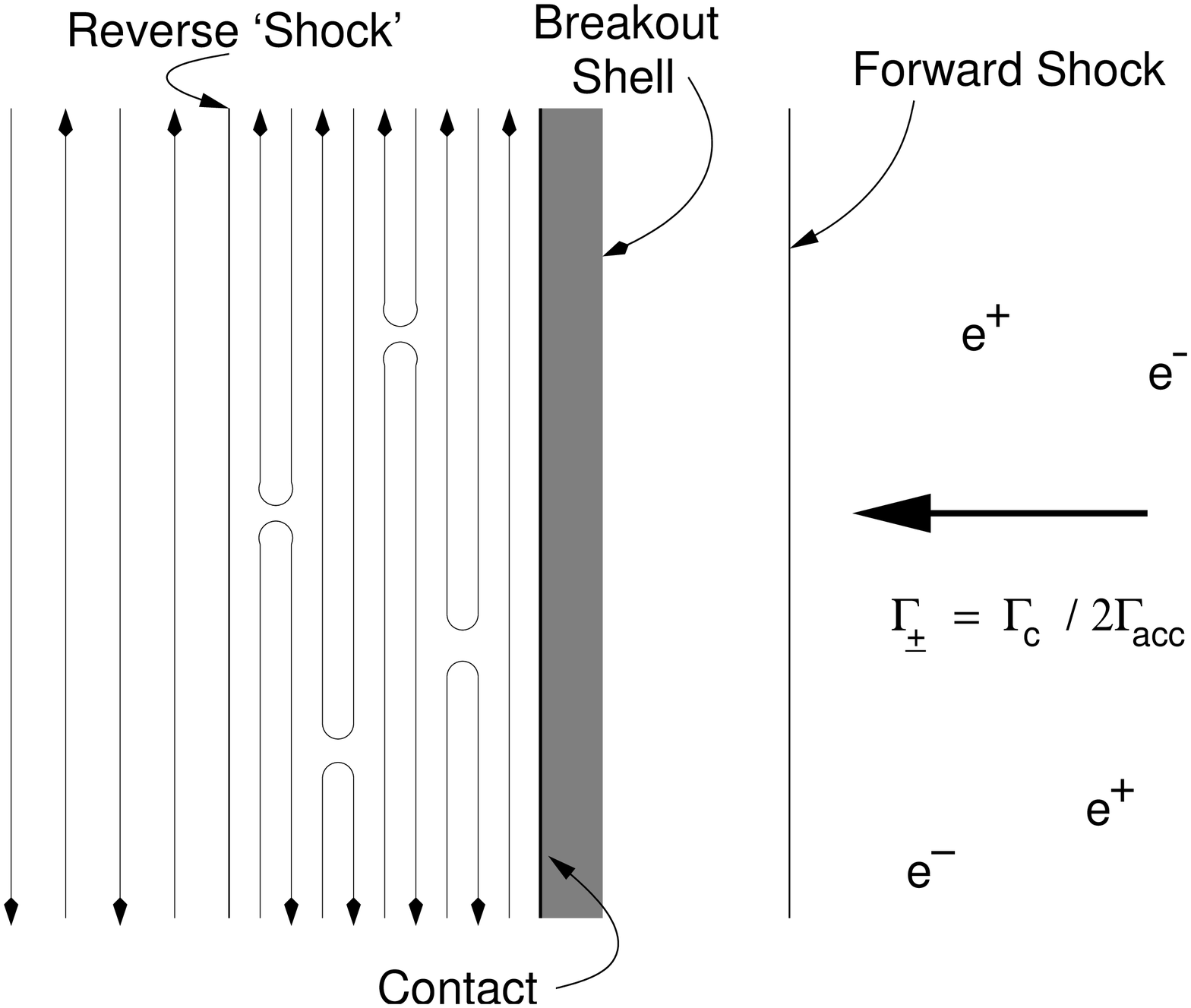}}}}
%\begin{figure}%[ht]
%\epsscale{.90}
%\plotone{fig1.eps}
\figcaption{Structure of the relativistic outflow between the forward
and reverse shock waves.  The ambient medium is loaded with 
electron-positron pairs that are created by side-scattering
of the gamma-ray photons.  These pairs and the ions from the
Wolf-Rayet wind collect in a shell behind the forward shock.
In the idealized one-dimensional flow problem, this 
shell is separated from the shocked shell of relativistic
eject by a contact discontinuity.  The magnetic field that is
advected across the reverse shock wave undergoes forced
reconnection and tangles up.  The reverse shock is, in fact, not
a true shock but has a finite thickness (\S \ref{shockrel}).  
The breakout shell is a thin layer of material from the 
Wolf-Rayet star that is advected out
by the jet head, and becomes transparent at a radius $\sim 10^{14}$
cm (\S \ref{sheath}).}
\label{figcon}
%\end{figure}
\vskip .2in

This result is modified when the effects of pair loading and pre-acceleration
of the Wolf-Rayet wind material are taken into account.  We are interested
in the regime where the wind material reaches a Lorentz factor 
$\Gamma_{\rm amb} \gg 1$ (but still less than 
the Lorentz factor $\Gamma_c$ of the contact). 
The Wolf-Rayet wind material is then swept up by the foward shock
at a radius close to where it was sitting at the moment of the explosion.
Each shell of mass $\Delta M_b$ passing through
the forward shock now deposits a momentum 
$\Gamma_c\Gamma_{\rm amb}(\beta_{\rm amb}-\beta_c)
({\cal M}_\pm \Delta M_b) c
\simeq - (\Gamma_c/2\Gamma_{\rm amb})({\cal M}_\pm \Delta M_b) c$ in 
the frame of the contact.  The left-hand side of eq. (\ref{mombal}) is 
therefore replaced by $(\Gamma_c^2/2\Gamma_{\rm amb})\,{\cal M}_\pm\rho_wc^2$,
where
\be
{\cal M}_\pm(\rho,n_{e^+}) = {2m_en_{e^+}\over \rho} + 1
\ee
corrects for the inertia of the electron positron pairs 
(density $n_{e^+}$) that have been created in the ambient medium
(Thompson \& Madau 2000; Beloborodov 2002).  The shell 
Lorentz factor becomes
\be\label{gamexp}
\Gamma_c = (2\Gamma_{\rm amb})^{1/4}\,\Gamma_{\rm eq}.
\ee
When pre-acceleration is efficient, $\Gamma_{\rm amb} \sim \Gamma_c/2$,
one has
\be\label{gameqb}
\Gamma_c \sim 
{\cal M}_\pm^{-1/3}\,\Gamma_{\rm eq}^{4/3} = 
{\cal M}_\pm^{-1/3}\,
\left({L_{\rm rel} V_w\over 4\dot M_w c^3}\right)^{1/3}.
\ee
It turns out that ${\cal M}_\pm \sim 1$ except very near the beginning
of the deceleration phase, and so we define a reference Lorentz factor
\be
\widetilde\Gamma_{\rm eq} = \Gamma_{eq}^{4/3}
= 1.1\times 10^2\,L_{\rm rel\,51}^{1/3}\,V_{w\,8}^{1/3}\,\dot M_{w\,-5}^{-1/3}.
\ee
with the dependence on ${\cal M}_\pm$ removed.  

The baryons provide
a useful set of markers for the flow:  after pre-acceleration their
number density is
\be
n_b \;=\; {\rho_w/m_p\over 1-\beta_{\rm amb}} \;\simeq\;
2\Gamma_{\rm amb}^2\,{\rho_w\over m_p} 
\;\;\;\;\;\;\;\;\;\;\;\;(\Gamma_{\rm amb} \gg 1).
\ee
In the frame of the contact one has
\be\label{barcon}
n_b' =
\Gamma_c\,{\rho_w\over m_p} \;\;\;\;\;\;\;\;\;\;\;\;(\Gamma_{\rm amb} \gg 1).
\ee
The rate at which baryons are swept up by the contact does not
depend on the speed of the ambient medium, as long as the ambient
material moves only a modest distance from its starting point at
the moment of the explosion.  (Pre-acceleration does still
alter the value of $\Gamma_c$ itself and the ram pressure of 
the oncoming flow.)

The outflow that powers the gamma-ray emission in a long burst
is concentrated\footnote{We suggest in \S \ref{spectrans} that
the end of the prompt gamma-ray burst coincides with the termination
of the thermal photon pulse, due to the explosion of the Wolf-Rayet star.
A broader relativistic wind may, therefore, continue to flow 
from the central engine after the GRB ends, but at a diminishing rate due
to the decreasing mass flux onto the central black hole.}
 within a time interval $\Delta t \sim 10$ s and therefore
carries a limited energy $E_{\rm rel} = L_{\rm rel} \Delta t$.
The energy which is swept up within a distance $r$
from the central engine is
\be
\Delta E_{\rm rel}(r) = L_{\rm rel} \left({r\over 2\Gamma_{\rm eq}^2c}\right)
= 2\Gamma_{\rm eq}^2\,M_w(r)c^2,
\ee
with nearly equal proportions being dissipated
behind the forward and reverse shocks.	
We are interested
in the dynamics of the outflow close enough to the engine that electrons
and positrons cool rapidly.  Nonetheless, the ions carry the bulk
of the energy flux across the forward shock in the latter part of
the prompt deceleration phase (\S \ref{taup}), and do not cool
effectively by simple incoherent processes.  About
${1\over 2}$ of the outflow energy is therefore
converted to radiation, and a fraction
$\varepsilon_+ \sim {1\over 4}$ is transported by radiation across
the forward shock,
\be\label{etot}
\Delta E_+ = \varepsilon_+\Delta E_{\rm rel}.
\ee
Above a critical value of the compactness
\be\label{compact}
\ell_+ =  {\sigma_{\rm T}\Delta E_+\over 4\pi r^2m_ec^2} =
\left({\varepsilon_+\over 2\Gamma_c^2}\right)
\,{\sigma_{\rm T} L_{\rm rel}\over 4\pi r m_ec^3},
\ee
the exterior medium is loaded with electron positron pairs and accelerated
to a high Lorentz factor comparable to $\Gamma_{\rm eq}$
(Thompson \& Madau 2000).  This critical
compactness is $\ell_{\rm crit} \sim 100$ when the gamma-ray spectrum
has a high energy tail that extends well above $\sim 1$ MeV 
(Beloborodov 2002).  A Lorentz factor 
$\Gamma_{\rm amb} \sim 10^2$ is attained at $\ell_+ \sim 10^3$. 
We make use of the approximate scaling
\be\label{gamacc}
\Gamma_{\rm amb} \simeq 
\left({\ell_+\over\ell_{\rm crit}}\right)^2\;\;\;\;\;\;(10^2 \la
\ell \la 10^3),
\ee
which agrees at the $\sim 20\%$ level  with the results obtained 
by Beloborodov (2002) for incident spectra with photon energy 
indices $-1.5$ to $-2$.

One can obtain the equilibrium Lorentz factor (\ref{gameqb}) in a simple
way by comparing the energy $\Delta E_+$ of the gamma-rays
passing across the forward shock, with the inertia of the ambient medium.
This implies a maximum Lorentz factor
\be
\Gamma_{\rm amb}({\rm max}) = {\Delta E_+\over {\cal M}_\pm M_w(r) c^2},
\ee
or equivalently,
\be
{\Gamma_{\rm amb}({\rm max})\over\widetilde\Gamma_{\rm eq}} 
= \left({\varepsilon_+\over 2{\cal M}_\pm}\right)\,
   {L_{\rm rel}V_w\over \Gamma_c^2 \widetilde\Gamma_{\rm eq} \dot M_w c^3}
= {2\varepsilon_+\over {\cal M}_\pm}\,
    \left({\Gamma_c\over\widetilde\Gamma_{\rm eq}}\right)^{-2}.
\ee
Substituting this into eq. (\ref{gamexp}), one sees that
$\Gamma_c \sim \widetilde\Gamma_{\rm eq}$ is the largest
Lorentz factor which can be maintained by a balance between momentum
deposition on the two sides of the contact.  

The first phase of deceleration corresponds
to the passage of the reverse shock through the shell.
We now show that this occurs before pre-acceleration
becomes ineffective.  The prompt emission of gamma rays is therefore
directly tied to the end of pair loading of the ambient medium.

The beginning of deceleration can be identified with the radius
$R_{\rm decel\,-}$ at which the ambient medium can no longer be pushed
to a Lorentz factor $\Gamma_{\rm amb} = {1\over 2}\widetilde\Gamma_{\rm eq}$
before hitting the forward shock.  The corresponding compactness 
is (eq. [\ref{gamacc}])
\be\label{elldec}
\ell_+(R_{\rm decel-}) \;\simeq\;
   \left({\widetilde\Gamma_{\rm eq}\over 2}\right)^{1/2}\, \ell_{\rm crit}.
\ee
Expressing $\ell_+$
in terms of $\widetilde\Gamma_{\rm eq}$ using eq. (\ref{compact}) gives
\be\label{rdecelmin}
R_{\rm decel-} = 1.1\times 10^{14}\,\varepsilon_+\,L_{\rm rel\,51}^{1/6}\,
  \dot M_{w\,-5}^{5/6}\,V_{w\,8}^{-5/6}\;\;\;\;{\rm cm}.
\ee
Outside this radius,
\be\label{gamcr}
\Gamma_c(r) = \widetilde\Gamma_{\rm eq}\,
 \left({r\over R_{\rm decel-}}\right)^{-1/4}\;\;\;\;\;\;(r > R_{\rm decel-}).
\ee
The prompt phase of deceleration is complete when
$\int (2\Gamma_c^2)^{-1}dr = r/3\Gamma_c^2 = c\Delta t$, 
which occurs at a radius
\be\label{fdecel}
{R_{\rm decel+}\over R_{\rm decel-}} = 
  \left({3\widetilde\Gamma_{\rm eq}^2c\Delta t\over 
   R_{\rm decel-}}\right)^{2/3} = 20\,\varepsilon_+^{-2/3}\,
\,L_{\rm rel\,51}^{1/3}\,\dot M_{w\,-5}^{-1}\,V_{w\,8}\,\Delta t_1^{2/3},
\ee
or, equivalently, at
\be\label{fdecelb}
R_{\rm decel+} = 2.5\times 10^{15}\,\varepsilon_+^{1/3}\,
L_{\rm rel\,51}^{1/2}\,\dot M_{w\,-5}^{-1/6}\,V_{w8}^{1/6}\,\Delta t_1^{2/3}
\;\;\;\;{\rm cm}.
\ee
Notice that $R_{\rm decel+}$ depends more strongly on the relativistic
outflow parameters ($L_{\rm rel}$, $\Delta t$) 
than it does on the pre-burst wind parameters.  
At this point, the Lorentz factor of
the contact has dropped to 
\be\label{gamrdecel}
\Gamma_c(R_{\rm decel+}) = 53\,\varepsilon_+^{1/6}\,L_{\rm rel\,51}^{1/4}
\dot M_{w\,-5}^{-1/12}V_{w\,8}^{1/12}\Delta t_1^{-1/6},
\ee
and the Lorentz factor of the ambient material to
\be\label{gamactrans}
\Gamma_{\rm amb} = 2.7\,\varepsilon_+^{2/3}\dot M_{w\,-5}^{2/3}\,
                         V_{w\,8}^{-2/3}\,\Delta t_1^{-2/3}.
\ee

Pre-acceleration has the smallest effect on deceleration for low values
of $\dot M_w$ and large burst durations $\Delta t$.
One can compare eq. (\ref{fdecelb})
with the deceleration radius of a relativistic shell propagating
in a static external medium.  Setting $c\Delta t = r/2\Gamma_{\rm eq}^2$ gives
\be\label{rdecelc}
R_{\rm decel} = \left({L_{\rm rel}\Delta t^2 V_w\over \dot M_w c}\right)^{1/2}
              = 7\times 10^{14}\,L_{\rm rel\,51}^{1/2}\,\Delta t_1
                        \dot M_{w\,-5}^{-1/2}\,V_{w\,8}^{1/2}\;\;\;\;{\rm cm}.
\ee
The relativistic motion of the external medium has died away before
the reverse shock completes its passage through the shell if
\be
\Delta t_1 \dot M_{w\,-5}^{-1}\,V_{w8}^{-1} \ga 12\varepsilon_+.
\ee
(The coefficient the right side of this equation is obtained by
substituting
$\ell_+ = \ell_{\rm crit}$ and $r = 2\Gamma_{\rm eq}^2c\Delta t$ in eq. 
[\ref{compact}].)

It has been suggested that, close enough to the central engine, 
collimation of the emitted gamma-ray shell (outside the emission
radius) would allow the ambient
medium to be accelerated to a Lorentz factor higher than $\Gamma_c$
(Beloborodov 2002).   This requires the compactness $\ell_+$ of the photons
escaping across the forward shock to exceed the minimum compactness
that is needed to push the ambient medium to a Lorentz factor
$\ga \Gamma_c$.  The further collimation of the photons outside
the emitting radius would then allow the ambient material to
`surf' to a higher Lorentz factor, so that it 
detaches from the ejecta.  This phenomenon can, indeed, be expected
if the source is not surrounded by a dense wind and the pair-loaded
medium has a small optical depth.  But in the present context,
the emission of gamma-rays cannot be decoupled from the pre-acceleration
process:  the optical depth of the pairs swept up from the Wolf-Rayet wind
is larger than unity at $r \la R_{\rm decel-}$, and the gamma-ray
photons escaping across the forward shock are in fact trapped in
the pair cloud that they create (\S \ref{taup}).

\subsection{Compactness and Speed of Cooling}\label{comp}

At this point, we can check our assumption that heat electrons and 
positrons will cool rapidly after passing through the forward and
reverse shocks.  Transforming to the frame of the contact,
and taking the radiation to be isotropic in this frame, the energy
radiated is
$\Delta E_{\rm rel}' = (3/4\Gamma_c)\Delta E_{\rm rel}$.
The radiation compactness is then
\be\label{ellbulk}
\ell'(r) = {\sigma_{\rm T}\Delta E_{\rm rel}'(r)\over 4\pi r^2 m_ec^2}.
\ee
It is related to the compactness $\ell_+$ of the radiation field 
outside the forward shock by 
\be
\ell' = {3\over 4\Gamma_c \varepsilon_+}\ell_+.
\ee
when the emission is isotropic in the bulk frame.  
The critical value of $\ell_+$ for pair loading and pre-acceleration
is significantly greater than unity (Beloborodov 2002), and so
one sees that $\ell'$ itself is not much greater than unity in the
frame of the flow.  
After substituting for $\Delta E_{\rm rel}'$ and then for
$\Gamma_c$ using eq. (\ref{gamcr}), we have
\ba\label{lrest}
\ell'(r) &=& 0.5\,
          {\ell_{\rm crit}\over \varepsilon_+\widetilde\Gamma_{\rm eq}^{1/2}}
         \,\left({r\over R_{\rm decel-}}\right)^{-1/4}\nn
  &=& 4.7\,{\dot M_{w\,-5}^{1/6} \over \varepsilon_+L_{\rm rel\,51}^{1/6}\,
               V_{w\,8}^{1/6}}\,\left({r\over R_{\rm decel-}}\right)^{-1/4}
            \;\;\;\;(r>R_{\rm decel-}).\nn
\ea
At the final deceleration radius $R_{\rm decel+}$, we have
\be\label{compfin}
\ell'(R_{\rm decel+}) = 2.3\, \varepsilon_+^{-5/6}\,L_{\rm rel\,51}^{-1/4}\,
\dot M_{w\,-5}^{5/12}\,V_{w\,8}^{-5/12}\,\Delta t_1^{-1/6},
\ee
with $\varepsilon_+ \sim {1\over 4}$.

We see that the compactness is regulated to a characteristic value that depends
weakly on the Wolf-Rayet wind luminosity and ambient density, and changes
very slowly with radius.  The compactness is large enough to ensure 
fast cooling of even mildly relativistic electrons and positrons.  
(The compactness is even higher in the portion the relativistic 
outflow that interacts with the fragmenting `breakout' shell; 
\S \ref{sheath}.)  This has important implications for the prompt 
gamma-ray spectrum, which we turn to in the following section.

\subsection{Scattering Depth and Energy of the Swept-up Pairs}\label{taup}

The scattering optical depth between the reverse and forward shocks
receives a significant contribution from electrons and positrons that
are swept up from the ambient medium.  This depth $\tau_{\pm\,\rm amb}$
is regulated to a value close to unity, for the same reason that
the radiative compactness is regulated.

It is useful first to relate the deceleration radius
to the radius $R_{\tau=1}$ at which the electrons in the pre-burst
wind ($Y_e \simeq {1\over 2}$) provide unit scattering depth,
\be\label{rtauone}
R_{\tau=1} = {Y_e\sigma_{\rm T} \dot M_w\over 4\pi m_p V_w} = 
    1.0\times 10^{11}\,\dot M_{w\,-5}\,V_{w\,8}^{-1}\;\;\;\;{\rm cm}.
\ee
When this inertia of the wind becomes dominated by
electron-positron pairs, the scattering photosphere increases 
significantly, to 
\be\label{rtaupm}
R_{\tau=1}^\pm \sim \left({m_p\over Y_e m_e}\right)\,R_{\tau=1} =
     3.7\times 10^{14}\,\dot M_{w\,-5}\,V_{w\,8}^{-1}\;\;\;\;{\rm cm}
\ee
during the first stages of deceleration.  Comparing this rescaled
photospheric radius with the deceleration radius, one has
\be
{R_{\rm decel-}\over R_{\tau=1}^\pm}
= 1.4{\varepsilon_+\,\widetilde\Gamma_{\rm eq}^{1/2}\over\ell_{\rm crit}}
= 0.15\,\varepsilon_+\,
    \left({L_{\rm rel\,51} V_{w\,8}\over \dot M_{w\,-5}}\right)^{1/6}.
\ee
One sees that the pairs swept up from the ambient wind provide
$\tau_{\pm\,\rm amb} 
\sim 3$-$10$ at $r \sim R_{\rm decel-}$.  The optical depth is
even larger closer to the central engine.  Pre-acceleration of
the ambient medium is in fact suppressed inside a radius $\sim 
\,R_{\rm decel-}$.

Further out in the flow,
\be
\tau_{\pm\,\rm amb}(r) = ({\cal M}_\pm-1)\,
                 \left({r\over R_{\tau=1}^\pm}\right)^{-1}.
\ee
Substituting for $R_{\rm decel-}$ using eq. (\ref{rtauone}), one has
\be\label{tauc}
\tau_{\pm\,\rm amb}(r) = 3.3\,{{\cal M}_\pm-1\over \varepsilon_+}\,
     \left({\dot M_{w\,-5}\over  L_{\rm rel\,51} V_{w\,8}}\right)^{1/6}\,
       \left({r\over R_{\rm decel-}}\right)^{-1}.
\ee
The further evolution of the scattering depth depends on the evolution
of the pair loading factor ${\cal M}_\pm-1$.  One has (Beloborodov 2002)
\be\label{mdecel}
{\cal M}_\pm -1 \simeq 0.1\,\left({\ell_+\over\ell_{\rm crit}}\right).
\ee
The compactness is given by eq. (\ref{elldec}) at the  beginning of 
deceleration, which implies ${\cal M}_\pm -1 = O(1)$.  Further out 
in the flow one has
\be\label{pairfrac}
{\cal M}_\pm -1 \simeq \left({r\over R_{\rm decel-}}\right)^{-1/2}.
\ee
The scattering depth accumulated from the freshly swept-up pairs therefore
decreases with radius as $\tau_{\pm\,\rm amb} \propto r^{-3/2}$ outside 
$r = R_{\rm decel-}$.  It drops below unity before the end
of the prompt deceleration phase, and reaches the value
\be\label{taurdecel}
\tau_{\pm\,\rm amb}(R_{\rm decel+}) = 0.1\,L_{\rm rel\,51}^{-2/3}\,
\dot M_{w\,-5}^{5/3}\, V_{w\,8}^{-5/3}\,\Delta t_1^{-1},
\ee
when the reverse shock has passed through the ejecta shell.

The optical depth that a gamma-ray photon sees ahead of the foward
shock (denoted by $\tau_{\pm\,\rm ex}$) is
significantly smaller than $\tau_{\pm\,\rm amb}$, because the
ambient material advances only a small distance beyond its
initial radius $r_0$ before being intercepted by the forward shock.
The wind material accelerates as it drifts back through the photon
shell.  The shell of radiation moving ahead of the shock at radius
$r \geq R_{\rm decel-}$ has, from eqs. (\ref{compact}) and (\ref{elldec}),
a compactness
\be
{\ell_+\over\ell_{\rm crit}}
= \left({\widetilde\Gamma_{\rm eq}\over 2}\right)^{1/2}\,
            \left({r\over R_{\rm decel-}}\right)^{-1/2}
\ee
and a width $\Delta r = r/3\Gamma_c^2$.
Having intercepted a radiation shell of compactness $\ell \leq \ell_+$, the
Lorentz factor of the wind material is
\be
\Gamma_{\rm amb}(\ell,r) 
= \left({\ell\over\ell_+}\right)^2\,\Gamma_{\rm amb}(r)
= {1\over 2}\widetilde\Gamma_{\rm eq}\,
 \left({\ell\over\ell_+}\right)^2\,\left({r\over R_{\rm decel-}}\right)^{-1}.
\ee
Starting at a radius $r_0$, the wind material advances a distance
\ba
{r - r_0\over r} 
&=& {\Delta r\over r} \int {d\ell\over\ell_+} 2\Gamma_{\rm amb}^2(\ell,r)\nn
        &=& {2\over 15}\left({\Gamma_{\rm amb}\over \Gamma_c}\right)^2
        = {1\over 30}\left({r\over R_{\rm decel-}}\right)^{-3/2}.
\ea
The scattering depth seen by a gamma-ray photon moving ahead of the
forward shock is, then,
\be
\tau_{\pm\,\rm ex} \simeq \left({r-r_0\over r}\right)\,\tau_{\pm\,\rm amb}
       = 0.1\,\varepsilon_+^{-1}\,
     \left({\dot M_{w\,-5}\over  L_{\rm rel\,51} V_{w\,8}}\right)^{1/6}\,
   \left({r\over R_{\rm decel-}}\right)^{-3}.
\ee
One sees that the photons moving ahead of the ejecta shell are
trapped inside a radius $\sim R_{\rm decel-}$. 
This brings us to the regime assumed by Thompson \& Madau (2000), where
the mean inertia per scattering charge behind the forward shock is
$\sim (0.1-1) m_e$.   Since the compactness
of the radiation emitted at $r > R_{\rm decel-}$ is too small to push
the ambient medium up to the speed of the contact, we conclude that
there is never any detachment between the pair-loaded wind material
and the ejecta shell.

Finally, let us calculate the characteristic Lorentz factor of
the pairs which are swept back across the forward shock.  This is
\be\label{gamdif}
\Gamma_\pm = {\Gamma_c\over 2\Gamma_{\rm amb}}
           = \left({\Gamma_c\over\widetilde\Gamma_{\rm eq}}\right)^{-3}
           = \left({r\over R_{\rm decel-}}\right)^{3/4}.
\ee
During the last stages of the prompt deceleration, one has
$\Gamma_\pm \sim 10$ (as may be seen by substituting eq. [\ref{fdecel}]
into eq. [\ref{gamdif}]).  The cooling time of these particles
is substantially shorter than the flow time, by a factor
$1/\ell' \Gamma_\pm$.  

Additional pairs are created when gamma-ray
photons are upscattered above the pair-creation threshold at
the forward shock, but the effect is only modest.
For example, photons of a bulk-frame energy $\sim m_ec^2$
have an optical depth $\sim 1$ to pair creation at radius
$R_{\rm decel+}$ (\S \ref{gamgam}).  If a power-law electron/positron
tail, $dN/d\gamma_e \propto \gamma_e^{-2}$, is generated at the
forward shock over the range of energies $\Gamma_\pm \la 
\gamma_e \la \gamma_{e\,\rm max}$, then the increase in optical
depth is $\Delta\tau_{\pm\,\rm amb}/\tau_{\pm\,\rm amb}
\sim 2/\ln\gamma_{e\,\rm max}$.
Pairs will also be created in the region behind the contact, as
a direct consequence of the gamma-ray emission.  Some important
subtleties which influence the pair-creation rate,
associated with beaming and the inertia of the breakout
shell, are addressed in \S \ref{gamgam}.

\section{Baryon Sheath from Jet Breakout}\label{sheath}

We have, so far, treated the interaction of the relativistic outflow
with the Wolf-Rayet wind under the assumption that the outflow emerges
with only a minimal baryon contamination.  The jet head must expand
sub-relativistically deep in the envelope of the star (Matzner 2003),
but a thin outer shell of stellar material can be accelerated to
relativistic speeds.  The existing analytic treatment of jet breakout
(Waxman \& M\'esz\'aros 2003) is one-dimensional and does not address
the sideways slippage of the the shocked stellar material at the head
of the jet.  We now give a simple estimate of the residual column
$\Sigma_\star$ of stellar material that 
survives sideways slippage away from the jet head, and compare this
with the column of material that is swept up from the Wolf-Rayet wind
(Fig. 2).

\vskip .2in
\centerline{{
\vbox{\epsfxsize=7.5cm\epsfbox{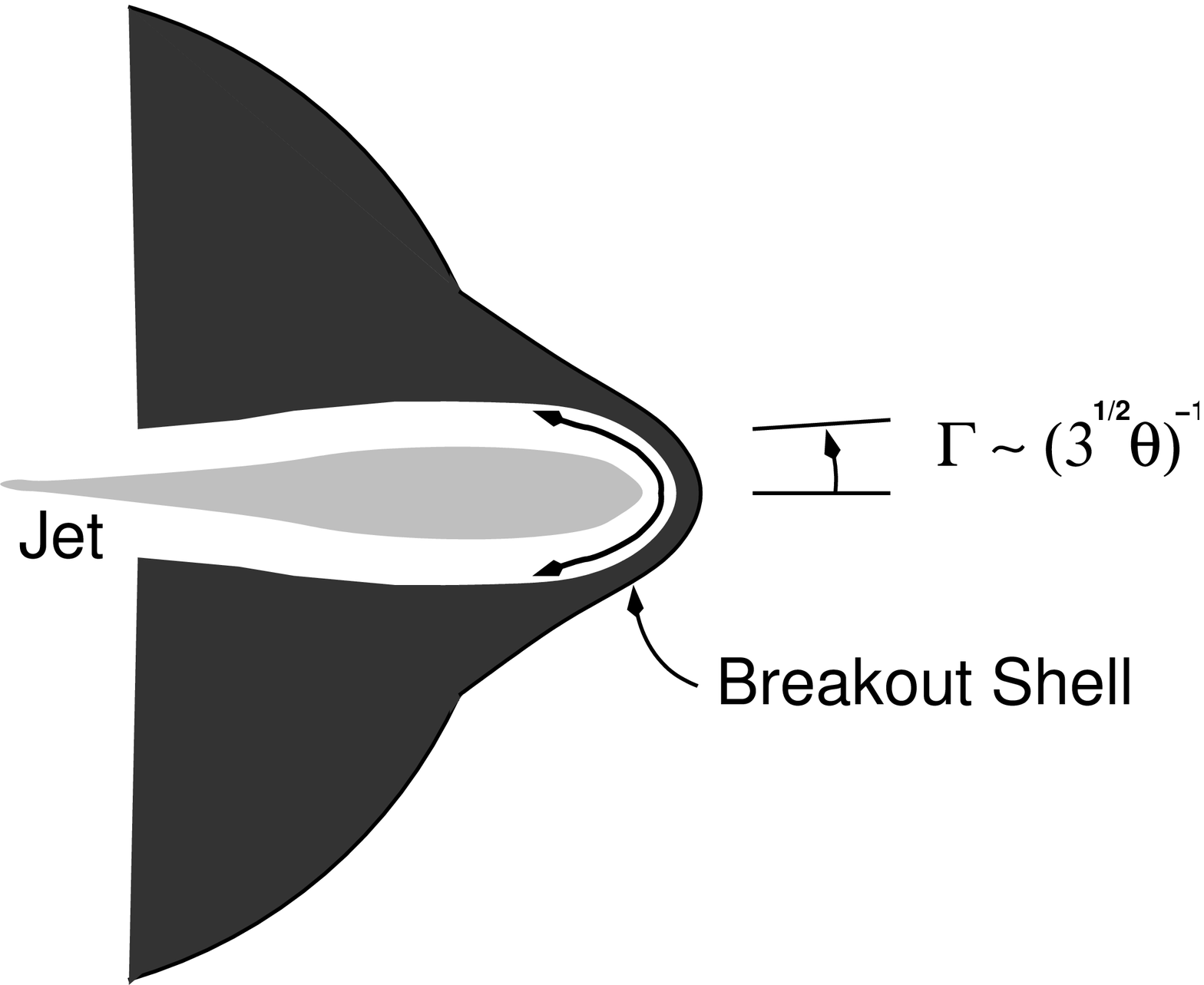}}}}
%\begin{figure}[ht]
%\epsscale{.90}
%\plotone{fig2.eps}
\figcaption{As the jet emerges from the star, the jet head
accelerates to a Lorentz factor $\Gamma \ga 1/\sqrt{3}\theta$.
A residual shell of Wolf-Rayet envelope material becomes
stuck at the jet head, and cannot flow to the side and out of the
path of the relativistic outflow.  This `breakout
shell' is optically thick close to the star, and is subject to
Rayleigh-Taylor instabilities when it begins to cool at
$\sim 10^{14}$ cm.}
\label{figbreak}
%\end{figure}
\vskip .2in

The forward shock wave accelerates close to the speed
of light as it approaches the photosphere of the star (Tan, Matzner,
\& McKee 2001).  Beyond this point, the sound speed of the shocked
stellar material is $c_s \simeq c/\sqrt{3}$.  The material in this
shocked shell can flow a distance $\sim c_s t'$ transverse to the jet axis 
in the time $t' \sim r/\Gamma c$ that it takes the jet head 
to cross a radius $r$ at Lorentz factor $\Gamma$.  
The material will therefore be lost from the jet
head if $\Gamma \la 1/\sqrt{3}\theta$, where $\theta$ is the opening angle.
To simplify matters, we refer to the isotropic
jet luminosity $L_{\rm iso}$, and do not separate out the energy
fluxes in thermal radiation, magnetic field, and baryons.
We denote by $\Delta M_\star = 4\pi R_\star^2 \Sigma_\star$ the equivalent
spherical mass
of stellar material riding at the head of the jet.  This shell intercepts
an energy $L_{\rm iso} (f\,r/\Gamma^2 c)$ from the jet, where the
coefficient $f$ must be determined self-consistently from the dependence
of $\Gamma$ on radius.  Neglecting radiative losses (the shell is 
initially optically thick) its Lorentz factor is
\be
\Gamma^3 = f\,{L_{\rm iso} r\over \Delta M_\star c^3}.
\ee
This gives $\Gamma \propto r^{1/3}$ and $f = {3\over 2}$.
The proportion of the relativistic outflow which has been swept
up by the shell grows slowly with radius;  its thickness is
\be
\Delta r = {3\over 2\Gamma^2} r.
\ee
Setting $\Gamma = 1/\sqrt{3}\theta$ at $r = R_\star$, and making
use of relation (\ref{gamvsl}), we obtain
\be
{\Delta M_\star c^2\over L_{\rm iso}\Delta t} = 
{3^{5/2}\,R_\star\over 2c\Delta t}\,\theta^3 = 
0.016\,L_{\rm iso\,51}^{-3/2}\,\Delta t_1^{-5/2}\,
       \left({R_\star\over 2\times 10^{10}~{\rm cm}}\right).
\ee
The shell continues to accelerate as it intercepts further energy
from the relativistic jet, so that
\be\label{gscale}
\Gamma \simeq {1\over \sqrt{3}\theta}\,\left({r\over R_\star}\right)^{1/3}
= 1.8\,L_{\rm iso\,51}^{1/2}\Delta t_1^{1/2}\,
    \left({r\over R_\star}\right)^{1/3}.
\ee

The reverse shock reaches the back of the ejecta
($\Delta r = c\Delta t$) at a radius
\be\label{rdecp}
R_{\rm decel}({\rm shell}) = 7\times 10^{14}\,L_{\rm iso\,51}^3\Delta t_1^6
\left({R_\star\over 2\times 10^{10}~{\rm cm}}\right)^{-2}\;\;\;\;{\rm cm}.
\ee
Referring to 
expression (\ref{fdecelb}) for $R_{\rm decel+}$ (the deceleration
radius in the absence of a breakout shell), one sees
that the inertia of the shell will speed up the passage of the reverse
shock through the relativistic wind.
Indeed, the shell mass $\Delta M_\star$ remains larger than the mass
(\ref{mwr}) of the swept-up Wolf-Rayet wind out to the radius
\be
3\times 10^{16}\,L_{\rm iso\,51}^{-1/2}\,\Delta t_1^{-3/2}\,
\dot M_{w\,-5}^{-1}\,V_{w\,8}\,
\left({R_\star\over 2\times 10^{10}~{\rm cm}}\right)\;\;\;\;{\rm cm}.
\ee
It should be noted that the normalization of $R_{\rm decel}$ is
very sensitive to the normalization of eq. (\ref{gscale}), scaling
as $R_{\rm decel} \propto [\Gamma(R_\star)]^6$.  If, for
example, one replaces $c_s = c/\sqrt{3}$ with $c$ when evaluating
the width of the jet, then $R_{\rm decel}$ grows by a factor $\sim 30$.

It is clear from the above
that Rayleigh-Taylor instabilities of the breakout shell can strongly
modulate the gamma-ray emission at larger radii.  
A homogeneous shell would become transparent to photons at the radius 
\be\label{rtrans}
R_{\tau=1}({\rm shell}) = 8\times 10^{13}\,Y_e^{1/2} L_{\rm iso\,51}^{-1/4}\,
      \Delta t_1^{-3/4}\,
\left({R_\star\over 2\times 10^{10}~{\rm cm}}\right)^{1/2}\;\;\;\; {\rm cm}
\ee
where $Y_e \sigma_{\rm T} \Delta M_\star/4\pi r^2 m_p = 1$.  Here $Y_e \simeq 
{1\over 2}$ is number of electrons per baryon in the outer layers of the
Wolf-Rayet star.  Before the breakout shell becomes optically thin,
the particles and photons will thermalize at a low temperature.  The
effective temperature at the radius $R_{\tau=1}$ is given by
\be
{4\over 3}\left[\Gamma(R_{\tau=1})\right]^2 a{T_{\rm eff}'}^4 
= {L_{\rm iso}\over 4\pi R_{\tau=1}^2 c}.
\ee
Making use of eq. (\ref{gscale}) to evaluate $\Gamma(R_{\tau=1})$,
one finds a temperature
\be
T_{\rm eff} = {4\over 3}\Gamma(R_{\tau=1})T_{\rm eff}'
      = 1.8\, L_{\rm iso\,51}^{7/12}\,\Delta t_1^{1/2}\,
\left({R_\star\over 2\times 10^{10}~{\rm cm}}\right)^{-1/3}\;\;\;\;{\rm keV}
\ee
in the frame of the central engine.

The pressure of the breakout shell is dominated by photons while it is
optically thick.  The shell therefore becomes much thinner as
it expands, and it becomes subject to a corrugation instability,
as we detail in \S \ref{rtcon}.  The thickness of a plume
which emerges at radius $r$ has a characteristic angular width 
$\delta\theta(r) \la 1/\Gamma(r)$.  The (isotropic) mass of baryons that is 
collected by such a newly formed plume is lower than $\Delta M_\star$,
but only by a numerical factor $\varepsilon_M$.  
Relativistic fluid that catches up with the shell at later times
will continue to flow through this opening, and the radius
$R_{\rm decel}$ will expand by a factor $\varepsilon_M^{-2}$.  
This process can be expected to repeat itself on a smaller angular
scale after $\Gamma$ has grown by one e-folding (that is, after 
the breakout shell has expanded a decade or so in radius beyond
the transparency radius [\ref{rtrans}]).

The last generation of Rayleigh-Taylor
plumes may therefore be clean enough that the final passage of the reverse
shock through the relativistic shell follows our previous
analysis.  The peak Lorentz factor of the contact does not, however,
attain the equilibrium value $\widetilde\Gamma_{\rm eq}$
(eq. [\ref{gameqb}]) in this case.
To find the peak Lorentz factor of the shell,
we equate expressions (\ref{gamcr}) and (\ref{gscale}).  The corresponding
radius sits between $R_{\rm decel-}$ (the beginning of deceleration
assuming no breakout shell; eq. [\ref{rdecelmin}]) and $R_{\rm decel+}$
(eq. [\ref{fdecelb}]).  One obtains
\be\label{gammax}
\Gamma_{\rm max} = 66\,L_{\rm iso\,51}^{3/7}\,\Delta t_1^{3/14}\,
       \varepsilon_+^{1/7}\,\dot M_{w\,-5}^{-1/14}\,V_{w\,8}^{1/14}\,
       \left({R_\star\over 2\times 10^{10}~{\rm cm}}\right)^{-1/7},
\ee
at the radius\footnote{It should be recalled that our derivation
assumes that the relativistic outflow moves at 
$\Gamma_{\rm rel} \gg \Gamma_c$.  In fact, $\Gamma_{\rm max}$ is 
close to the maximum jet Lorentz factor (\ref{gambcrit}) derived previously.
This approximation has the effect of reducing $\Gamma_{\rm max}$
by 10-$20\,\%$.}
\be\label{rgammax}
R_{\Gamma_{\rm max}} = 1\times 10^{15}\;
   {\varepsilon_+^{3/7}\,V_{w\,8}^{3/14}\over
  L_{\rm iso\,51}^{3/14}\,\Delta t_1^{6/7}\,
  \dot M_{w\,-5}^{3/14}}\,
  \left({R_\star\over 2\times 10^{10}~{\rm cm}}\right)^{4/7}\;\;\;\;{\rm cm}.
\ee
The compactness of the outflow in the bulk frame can be obtained by
substituting these expressions for $\Gamma_{\rm max}$ and 
$R_{\Gamma_{\rm max}}$ into eq. (\ref{ellbulk}),
\be
\ell'(R_{\Gamma_{\rm max}}) 
    = 9\,L_{\rm iso\,51}^{-1/14}\,\Delta t_1^{3/14}\,\varepsilon_+^{-6/7}\,
    \dot M_{w\,-5}^{3/7}\,V_{w\,8}^{-3/7}\,
     \left({R_\star\over 2\times 10^{10}~{\rm cm}}\right)^{-1/7}.
\ee
Note the very weak dependence of the compactness on the energy
of the relativistic outflow.  The compactness at smaller radii is
\be\label{lscale}
{\ell'(r)\over\ell'(R_{\Gamma_{\rm max}})} =
  \left({\Gamma\over\Gamma_{\rm max}}\right)^{-3}\,
  \left({r\over R_{\Gamma_{\rm max}}}\right)^{-1} = 
  \left({r\over R_{\Gamma_{\rm max}}}\right)^{-2}.
\ee

\section{Dynamics of the Magnetic Field near the 
               Contact Discontinuity}\label{fieldyn}

There are several reasons to expect that the magnetic field will
contribute a significant fraction of the pressure
between the contact discontinuity and the reverse shock.
First, most of the particle pressure behind the contact is supplied by
electrons and positrons (\S \ref{gamgam}) which lose energy rapidly to cooling.
Second, simulations of relativistic jets show that magnetic and
kinetic pressures remain in approximate equipartition out a
large distance from the engine (McKinney 2005a,b). Third,
the magnetic pressure increases by an order of magnitude across a
reverse shock that is only weakly magnetized (Kennel \& Coroniti 1984;
Zhang \& Kobayashi 2005).
One caveat here is that strong pulsations in the outflow 
can force a reduction in $\sigma$ at a large radius: after the
individual pulses spread and merge,  the
ratio of $L_P$ to $\dot M_bc^2$ drops by a factor of 
the pulse duty cycle.

We found that the energy density in the relativistic outflow is approximately
independent of its luminosity at the point where the reverse shock
has completed its passage through the ejecta shell
(when the inertia of the swept up material is dominated by the Wolf-Rayet
wind).  This result follows from the 
scaling $R_{\rm decel+} \propto L_{\rm rel}^{1/2}$ in equation 
(\ref{fdecelb}).  
Setting the Poynting flux carried by a toroidal magnetic field $B$ equal
to a fraction $\varepsilon_B$ of the wind energy flux gives
\be
{B^2\over 4\pi} = \varepsilon_B {L_{\rm rel}\over 4\pi r^2 c}.
\ee
There is a characteristic magnetic field in the outflow,
\be\label{magval}
B = 1\times 10^5\;{\varepsilon_B^{1/2}\,\dot M_{w\,-5}^{1/6}\over
      \varepsilon_+^{1/3}\,V_{w8}^{1/6}\,\Delta t_1^{2/3}}\;\;\;\;{\rm G},
\ee
in the frame of the central engine.  This field is
independent of $L_{\rm rel}$, and depends only weakly on $\dot M_w$ 
and $V_w$.  (The field in the bulk frame
is weaker by a factor $\sim \Gamma_c^{-1}$.)
The synchrotron energy of a relativistic electron gyrating with
a lorentz factor $\gamma_e$ in the bulk frame is
\be\label{esync}
E_{\rm sync} \sim 0.3\,\gamma_e^2\,{\hbar eB\over m_ec}
= 3\;\left({\gamma_e\over 10^2}\right)^2\,
{\varepsilon_B^{1/2}\,\dot M_{w\,-5}^{1/6}\over
      \varepsilon_+^{1/3}\,V_{w8}^{1/6}\,\Delta t_1^{2/3}}\;\;\;\;{\rm eV}
\ee
(as measured by the observer).

\subsection{Reverse Shock Wave in an Outflow with a 
Stochastic Magnetic Field}\label{shockrel}

We now examine the significant changes to the structure of the
reverse shock that can result from an advected magnetic field.
The fast-mode speed in the fluid upstream of the reverse shock 
can be approximated by the cold fluid formula,
\be\label{vfast}
{V_F'\over c} = {B'\over[4\pi\rho' c^2 + (B'^2)]^{1/2}} = 
\left(1+{1\over 2\sigma}\right)^{1/2}.
\ee
(The fast mode is nearly isotropic in this regime.)
Here $\rho' = \rho'_ic^2 + n_\pm' m_ec^2$ is the total particle rest
energy density in the bulk frame, $\rho_i'$ is the proper rest density
of ions, $n_\pm' = n_{e-}' + n_{e^+}'$ is the proper density 
of electrons and positrons, and
\be
\sigma = {(B')^2\over 8\pi \rho' c^2} = 
     {\varepsilon_B'\over 1-\varepsilon_B'}.
\ee
As previously,
$\varepsilon_B'$ is the fraction of the energy in the magnetic field
in the bulk frame.  Transforming to a relativistically moving 
frame,\footnote{A number
of authors use $\sigma$ to denote the ratio of Poynting and kinetic 
energy fluxes in the frame of the star
(e.g. Kennel \& Coroniti 1984; Drenkhahn \& Spruit 2002), but we
are more concerned here with the properties of the magnetofluid in
the bulk frame.}  
the ratio of of Poynting flux to (cold) kinetic energy flux is $2\sigma$.  

It is possible for $V_F'$ to exceed the speed of the relativistic outflow
with respect to the contact, so that
a reverse shock wave does not form  (Lyutikov \& Blandford 2003).
The external medium will, instead, transmit
momentum to the ouflow through a gradual compressive disturbance.
This happens if
\be\label{gamrelmin}
{\Gamma_{\rm rel}\over 2\Gamma_c}
\;<\; {1\over [1-(V_F')^2/c^2]^{1/2}} \;=\; (1+2\sigma)^{1/2}.
\ee
We have seen that $\Gamma_{\rm rel}$ (eq. [\ref{gambcrit}])
and $\Gamma_c$ (eqs. [\ref{gamcr}], [\ref{gscale}], [\ref{gammax}])
may be comparable in the deceleration zone at $\sim 10^{14}-10^{15}$ cm.
This means that $\sigma$ does not have to be much greater than unity for the
reverse shock to be suppressed in an outflow with an ordered toroidal
magnetic field.   

It should be emphasized
that the wave speed (\ref{vfast}) applies to disturbances propagating in
a {\it uniform} magnetofluid.   The response of the fluid to 
compression is altered substantially if the field undergoes
high-frequency reversals, over a distance much smaller than $c\Delta t$.
Stochasticity in the dynamo operating in the central engine will
produce flips in the sign of the field that drives the outflow.
Some further tangling of this field can occur as the 
jet pushes out through the envelope of the Wolf-Rayet star (\S \ref{param}).
The weakened magnetic field is swept outward by the jet beyond 
the stellar photosphere.  Starting in a disordered state, the field remains
disordered in the bulk frame
as long as $\Gamma_{\rm rel}$ continues to increase linearly with
radius.  That is, the two non-radial components $B_\phi'$ and
$B_\theta'$ are comparable in magnitude and
\be
{B_\phi'}^2 + {B_\theta'}^2 \sim {1\over \Gamma_{\rm rel}^2}\,
\left({r\over R_\star}\right)^2\,B_r^2.
\ee
The field becomes predominantly non-radial in the bulk frame
after the fluid Lorentz factor stops growing 
(e.g. after $\Gamma_{\rm rel}$ saturates at the value [\ref{gambcrit}]).

The deceleration of the outflow
involves a compression of the fluid on a scale $\sim c\Delta t$,
much larger than the field reversal scale.
The reversing field is susceptible to 
reconnection and tangling upon compression (Thompson 1994).
After tangling, the fluid becomes more
nearly isotropic and its sound speed closer to $c/\sqrt{3}$.  
This makes a crucial difference to the propagation of signals in
the outflow.  In this situation, the equation of state varies smoothly over
a macroscopic distance within the magnetofluid.
A localized jump in density can therefore appear if
the flow speed is higher than $\sim c/\sqrt{3}$ 
upstream of the contact.   

In the GRB emission model discussed in this paper, the plasma is pair-rich 
and the mean energy per particle is a modest multiple of $m_ec^2$.
A fluid approach is therefore appropriate to the dynamics of the
electromagnetic field and the entrained particles.  This contrasts with
a pulsar wind, where the rest energy of the particles 
comprises a much smaller fraction of the
total wind energy.  In that case, the mean Larmor
radius of the particles in a neutral sheet can approach the sheet thickness,
thereby facilitating reconnection.  Lyubarsky (2003) has argued that
the reversing component of the magnetic field will be erased at the reverse
shock in a pulsar wind, so that the shock jump condition is equivalent 
to that of a flow containing only the mean magnetic field.

We now consider the thickness of such a mock shock, in the case
where the sign of the non-radial magnetic field 
varies stochastically on a scale $\Delta r_B'$ 
(as the result of a dynamo
process in the central engine; \S \ref{tangle}).  
The mean flux density averaged over a radial
scale $\Delta r' \gg \Delta r_B'$ is $\langle B'\rangle \simeq
(2L_P/\Gamma_c^2r^2c)^{1/2}(\Delta r'/\Delta r_B')^{-1/2}$.  The
magnetization parameter associated with the smoothed magnetic field is
\be
\langle\sigma\rangle \simeq {\langle B'\rangle^2\over 8\pi\rho' c^2}
\simeq \left({\Delta r'\over \Delta r_B'}\right)^{-1}.
\ee
In this situation, the
radially-averaged magnetic field decays much more slowly with smoothing
length than it would in a pulsar wind, where the field structure is
nearly periodic (e.g. Coroniti 1990).
A reduction in $\sigma$ from a value $\sim 1$ down to $\sim 0.03$
takes place over a lengthscale $\Delta r \sim 30\,\Delta r_B
= 3\,(\Delta r_B/0.1~{\rm s})$ light seconds.  The degree of compression
at the reverse shock will therefore depend in a non-trivial way on
the dynamo timescale in the engine, relative to the duration of the burst.
The case where the reverse shock is spread out over a lengthscale
comparable to the thickness of the prompt ejecta shell deserves
examination, but will not be addressed in this paper.

This model of a shock in a flow with a reversing
magnetic field implies several modifications of the 
mechanism of first-order shock acceleration that is commonly
employed in GRB emission models.   First,
the heating of even mildly relativistic 
particles is de-localized.  A distribution
of particle energies will result from the finite rate of creation
of electron-positron pairs per unit volume 
(\S \ref{ic}).  Second, this means that rapid Compton
cooling can significantly suppress the upper bound on the particle
energy behind the shock, in comparison with a fast heating
mechanism such as shock acceleration or resonant absorption of
ion cyclotron waves by positrons.
And, third,
the heated light charges are not isotropic, being electrostatically
accelerated along the background magnetic field (\S \ref{aldamp}).

Even when a standard magnetosonic shock can form, its compression
is significantly weaker at large $\sigma$ than it would be in a cold
fluid without magnetic field.
Strong compression is essential for the formation of a hard particle
spectrum  by first-order Fermi acceleration (Blandford \& Eichler 1987).
The Lorentz factor behind a static shock 
is $\Gamma_{\rm ps} \simeq (2\sigma)^{1/2}$ when the magnetic field
runs parallel to the shock.   The field is only compressed
by a factor $1 + (4\sigma)^{-1}$ (Kennel \& Coroniti 1984).  The
compression ratio is essentially unaltered at small $\sigma$ -- but only
if $\sigma \la 0.05$, that is, if the energy carried by the magnetic
field is less than $\sim 10\,\%$ of the kinetic energy of the entrained
baryons.  In particular, the value of
$\sigma$ estimated in eq. (\ref{epsb}) is large enough that 
the magnetic field will have a significant softening effect on the 
particle spectrum generated by shock acceleration.  

Calculations of the synchrotron emission from
shock-accelerated pairs (M\'esz\'aros \& Rees 1997;
Kobayashi et al. 2005, and references therein) will, therefore, 
require significant modification if the magnetic field
carries 10 percent or more of the energy flux.
In the rest of the paper, we will generally 
refer to the reverse shock without 
qualifiers, but it should be understood that it is likely 
to behave quite differently than a standard shock in an ideal magnetofluid.

\subsection{Shocked and Pair-Enriched Wolf-Rayet Wind}\label{shockwr}

The medium in between the contact discontinuity and the forward
shock contains a relatively weak magnetic field compared with the
relativistic outflow.  The field that
is swept up from the Wolf-Rayet wind is predominantly toroidal
and carries a fraction
\be\label{fnorm}
\varepsilon_{B,w} = {B_w^2\over 8\pi \rho_w V_w^2}
\ee
of the kinetic energy density in the wind.  The progenitors
of GRBs must differ in some key respect from the
large majority of Wolf-Rayet stars.  Rapid rotation has been invoked
(Woosley 1993).  If the entire star rotated rapidly, then a relatively
strong magnetic field would be advected into the wind.  In the regime
where the magnetic stresses themselves play a negligible role in driving
the wind, one has
$\varepsilon_{B,w} \simeq 
B_\star^2R_\star^2 V_{\rm rot\,\star}^2/\dot M_wV_w^3 = 
10^{-2}\,B^2_{\star\,4}\,(R_{\star\,10}/2)^2\,
V_{\rm rot\,\star\,7}^2\,\dot M_{w\,-5}^{-1}\,V_{w8}^{-3}$.
Here $B_\star = B_{\star\,4}\times 10^4$ G is the surface magnetic
field and $V_{\rm rot\,\star} = V_{\rm rot\,\star\,7}\times 10^7$ cm s$^{-1}$
is the surface rotation speed of the Wolf-Rayet progenitor. (The chosen
normalization of $B_\star$ corresponds to a net flux $\simeq
10^{25}$ G-cm$^2$, comparable to that threading an ordinary radio pulsar
but a few orders of magnitude less than what is ultimately needed to drive a 
Poynting-dominated jet with a luminosity of $\sim 10^{51}$ ergs s$^{-1}$.)

 During the last
stages of prompt deceleration, the external medium develops
a relativistic motion (\ref{gamdif}) with respect to the 
ejecta shell.  The magnetic field upstream of the forward shock can
be obtained by noting that both the toroidal flux density and the 
baryon density are compressed by the same factor (eq. [\ref{barcon}]),
\be
B_w' = {n_b'\over \rho_w/m_p} B_w = \Gamma_c B_w.
\ee
This field is further compressed behind the forward shock by
a factor up to $3$ (when $\Gamma_\pm \gg 1$).   

Slow cooling of the shocked ions would prevent much additional
compression of the fluid behind the forward shock.  We argue in
\S \ref{hetail} that resonant absorption of ion cyclotron
waves by positrons (Hoshino et al. 1992) is not likely
to prevent effective thermalization of the protons behind the shock,
due to the relatively small fraction of the kinetic energy carried
by the positrons.

The energy density behind the forward shock is 
\be
U_w' = \Gamma_\pm (\Gamma_c \rho_w c^2) =
\left({\Gamma_c^2\over 2\Gamma_{\rm amb}}\right)\,\rho_w c^2
\ee
and so the fraction of the energy density carried by
the immediate post-shock field is
\be\label{varbfor}
\varepsilon_B' = {(3B_w')^2/8\pi\over U_w'} = 
18\Gamma_{\rm amb}\varepsilon_{B,w}\,\left({V_w\over c}\right)^2
= 2\times 10^{-4}\,\Gamma_{\rm amb}\,\varepsilon_{B,w}\,V_{w8}^2.
\ee
We have found 
$\Gamma_{\rm amb} \sim 2$ during the final stages
of the prompt deceleration.  The linearly compressed magnetic field
can therefore attain a pressure in excess of $10^{-5}$ of the thermal
pressure behind the forward shock.  This is close to some estimates
of the magnetic energy density from the observed optical 
of some GRBs (e.g. Panaitescu \& Kumar 2004, and references therein).

The structure of the magnetic field in the Wolf-Rayet wind can
be contrasted with that expected in the relativistic outflow.  The 
Wolf-Rayet wind takes a year or so to flow out to the radius $R_{\rm decel+}
\sim 10^{15}$ cm (eq. [\ref{fdecelb}]).
Even if the progenitor supports an active dynamo, it will be located
deep in the star.   The magnetic field anchored at the base of the 
Wolf-Rayet wind is therefore likely to be fairly constant over the
time the wind expands to the deceleration zone of the relativistic
GRB ejecta.

It has been noted that the
Weibel instability will be excited behind a relativistic shock
wave that is moving into a weakly magnetized medium (Kazimura et al. 
1998; Medvedev \& Loeb 1999).  This electromagnetic instability
is generated by the two counterstreaming populations of electrons 
(and positrons) behind
the shock, and creates strong current fluctuations on the plasma
scale $\sim c/\omega_{Pe}$.  The instability can be modified or suppressed
by a strong seed magnetic field $B_w' = \Gamma_c B_w$ if the associated
electron gyrofrequency $\omega_{ce} = eB'_w/m_ec\gamma_e$ is
larger than the plasma frequency $\omega_{Pe} = 
(4\pi n_\pm' e^2/m_e\gamma_e)^{1/2}$
(Hededal \& Nishikawa 2005).    Taking 
$\gamma_e = \Gamma_\pm = \Gamma_c/2\Gamma_{\rm amb}$ one finds
\be\label{freqrat}
{\omega_{ce}\over\omega_{Pe}} = \left[{{B_w'}^2\over 
    4\pi(\Gamma_c/2\Gamma_{\rm amb}) n_\pm' m_ec^2}\right]^{1/2}
  = \left({4\Gamma_{\rm amb}\varepsilon_{B,w}\over
         {\cal M}_\pm -1}\right)^{1/2}\,\left({V_w\over c}\right).
\ee
Substituting $({\cal M}_\pm-1) \sim n_\pm' m_e/(\Gamma_c\rho_w)
\sim 0.2$ (the pairs carry 20 percent of the particle inertia)
and $\Gamma_{\rm amb} \sim 2$, this expression becomes
\be
{\omega_{ce}\over\omega_{Pe}} \simeq 0.02\,\varepsilon_{B,w}^{1/2}\,V_{w8}.
\ee
The pre-existing magnetic field is therefore not strong enough to
suppress the Weibel instability.

Nonetheless, the small grow scale
$\sim c/\omega_{Pe}$ of the Weibel instability suggests that both
the mean value and the r.m.s. value of the magnetic field will relax
to very small values downstream of the shock, leaving the flux
that is swept up from the Wolf-Rayet wind.  Particle-in-cell
simulations of the aftermath of the Weibel instability 
(Medvedev et al. 2005) have demonstrated
that the coherence length of the magnetic field will grow with time;
but these simulations cover only a limited dynamic range.
On very long wavelengths, the smoothing out of the magnetic field
is limited by the speed $V_A = B'/\sqrt{4\pi\rho'}$ with which
the field can move the entrained charged particles (Gruzinov 2001).
To estimate the limiting magnetic field strength, 
we suppose that the field has reached equipartition with the kinetic
energy density of the pairs on the plasma scale
\be
\ell_p \sim \left({\gamma_em_ec^2\over 4\pi e^2 n_\pm'}\right)^{1/2},
\ee
namely
\be
B'(\ell_p) \sim B_{\rm eq}' = \left(8\pi \gamma_e n_\pm' m_ec^2\right)^{1/2}.
\ee
The conservation of magnetic flux implies that the smoothed field 
$\bar B'$ is
\be
\left[{\bar B'(L)\over B_{\rm eq}'}\right]^2 \simeq 
\left({\ell_p\over L}\right)^2
    \simeq {\alpha_{\rm em}\gamma_e\over \sigma_{\rm T} n_\pm L}
                \left({\hbar\over m_e c L}\right).
\ee
on a scale $L > \ell_p$.
The associated Alfv\'en speed is $V_A(L) = \bar B'(L)/(4\pi\rho')^{1/2}$.
Setting $L/V_A(L)$ equal to the post-shock flow time $\sim r/\Gamma_c c$
(as measured in the bulk frame) gives
\ba
\left[{\bar B'(L)\over B_{\rm eq}'}\right]^2 &\simeq&
   {\Gamma_\pm m_ec^2\over ({\cal M}_\pm-1)^{1/2} eB_{\rm eq}'(r/\Gamma_c)}\nn
 &=& 5\times 10^{-13}\,{(\Gamma_c/50)^3\over r_{15}({\cal M}_\pm-1)^{1/2}}
     \,\left({\Gamma_{\rm amb}\over 2}\right)^{-1}\,
   \left({\Gamma_c B_{\rm eq}'\over 10^5~{\rm G}}\right)^{-1}.\nn
\ea
This shows that the smoothed field that is left behind from the Weibel
instability is weaker
than the field that is swept up from the Wolf-Rayet wind
(eq. [\ref{varbfor}]), even if the rate of smoothing is only
$\sim 10^{-4}$ of the maximal rate $V_A/L$.

\subsection{Feedback of Particle Cooling on Magnetic Reconnection}\label{recon}

The type of dissipation that is triggered by magnetic reconnection
requires careful clarification.  In the equatorial region of
a pulsar wind, it has been suggested that the reversing magnetic field
will largely cancel out (Coroniti 1990).  In this manner, the
fluctuating component of the non-radial field may be erased
whilst preserving the mean field (Rees \& Gunn 1974; Lyubarksy 2003).
The alternative, which we favor in a relativistic outflow with a much
larger particle density, is that
reconnection changes the {\it topology} of the magnetic 
field lines, so that the non-radial field is partly converted to radial field.
In an expanding outflow, this allows the magnetic energy to decrease more
rapidly than it otherwise would (Thompson 1994).  

A minimal particle density
$n_{e\,\rm min} = |J_r|/ec$ is required to supply the radial electric current
in an outflow carrying a toroidal magnetic field,
\be
|J_r|  = {c\over 4\pi r}\left|{\partial B_\phi\over\partial\theta}\right|
  \sim {(L_{\rm rel} c)^{1/2}\over 4\pi r^2}.
\ee
The actual pair density in a pulsar wind may exceed $n_{e\,\rm min}$
by 4-5 orders of magnitude (e.g. Hibschman \& Arons 2001).  
In a gamma-ray burst outflow there is
generally a radius at which the Thomson depth 
is unity, outside of which the electrons (and pairs) are largely frozen
into the flow.   At this photosphere (radius 
$R_{\tau=1}$) one has $n_e = (\sigma_{\rm T} R_{\tau=1}/2\Gamma^2)^{-1}$, and 
\ba\label{nenemin}
{n_e\over n_{e\,\rm min}} &\sim&
 {8\pi e [\Gamma(R_{\tau=1})]^2\,R_{\tau=1}\over 
      \sigma_{\rm T} (L_{\rm rel}/c)^{1/2}}\nn
   &=& 2\times 10^6\,{[\Gamma(R_{\tau=1})]^2\over L_{\rm rel\,51}^{1/2}}\,
       \left({R_{\tau=1}\over 2\times 10^{10}~{\rm cm}}\right).\nn
\ea

In a gamma-ray burst outflow, one can also expect $B_\phi$
to switch sign on a scale $\Delta r_B \ll r$, due to the intrinsic
stochasticity of the dynamo in the central engine (\S \ref{shockrel}).
The current then has a dominant component $|J_\theta|'
\sim cB_\phi'/4\pi\Delta r_B' = cB_\phi/4\pi \Gamma^2 \Delta r_B$
in the bulk frame.  Comparing $|J_\theta|'$ with $|J_r|' = 
|J_r|/\Gamma$, one sees that the minimum charge density needed
to support the current increases by a factor $\sim r/\Gamma\Delta r_B$. 
For realistic values of $\Delta r_B$, this factor is
$\sim 3\times 10^3\,r_{15}\,\Gamma_2^{-1}\,[(\Delta r_B/c)/0.1~{\rm s}]^{-1}$,
which is significantly smaller than the pre-factor in eq. (\ref{nenemin}).
The charges advected by the outflow are therefore capable of maintaining
the non-radial current out to a radius where photon collisions will
raise the pair density above the freezeout value (\S \ref{gamgam}).

The importance
of pair creation has sometimes been forgotten when considering
{\it macroscopic} breakdowns of MHD in gamma-ray burst outflows.
For example, strong wave acceleration at the forward shock 
(Smolsky \& Usov 1996) will be suppressed by 
pair creation outside the forward shock; and similarly 
long-wavelength electromagnetic disturbances (Lyutikov \&
Blackman 2001) will behave in an essentially hydromagnetic manner.
The electromagnetic field cannot develop an essentially electric
character ($E^2 > B^2$) because the entrained electric charges
would have to move at a velocity exceeding the speed of light.
Nonetheless, a localized breakdown of the MHD approximation is 
possible where the magnetic field in the outflow has discontinuities.

We now consider the effect that radiative cooling has on the speed
with which magnetic flux can be advected toward a neutral sheet
(given that such a field configuration is present in a GRB outflow).  
A pure electron-positron plasma 
trapped in a neutral sheet can cool rapidly, which has the effect of
reducing the volume integral of the particle pressure across the sheet
and greatly increasing the compressibility of the medium.  
By contrast, ions will cool very slowly by incoherent synchrotron or
inverse-Compton processes in the emission region of a gamma-ray burst.
The ions can therefore supply a significant back pressure (Fig. 3).

The relativistic expansion of the outflow has an important influence
on the geometry of the reconnecting magnetic field.  In contrast with
the Solar magnetosphere, the
magnetic field lines are nearly homogeneous away from the 
neutral sheet, and extend far beyond the causal scale $r/\Gamma_c$.
As a result, one expects multiple X-type points to form along
the sheet (Thompson 1994; Lyubarsky 2003). The rate of particle cooling
implies a characteristic thickness $\Delta$ for the layer
within which the magnetic field becomes disordered.

\vskip .2in
\centerline{{
\vbox{\epsfxsize=7.5cm\epsfbox{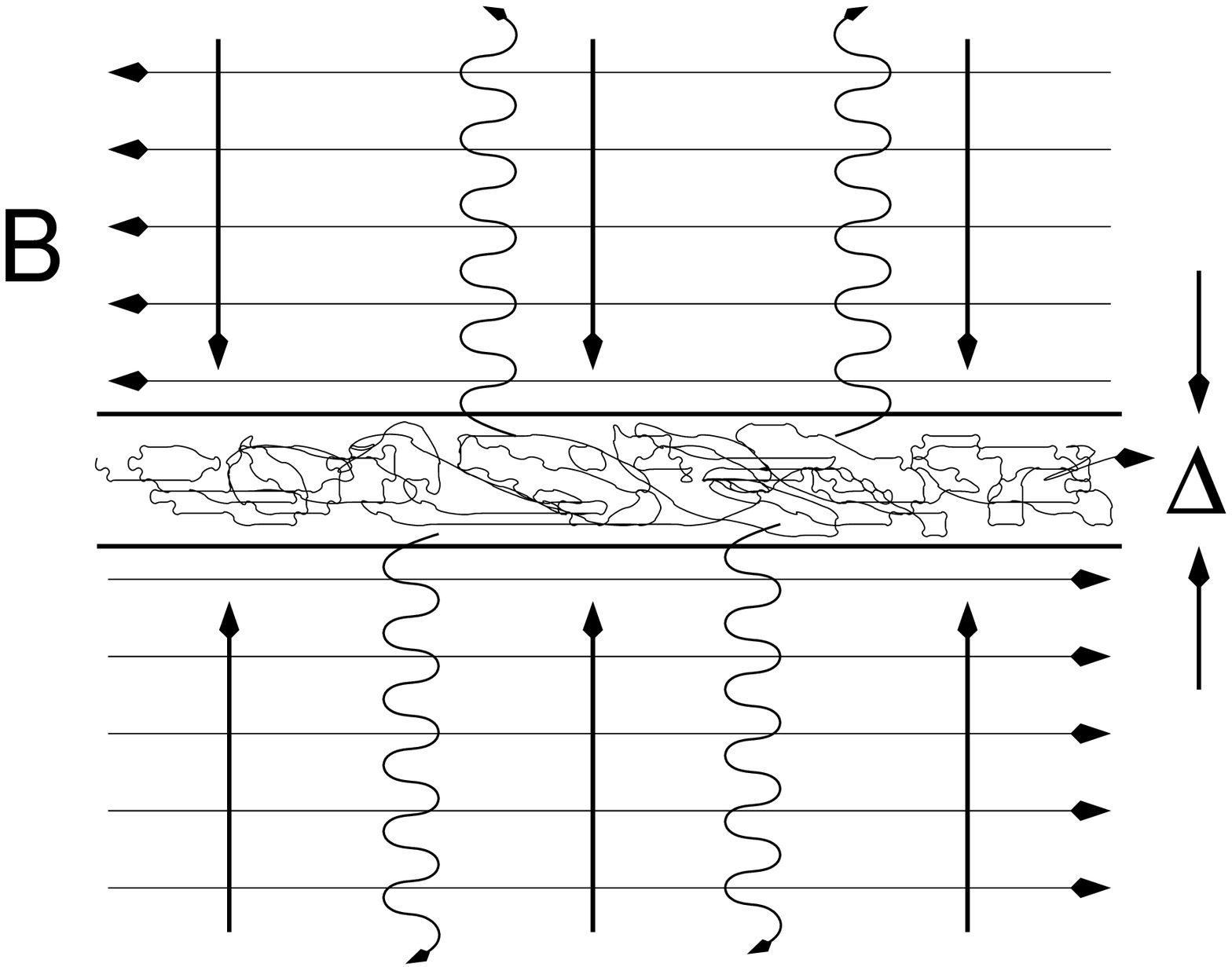}}}}
%\begin{figure}[ht]
%\epsscale{.75}
%\plotone{fig3.eps}
\figcaption{Distribution of magnetic field lines around an
idealized neutral sheet.  The magnetic field is turbulent within
the sheet, but all three components of the spatially-averaged
field vanish at the mid-plane.   When the entrained charges cool rapidly,
it is possible for the oppositely directed field lines to reconnect
and cancel out, without any sideways fluid motion.  The thickness
$\Delta$ of the sheet is determined self-consistently by the balance
between the Poynting flux toward the sheet,
and the outward radiative flux.
However, a small contamination by slowly cooling ions
will induce a significant back pressure which chokes off
the transport of magnetic flux toward the neutral sheet.}
\label{figrecon}
%\end{figure}
\vskip .2in

In what follows, we assume that the magnetic field becomes strongly
turbulent within this layer, and that the energy in the fluctuating
component of the field is transfered to the particles by a Kolmogorov-type
cascade on a timescale $\sim \Delta/c$.  As a result,
the magnetic field has annihilated in a layer of thickness $\sim ct'$
after the lapse of a time $t' < r/c\Gamma$.
The buildup of particles in the sheet forces a decrease in the
magnetization parameter\footnote{Throughout this paper, we
define the magnetization
parameter in terms of the {\it rest} energy density of the particles
in the bulk frame, $\sigma_\pm = B^2/4\pi n_\pm' m_ec^2$.},
\be
\sigma_\pm \sim \sigma_\pm^0 \,\left({\Delta\over ct'}\right).
\ee
Here $\sigma_\pm^0$ is the ambient value, as determined self-consistently
by the gamma-ray spectrum.
Balancing the advection time $\Delta/c$
with the synchrotron cooling time 
$\sim (r/c\Gamma)(\ell_B' \sigma_\pm)^{-1}$,
where
\be
\ell_B' = {\sigma_{\rm T} (B')^2 \over 8\pi m_ec^2}\,{r\over \Gamma_c},
\ee
gives
\be
{\Delta^2\over (r/\Gamma)(ct')} \sim {1\over \ell_B'\sigma_\pm^0},
\ee
and
\be\label{sigeq}
\sigma_\pm \sim {(\sigma_\pm^0)^{1/2}\over (\ell_B')^{1/2}}\,
\left({ct'\over r/\Gamma}\right)^{-1/2}.
\ee

The scattering depth across the sheet is smaller than the
total depth of the outflow ($\tau_{\rm T} \sim \ell_B'/\sigma_\pm^0$)
by a factor $\Gamma ct'/r$,
\be
\tau_{\rm T}(\Delta) \sim {\ell_B'\over \sigma_\pm^0}\,
       \left({ct'\over r/\Gamma}\right).
\ee
The optical depth to annihilation (over a timescale $t'$) is
\be 
\tau_{\rm ann}(t) \;=\; {\langle \sigma_{\rm ann} v\rangle\over\sigma_{\rm T} c}\,
{\tau_{\rm T}(\Delta)\over \Delta/ct'}
     \; \sim \; \,{(\ell_B')^2\over \sigma_\pm^0}\,
   \left({ct'\over r/\Gamma}\right)^2.
\ee
Here we have made use of $\langle \sigma_{\rm ann} v\rangle
\sim \sigma_{\rm T} c/\sigma_\pm$ for relativistic pairs of energy 
$\sim \sigma_\pm m_ec^2$.  One sees that
the annihilation of the collected pairs can be neglected as long as 
$\sigma_\pm^0 \gg (\ell_B')^2$.    

Now consider the effects of a small contamination by ions, with
a magnetization parameter $\sigma_p = B'^2/4\pi n_p' m_pc^2 
\gg 1$.  The ions collect in the neutral
sheet.  
Reconnection can be choked off if the ions acquire a significant
fraction of the wave pressure.  Cooling of 
the ions by an incoherent process like synchrotron emission is 
very slow, and can be neglected as long as $\sigma_p < (m_p/m_e)^3\,\ell_B'$.  
The dominant heating mechanism of the pairs is likely to be electrostatic
when $\sigma_\pm^0 \ga 1$ (\S \ref{aldamp}).  The
ions will be accelerated to relativistic energies by the same electric
field if $\sigma_\pm \ga m_p/m_e$ in the neutral sheet, but will absorb
a significant fraction of the wave energy only if the numbers of
ions and positrons are comparable, i.e., only if 
$\sigma_p^0 \ga (m_p/m_e)\sigma_\pm^0$.   It should be noted that
requiring $\sigma_\pm \ga m_p/m_e$ translates into a stringent constraint
on the magnetization of the pairs outside the sheet,
$\sigma^0_\pm \ga (m_p/m_e)^2\ell_B'$ (from eq. [\ref{sigeq}]).
More effective ion heating would result from resonant absorption of
high-frequency fast waves (e.g. Yan \& Lazarian 2002).  

Finally, it should be noted that two-dimensional models of reconnection
at a neutral sheet oversimplify the field geometry that is
expected in a GRB outflow.
If the magnetic field starts in a disordered state, then the
field at the boundary of a fluid element in the jet will typically be
inclined by a large angle with respect to the field in a 
neighboring fluid element.
(In other words, the two non-radial components $B_\phi'$ and $B_\theta'$
are comparable in magnitude, in a spherical coordinate system
aligned with the jet axis.)  Near the boundary between two fluid
elements, this means that one tangential component of the field will 
keep a constant sign and will not vanish in the current sheet at the 
boundary.  The uniform advection of magnetic flux toward the fluid boundary
will therefore be suppressed by the pressure of this component of the field.

\subsection{Rayleigh-Taylor Instability}\label{rtcon}

We now consider the Rayleigh-Taylor stability of the contact
discontinuity that sits at the front of the relativistic ejecta shell.
We consider two separate cases:  where the inertia of the material
that is swept up in front of the contact is dominated by
the Wolf-Rayet wind (\S \ref{shockdec}), 
and where it is dominated by a `breakout shell' that is
collected during the emergence of the jet from the
outer layers of the star (\S \ref{sheath}).  We then consider the
influence of a magnetic field on the instability.

The effective gravity felt by material near the contact discontinuity is 
\be
g' = {d^2r'\over (dt')^2} = -c^2{d\Gamma_c\over dr},
\ee
where $dr'$ is the radial displacement in the bulk frame.
This is directed outward when the effect of the breakout shell
is negligible and the dynamics of the contact is dominated by 
the interaction with the Wolf-Rayet wind.  As radiative pre-acceleration
is turning off, the Lorentz factor of the contact decreases with radius, as
$\Gamma_c \propto r^{-1/4}$.  The rest-frame time coordinate is 
$t' = \int (dr/\Gamma_cc) \simeq {4\over 5}r/\Gamma_c c$, and so 
\be\label{gdec}
g' = +{c\over 5t'}.
\ee
On the other hand, the Lorentz factor is still growing close
to the central engine, $\Gamma_c \propto r^{1/3}$,
where the breakout shell dominates the
inertia of the contact.  The effective gravity is
inward in this regime, and also somewhat stronger,
\be\label{gacc}
g' = -{c\over 2t'}.
\ee
Here we have made use of $t' \simeq {3\over 2}r/\Gamma_c c$.

The forward shock is moderately relativistic near the end of
the  prompt deceleration phase (eq. [\ref{gamdif}]).  The pressure 
$P_+'$ on the front side of the contact discontinuity is dominated 
by slow-cooling ions of an energy $\Gamma_\pm \sim 10$. 
The gravitating mass density $\rho_+'$ can
be expressed in terms of $P_+'$ and the energy density
$e_+' \simeq 3P_+'$ by
\be\label{eosa}
\rho_+' \simeq {1\over c^2}\left(e_+' + P_+'\right) \simeq {4P_+'\over c^2}.
\ee
On the back side of the contact, the 
particles (mainly pairs) are only mildly relativistic when the effects
of Compton cooling are taken into account.  Their pressure is 
$P_-' < {1\over 3}\rho_-'c^2$ in that case.  The two sides
of the contact will be in approximate pressure equilibrium,
$P_-' = P_+'$, and so one deduces that the fluid behind the contact
is denser, $\rho_-'> \rho_+'$.  The effective gravity is
outward during the last stages of pair-loading in the external medium, and
so one expects the contact to be Rayleigh-Taylor unstable. 

The fluid on either side of the contact suffers from a strong corrugation
instability when its thickness  $\Delta r'$
is much smaller than $r/\Gamma_c$ in the bulk frame.  
This instability is demonstrated by Vishniac (1983) in the case 
of a thin cooling shell in a non-relativistic blast wave, but the
derivation is not altered in the relativistic case when
$\Delta r' \ll r/\Gamma_c$.  In this regime, the growth rate
is determined by the net acceleration of the shell material
(see eqs. [2.19]-[2.21] of Vishniac 1983).  

The peak growth rate is
$\gamma' \sim (g' k_\parallel)^{1/2} \sim (2\pi g'/\Delta r')^{1/2}$ at
a wavenumber $k_\parallel \sim 2\pi/\Delta r'$ parallel to the contact. 
When the contact is decelerating, the ratio of flow time to growth time
is $\gamma't' \sim (2\pi/5)^{1/2} (\Delta r'/ct')^{-1/2}$.
This shows that
the growing Rayleigh-Taylor fingers will
induce only a limited amount of mixing when both fluid shells
(on either side of the contact) have a thickness $\Delta r' \sim ct'$.  
The corrugation instability of an {\it adiabatic}
relativistic shell has been analyzed by Wang, Loeb, and Waxman (2002),
who find no growing modes.  But growing modes do definitely exist
in the thin-shell limit.  

The breakout shell, in particular, suffers
from a strong corrugation instability.  The direction of
gravity and the density
contrast are $g' < 0$, $\rho_+' > \rho_-'$ in this case, since the shell
material is cooled by adiabatic expansion.
The shell also becomes relatively
thin as it is pushed outward.  The pressure inside
the shell decreases with radius as $P' \sim L_{\rm rel}/4\pi\Gamma_c^2 r^2c
\propto r^{-8/3}$, given the scaling
$\Gamma_c \propto r^{1/3}$ (eq. \ref{gscale}).  
Before the shell becomes optically thin, $P'$
is dominated by the entrained photons.  The pressure 
therefore scales
with the shell thickness $\Delta r'$ as $P' \propto (\Delta r' r^2)^{-4/3}$
under adiabatic expansion,
and the aspect ratio decreases as $\Delta r'/ct' \propto r^{-2/3}$.  The shell 
suffers from a strong corrugation instability even before it
becomes optically thin to scattering (at radius [\ref{rtrans}]).

The details of the Rayleigh-Taylor instability change when the
magnetic field dominates the pressure behind the contact.  
If the field were predominantly toroidal and of a constant
sign (running parallel to the contact), then
a Rayleigh-Taylor instability would be suppressed. 
This can be seen from the dispersion relation (Chandrasekhar 1981)
\be
\omega'^2 = 
    g'k_\parallel\left({\rho_+' - \rho_-'\over\rho_+'+\rho_-'}\right)
+ 2V_{A-}^2k_\parallel^2\left({\rho_-'\over\rho_+'+\rho_-'}\right)
\ee
which is the sum of the usual Rayleigh-Taylor piece and a second
positive piece representing the restoring force of the bent magnetic
field lines behind the contact.  
The growth rate $\gamma' = {\rm Im}\omega'$ takes the maximum value
\be
\gamma'_{\rm max} = \left|\left({g'\over 2^{3/2}V_{A-}}\right)
   {\rho_+' - \rho_-'\over(\rho_-')^{1/2}
              (\rho_+'+\rho_-')^{1/2}}\right|
\ee
at a wavevector
\be
k_{\parallel\,\rm max} = \left|\left({g'\over 4V_{A-}^2}\right)
   {\rho_+' - \rho_-'\over\rho_-'}\right|.
\ee

Effective growth of the Rayleigh-Taylor mode requires 
$\gamma't' \gg 1$.  When the contact is decelerating 
($g'$ is given by eq. [\ref{gdec}]), this is possible if
$V_{A-}/c \ll 0.1\,|(\rho_+'-\rho_-')/\rho_-'|$.
The particles behind the contact are mildly relativistic
in this situation, and so a large density ratio 
$\rho_-' \gg \rho_+'$ is inconsistent with pressure equilibrium
across the contact.  We conclude that a modest magnetic
field, which supplies $\sim 10^{-2}$ of the pressure
behind the contact, can suppress the Rayleigh-Taylor instability.

On the other hand, a reversing magnetic field
is susceptible to reconnection and tangling after it passes through
the reverse shock wave.  After the field has broken up
into small-scale loops, it then behaves more like an isotropic
fluid on larger scales.  The stress-tensor of the magnetofluid must
still be somewhat anisotropic, with a non-radial component than
is larger than $2^{1/2}$ times the radial component,
${B'_\phi}^2 + {B'_\theta}^2 = 2(1+\varepsilon){B'_r}^2$.  We can
define an average pressure $P_-' = 
({B'_\phi}^2+{B'_\theta}^2-{B'_r}^2)/8\pi$ normal
to the contact.  The equation of state of the magnetofluid is then
\be\label{eosb}
e'_- = {{B'_\phi}^2\over 8\pi} + 
 {{B'_r}^2\over 8\pi} +   {{B'_\theta}^2\over 8\pi}
= {3+2\varepsilon\over 1+2\varepsilon}P_-';\;\;\;\;\;\;\;\; 
\rho_-' \simeq {1\over c^2}\left(e_-' + P_-'\right).
\ee
Equating the pressures on either side of the contact and
relating them to the densities through equations (\ref{eosa}), (\ref{eosb}),
gives
\be
{\rho_+'\over\rho_-'} \simeq {1+2\varepsilon\over 1+\varepsilon}.
\ee
This expression assumes that $\varepsilon$ is larger than
$1/\gamma$ of the relativistic particles forward of the contact,
so that $\rho_+' \simeq 4P_+'/c^2$.
In this case, the anisotropic magnetic field is effectively
the {\it lighter}  \hskip .02in
fluid.  The Rayleigh-Taylor mode is therefore
stabilized while the contact is decelerating.

\section{Damping of Alfv\'en Turbulence in a Magnetically Dominated
            Medium}\label{aldamp}

Long-wavelength magnetohydrodynamic turbulence is a generic outcome of 
various instabilities in a GRB outflow, which include
the tangling of the stretched non-radial field lines
by reconnection (Thompson 1994) and kink instabilities in a jet
(Lyutikov \& Blandford 2003). 
The outflow retains enough charges outside its photosphere to enforce 
the MHD condition on very small scales compared with the causal scale
(\S \ref{recon}).  We now examine how energy is transferred from
waves to particles, and thence to the photons,
when the magnetic energy density approaches or exceeds the
rest energy density of the entrained charges.  

We should emphasize that we
are examining the regime where the energy density in the 
background magnetic field is much larger than the energy density
in high-frequency torsional waves.  The conservation of Kolmogorov energy
flux, from some forcing scale  (eq. [\ref{casflux}]) to higher
frequencies, implies that the r.m.s. wave amplitude decrease 
with frequency as $(\delta B^2)^{1/2}_\omega /B_0 \propto \omega^{-1/2}$.  
Models of `jitter' radiation (Medvedev 2000) focus on the 
opposite limit where the magnetic field has only weak long-range order,
and the gyroradius of a particle is much larger than the coherence
length of the field.

Our focus here is on the torsional MHD waves which
are excited when a reversing magnetic field in the outflow is
forced to reconnect.  These waves transport energy along 
the background magnetic field.  
The excited modes have a characteristic wavevector
$k_{\perp 0} \sim 1/\Delta r_B'$ perpendicular to the background magnetic
field.    Here $\Delta r_B'$ is the reversal scale of the magnetic field
(due to the stochasticity of the dynamo
operating in the central engine; \S \ref{param}).
The component of the wavevector parallel to the background field
is $k_{\parallel 0} \sim (V_{\rm rec}/c)\,k_{\perp 0}$, where $V_{\rm rec}$
is the speed of transverse motions excited by reconnection.
The excited turbulence will therefore be anisotropic even at
the outer scale $L_0 \sim 1/k_{\parallel 0}$.  In what follows, this
forcing scale will be normalized to the causal scale of the outflow,
\be\label{louter}
L_0 = \varepsilon_0\,{r\over \Gamma_c}.
\ee
The amplitude of the turbulence at the forcing scale is $\delta B_0$.
Notice that 
\be\label{louterb}
\varepsilon_0 \sim 
\left({V_{\rm rec}\over c}\right)^{-1}\,
\left({\Delta r_B'\over r/\Gamma_c}\right) = 0.3\,
{\Gamma_{c\,2}^2\over r_{15}}\left({V_{\rm rec}/c\over 0.1}\right)^{-1}\,
\left({\Delta r_B'/c\over 0.1~{\rm s}}\right),
\ee
is characteristically $\sim 0.1$ if the magnetic field reverses on
a timescale of $\sim 0.1$ s.
In this section, we work consistently in the bulk frame, and so we 
drop the $'$ on all quantities for notational convenience.

Torsional MHD waves can be damped
directly by bulk Compton drag (Thompson 1994); or indirectly when non-linear
couplings between oppositely propagating waves
create higher wavenumber modes (Goldreich \& Sridhar 1995).  
The current density increases with wavenumber in such a cascade,
and an upper bound to the frequency of the waves is obtained
by  balancing this fluctuating current with the maximum
conduction current available from the embedded charges
(Thompson \& Blaes 1998; Lyutikov \& Thompson 2005).  
Damping of the wave motions occurs primarily through electrostatic
acceleration of these charges along the background magneic field.
The accelerated charges cool primarily by inverse-Compton scattering
the ambient photon field.

In a fluid with $\tau_\pm \la 1$, bulk Compton drag is most effective 
at the outer scale $L_0$ of the turbulent spectrum.  The drag timescale is
\be\label{tdrag}
t_{\rm drag}
\sim {(\delta B)^2/8\pi\over \sigma_{\rm T} n_e c (\delta V_e/c)^2 \,U_\gamma}.
\ee
Here $\delta V_e/c \sim \delta B/B$ is the velocity with which the
electrons are advected past the ambient photon gas by the MHD wave motions;
and $U_\gamma$ is the energy density in the photons.
Normalizing eq. (\ref{tdrag}) to the flow time in the bulk frame,
and re-expressing the magnetization parameter of the pairs in terms of the 
bulk frame compactness $\ell_B$ and scattering depth $\tau_\pm$,
\be
\sigma_\pm = {\ell_B\over \tau_\pm} \sim 3-10,
\ee
one has
\be
{t_{\rm drag}\over r/\Gamma_c c} \sim {\sigma_\pm\over\ell_\gamma}
\sim {1\over\tau_\pm}\left({\ell_B\over\ell_\gamma}\right).
\ee
We see that $t_{\rm drag}$ is comparable to the flow time
if $\tau_\pm \sim 1$ and most of the energy
in the magnetic field has already been transferred to the photons.

The timescale for 3-mode wave interactions can, however, be significantly
shorter than the flow time.  
In the remainder of this section, we examine in more detail how
energy is transferred to the electrons and positrons through the formation
of a spectrum of high-frequency waves, and how the heated particles cool.

\subsection{Electrostatic Acceleration vs. Resonant Cyclotron Heating of 
Particles}\label{electro}

A turbulent spectrum of torsional
MHD waves is created through non-linear couplings
between the waves on the outer scale $L_0$.  Both compressive
(fast) and  torsional (Alfv\'enic) waves are created in the fluid,
but the torsional waves have stronger couplings with each other than
do fast modes of a similar frequency (Troischt \& Thompson 2004).  
Alfv\'en turbulence has a strong tendency to anisotropy, the wavepackets
becoming elongated along the magnetic field at high frequencies
(Higdon 1984; Goldreich \& Sridhar 1995).  This property is preserved
in the relativistic regime (Thompson \& Blaes 1998; Cho 1995).  

The energy that cascades from 
the outer scale $L_0 = \varepsilon_0(r/\Gamma_c)$ 
(eq. [\ref{louterb}]) can be written as
\be\label{ecas}
E_{\rm cas} = \varepsilon_{\rm cas}{B^2\over 8\pi}
\ee
per unit volume.  The coefficient $\varepsilon_{\rm cas}$ depends
on the strength of the turbulence.  It takes the maximum
value $\varepsilon_{\rm cas} \sim \varepsilon_0^{-1}(\delta B_0/B)^2$ 
when the coupling between colliding wavepackets is strong on the
timescale $\sim L_0/c$.   The three-mode coupling is strong when
\be\label{coup}
{\Delta(\delta B)\over\delta B} \sim  {k_\perp\over k_\parallel}
    {\delta B\over B} \sim 1
\ee
(Goldreich \& Sridhar 1995).
Here $k_\parallel^{-1}$ is the the length of a wavepacket parallel
to the background magnetic field, and $k_\perp^{-1}$ is the transverse
size.  We can then set a lower bound on this coupling parameter from the
requirement that a cascade have time to develop on the (bulk frame) timescale
$r/\Gamma_c$.  When the parameter is small, the effects of subsequent 
collisions will accumulate as a random walk.  A wavepacket of length
$\sim L_0$ accumulates a distortion $[\Delta(\delta B)/\delta B]^2
\sim \varepsilon_0^{-1}(k_\perp\delta B/k_\parallel B)_0^2$
in a time $r/\Gamma_c c$.  This
distortion is of the order of unity if $(k_\perp \delta B/k_\parallel B)_0
\ga \varepsilon_0^{1/2}$.  
Taking $(k_\perp/k_\parallel)_0 \sim (V_{\rm rec}/c)^{-1}$, and
$\varepsilon_0$ from eq. (\ref{louterb}),
one sees that a significant fraction of the total magnetic field energy
will transfer to high-frequency torsional waves if
\be
\left({\delta B_0\over B}\right)^4
\ga 3\times 10^{-3}\,{\Gamma_{c\,2}^2\over r_{15}}\,
     \left({V_{\rm rec}/c\over 0.1}\right)\,
     \left({\Delta r_B/c\over 0.1~{\rm s}}\right).
\ee
This inequality is satisfied if $\delta B_0/B \ga 0.2$.  

If coupling parameter (\ref{coup}) is smaller than unity at the 
forcing scale, it grows with wavenumber and the cascade quickly
enters the 
`critically-balanced' regime where $k_\perp \delta B/k_\parallel B \sim 1$
(Goldreich \& Sridhar 1995; Cho 2005).  
One then is led to a simple result that has considerable importance
for the radiative emission from the turbulent medium (Thompson \& Blaes
1998; Lyutikov \& Thompson 2005):  the
turbulent spectrum must be cut off at a frequency that lies well
below the gyrofrequency of the electrons and positrons.

A torsional wave transports charge along the 
background magnetic field at a rate $\delta J = c k_\perp \delta B/4\pi$.
Balancing this against the
maximum conduction current that can be supplied by the ambient electrons
and positrons, one has
\be\label{inner}
en_\pm \sim {k_\perp \delta B\over 4\pi} \sim {k_\parallel B \over 4\pi}.
\ee
At higher current densities, a displacement current develops through
the Maxwell equation
\be
{\partial E_\parallel \over \partial t} = ck_\perp\delta B -
4\pi en_\pm v_\pm \simeq ck_\perp\delta B -
4\pi en_\pm c.
\ee
Only a small imbalance on the right hand side is sufficient to
generate an accelerating electric field that deposits energy
in the particles.  

The charge-starvation scale (\ref{inner}) is therefore
\be\label{omin}
k_{\parallel\,\rm starve} = {4\pi en_\pm \over B} \simeq
(2\sigma_\pm)^{-1}{eB\over m_ec^2}.
\ee
Referenced to the outer scale $L_0$, this is
\be\label{kstarve}
k_{\parallel\,\rm starve}\, L_0 \sim {\tau_\pm\over \alpha_{\rm em}
  (B/B_{\rm QED})}.
\ee
Here $\tau_\pm = \sigma_{\rm T} n_\pm L_0$ is the scattering depth of the
pairs across a distance $L_0$, 
 $\alpha_{\rm em} = 1/137$ is the fine structure constant, and
$B_{\rm QED} = m_ec^3/e\hbar = 4.41\times 10^{13}$ G is the QED magnetic
field.  The field in the dissipation zone\footnote{Measured in the bulk
frame of the outflow.} is weaker than $B_{\rm QED}$ by
more than 10 orders of magnitude (eq. [\ref{magval}]).  One therefore
has $k_{\parallel\,\rm starve} \sim 10^{12}\,\tau_\pm\,L_0^{-1}$.  
One sees that torsional wave turbulence can be supported in a GRB 
outflow over a very wide range of frequencies.

In general, we will be considering the case where the rest energy
density in electrons and positrons smaller than the energy
density of the background magnetic field,
\be\label{sigpm}
\sigma_\pm^{-1} \equiv {n_\pm m_ec^2\over B^2/8\pi}
\sim {\tau_\pm\over \ell_B} \sim 0.1-0.3.
\ee
In this regime, the velocity of a torsional wave 
(the `Alfv\'en' velocity $V_A$) is generally close to the speed of light.
However, $V_A$ is significantly smaller
than $c$ when the wave is so strongly sheared that
$k_\perp \ga \omega_{Pe}/c$.  The cold plasma dispersion relation reads
\be\label{alfrev00}
\omega \simeq V_A k_\parallel,
\ee
at low frequencies $\omega \ll \omega_{Pe} = 
(4\pi n_\pm e^2/m_e)^{1/2}$.  The Alfv\'en speed is
\be\label{alfrev0}
{V_A\over c} = \left(1+{1\over 2\sigma}\right)^{-1/2}\,
\left(1+{k_\perp^2c^2\over \omega_{Pe}^2}\right)^{-1/2},
\ee
including the effects of
particle inertia and a strong gradient perpendicular to the background
magnetic field.
Here $\sigma^{-1} = \sigma_\pm^{-1} + \sigma_p^{-1}$ is the total
magnetization parameter, where
\be
\sigma_p \equiv {B^2\over 8\pi n_p m_pc^2}.
\ee
refers to the ions.
(The derivation of equation [\ref{alfrev0}] is similar to that
given by Arons \& Barnard 1986 for sheared Alfv\'en waves in the
infinite-$B$ limit.)  

In this regime, the energy in the wave is divided nearly
equally between the transverse fluctuation in the electromagnetic field,
and a {\it longitudinal}
oscillation of the electrons and positrons.  Namely,
\be\label{alfrev}
{\langle{1\over 2}n_\pm m_e v_{\pm\parallel}^2\rangle
\over \langle \delta B^2/8\pi\rangle}
\simeq {k_\perp^2 c^2 \over \omega_{Pe}^2}
\ee
and 
\be
{\langle\delta E^2\rangle\over \langle\delta B^2\rangle} = 
\left(1+{k_\perp^2 c^2\over\omega_{Pe}^2}\right)\,
\left(1+{1\over 2\sigma}\right)^{-1}.
\ee
At a perpendicular wavenumber $k_\perp \ga V_e/\omega_{Pe}$, the
phase speed of the mode along the magnetic field drops below
the speed of light, and the mode is Landau damped on the longitudinal 
motion of the electrons.

When the ions are warm, the excited modes are electron-supported
at perpendicular wavenumbers larger than the inverse ion gyroscale,
$k_\perp \gg eB/m_pcV_p$.  It should be emphasized that the dispersion
relation of these electron-supported modes is not the usual whistler
dispersion relation if the kinetic energy density
of the ions is smaller than the magnetic energy density,
${1\over 2}m_i n_i V_i^2 \ll B^2/8\pi$.  The dispersion relation is
instead Alfv\'en-like, and differs from (\ref{alfrev00}), (\ref{alfrev0}) 
only in the absence of the factor 
$(1+\sigma^{-1})^{-1/2}$.  Another difference with non-relativistic
fluids is that modes of both (circular) polarizations are
supported, and so we conjecture that the mode-mode interactions are
similar to those of lower-frequency magnetohydrodynamic waves.   
(More generally, the electron-supported
modes will have an Alfv\'en-like dispersion relation over some range
of wavenumbers $k_\perp \la \omega_{Pe}/c$ as long as $n_e m_ec^2
\ll B^2/\pi$.  Our considerations are therefore relevant to a
non-relativistic medium in which $\rho c^2$ exceeds $B^2/8\pi$, but
not by a factor exceeding $\sim m_p/m_e$.)

We will generally be interested in the case where there is a
constant energy flux through a critically-balanced spectrum
of torsional waves,
\be\label{casflux}
{dE_{\rm cas}\over dt} =  {V_A\over L_0}\,
\left({\delta B_0^2\over 8\pi}\right)
  = k_\parallel V_A\,\left({\delta B^2\over 8\pi}\right).
\ee
If the coupling parameter (\ref{coup}) is unity at the outer
scale $k_\parallel \sim L_0^{-1}$, one has $k_\perp L_0 = 
(\delta B_0/B)^{-1}(k_\parallel L_0)^{3/2}$  (Goldreich \& Sridhar
1995).  The scale $k_\perp \sim \omega_{Pe}/c$ is therefore reached
at a parallel wavelength
\be\label{kohm}
{k_{\parallel\,\rm shear}\over k_{\parallel\,\rm starve}} \sim
  {(\alpha_{\rm em}\sigma_\pm)^{1/3}\over \tau_\pm^{1/3}(L_0)}\,
    \left({B\over B_{\rm QED}}\right)^{1/3}\,
    \left({\delta B_0\over B}\right)^{2/3}.
\ee 
Note that the dependence of $V_A$ on the total magnetization
parameter $\sigma$ cancels here.  
Equation (\ref{kohm}) can be re-written as 
\be
k_{\parallel\,\rm shear}\, L_0  \sim {\sigma_\pm^{1/3}\tau_\pm^{2/3}(L_0)
       \over \alpha_{\rm em}^{2/3}}\,\left({B\over B_{\rm QED}}\right)^{-2/3}\,
    \left({\delta B_0\over B}\right)^{2/3}
\ee
using eq. (\ref{kstarve}).   
This transition occurs at a wavenumber
\be
k_{\parallel\,\rm shear} \sim 10^{-4}\,k_{\parallel\,\rm starve}
\ee
when $\tau_\pm \sim 1$, $\sigma_\pm \sim 10$, $\delta B_0/B \sim 1$,
and the magnetic field is given by eq. (\ref{magval}).  

A torsional wave of frequency $\omega$
will resonate with the gyromotion of an
electron (Lorentz factor $\gamma_e$ and speed $\beta_{e\,\parallel}$
parallel to the background magnetic field) only if
\be
\gamma_e (\omega - \beta_{e\,\parallel}c
   k_{\parallel\,\rm starve}) = \pm {eB\over m_ec}.
\ee
We have found that $V_A \ll c$ at such high frequencies, and so
the resonant energy is 
\be
\gamma_e|\beta_{e\,\parallel}| \sim 2\sigma_\pm.
\ee
If all the charges obtained this Lorentz factor and did not lose
energy to radiation, then the kinetic energy density in particles
would be comparable to the energy in the {\it background} magnetic field,
\be
\langle\left(\gamma_e^2-1\right)^{1/2}\rangle m_ec^2 n_\pm  \sim  
{B^2\over 8\pi}.
\ee
This shows that resonant heating of the particles will be prevented
if one of two conditions are satisfied:  
\begin{enumerate}
\item The particles
cool significantly in the time required for the MHD wave energy
to cascade to small scales;  

\item The net wave energy deposited in particles
is small compared with the background field energy, 
$(\delta B)^2 \ll B^2$.
\end{enumerate}

\subsection{Heating of Ions by Alfv\'en Waves}\label{ionheat}

It is possible in principle for {\it resonant} heating of the ions
to limit the range of wavenumbers over which left-handed torsional
waves can be supported.  This effect can be neglected if the wave
frequency at the scale (\ref{omin}) is smaller than the ion gyrofrequency,
\be\label{kparmax}
\left(k_\parallel V_A\right)_{k_{\parallel\,\rm starve}}
 \;\la\; {eB\over m_p c} \;=\; 2\sigma_\pm \left({m_e\over m_p}\right)\,
         k_{\parallel\,\rm starve}\,c.
\ee
We take the ions to be protons for simplicity.  In fact some
significant admixture of heavier ions with a charge/mass ratio
$\simeq {1\over 2}$ is expected in most circumstances.  This additional
component is mainly
helium if the baryons are derived from neutronized material in the central
engine (e.g. Beloborodov 2003); but could be 
carbon and oxygen if there is mixing between a relativistic jet 
and the outer layers of a Wolf-Rayet star.

%The dispersion relation
%of transverse waves propagating parallel to ${\bf B}$ is
%\be
%{c^2k_\parallel^2\over\omega^2} = 
%1 + {1\over 2\sigma_\pm} + \left(1\pm{\omega\over\omega_{c,p}}\right)^{-1}
%{1\over 2\sigma_p}
%\ee
%at frequencies below the electron cyclotron frequency.  Here the sign
%$+$ ($-$) inside the parentheses refers to right-handed (left-handed) waves,
%which resonate respectively with the gyromotion of 
%negative (positive) charge carriers.

To evaluate this expression, it suffices to note that
$V_A \propto k_\perp^{-1}$ when $k_\perp \ga \omega_{Pe}/c$
(eq. [\ref{alfrev0}]).  We assume that the cascade remains
critically balanced (i.e., set the coupling parameter
$k_\perp \delta B/k_\parallel B \sim 1$), and that the
Kolmogorov energy flux (\ref{casflux}) is conserved.  Then we find
$\delta B \sim$ constant and $k_\parallel \propto k_\perp$
when $k_\perp \ga \omega_{Pe}/c$.    The left-hand side of
eq. (\ref{kparmax}) is therefore equal to its value 
at the transition scale $k_{\parallel\,\rm shear}^{-1}$.   Defining
\be
k_{\parallel\,\rm ion} 
= {eB\over m_p c^2}%\,\left(1+{1\over 2\sigma}\right)^{1/2},
\ee
we have
\be\label{kion}
{k_{\parallel\,\rm shear}\over k_{\parallel\,\rm ion}} \sim
  {\alpha_{\rm em}^{1/3}(m_p/m_e)
    \over \tau_{\rm T}^{1/3}(L_0)\,\sigma_\pm^{2/3}}\times
%   \left(1+{1\over 2\sigma}\right)^{-1/2}
    \left({B\over B_{\rm QED}}\right)^{1/3}
    \left({\delta B_0\over B}\right)^{2/3}.
\ee
This expression  simplifies in a non-relativistic medium where
pairs are absent.  After expressing the free-electron density in terms of the
magnetization parameter $\sigma_e = B^2/8\pi n_e m_ec^2 =
(m_p/2m_e)(V_A/c)^2$, one has
\be\label{kionb}
{k_{\parallel\,\rm shear}\over k_{\parallel\,\rm ion}} \sim
  {\alpha_{\rm em}^{1/3}(m_p/m_e)^{1/3}
    \over \tau_{\rm T}^{1/3}\,(V_A/c)^{1/3}}\,
    \left({B\over B_{\rm QED}}\right)^{1/3}\,
    \left({\delta B_0\over B}\right)^{2/3}.
\ee 

We reach the interesting conclusion that {\it the electrons and positrons
will experience non-resonant, electrostatic heating even
in a fluid where $\rho c^2 > B^2/8\pi$} (but not by a factor exceeding
$\sim m_p/m_e$).  Non-resonant heating occurs
generically if the magnetic field carries even a modest fraction of the
outflow luminosity.

\subsubsection{Implications for Black Hole Coronae}

We now consider the implications of these results for the damping
of magnetohydrodynamic turbulence in black hole coronae.  
An additional damping mechanism,
Compton drag by the ambient radiation field, is
available to MHD waves in a black hole accretion disk.  We
argue that, in this case, the dominant damping mechanism remains
the  turbulent cascade.
For the turbulent layer at the base of the black hole corona, we take 
$V_A/c \sim 0.1$ and $B \sim 7\times 10^6\,(M_{\rm BH}/10~M_\odot)^{-1/2}$ G
at $r \sim 10\,GM_{\rm BH}/c^2$ 
(the magnetic field which supplies the accretion torque:
eq. 5.9.10, Novikov \& Thorne 1973).  This gives
\be
{k_{\parallel\,\rm shear}\over k_{\parallel\,\rm ion}} 
    \sim 0.02\,\tau_{\rm T}^{-1/3}\,
    \left({M_{\rm BH}\over 10~M_\odot}\right)^{-1/6}\,
    \left({\delta B_0\over B}\right)^{2/3}.
\ee
Thus,  electrostatic
heating becomes effective before resonant absorption on the ions
(see also Thompson \& Blaes 1998).  (A relatively small proportion of
the cascade energy can be transferred to the ions by Landau damping
on the gyrational motion of the ions\footnote{I thank Yoram Lithwick
for a discussion of this point.} at an intermediate wavenumber
$k_\perp \sim eB/m_pcV_p$:
 Quataert \& Gruzinov 1999).

Now let us examine the Compton drag
of the X-ray (and UV) radiation field acting on the bulk magnetohydrodynamic
motions.  The drag timescale is 
$t_{\rm drag} \sim 3m_pc/4\sigma_{\rm T} U_\gamma$
in a non-relativistic medium whose inertia is dominated by the ions.  
Comparing $t_{\rm drag}$ with the non-linear damping timescale 
$t_{\rm NL} = 1/k_\parallel V_A$ gives
\be\label{dragrat}
{t_{\rm drag}\over t_{\rm NL}} \sim  {3(c/V_A)\over 2\tau_{\rm T}(\lambda_\parallel)}
     \,\left({B^2\over 8\pi U_\gamma}\right),
\ee
where we have defined
$\tau_{\rm T}(\lambda_\parallel) = n_e\sigma_{\rm T} k_\parallel^{-1}$. 
 It has been suggested that the X-ray photons emerging from
an optically thick, radiation-dominated disk could be upscattered
into a non-thermal tail by these bulk Alfv\'en motions even close
to the disk mid-plane,
where the scattering depth $\tau_{\rm T} \gg 1$ (Socrates, Davis, \&
Blaes 2004).  Setting aside the question of whether the radiation
density will be as high as the thermally unstable Shakura-Sunyaev
disk solution would imply, we note that the formation of a non-thermal
X-ray continuum requires a significant fraction of the energy flux to be 
carried by the magnetic field.  That is, one requires
\be
{B^2\over 8\pi}V_A \ga {c\over 3\tau_{\rm T}}U_\gamma
\ee
at a vertical Thomson depth $\tau_{\rm T} > 1$.  Substituting this expression
into eq. (\ref{dragrat}) gives
\be
{t_{\rm drag}\over t_{\rm NL}} \ga  {1\over 2\tau_{\rm T}(\lambda_\parallel)
\,\tau_{\rm T}}\,\left({V_A\over c}\right)^{-2}.
\ee
To evaluate this expression, note that
\be
{1\over \tau_{\rm T}}\,\left({V_A\over c}\right)^{-2} = 
{4\pi m_pc^2\over \sigma_{\rm T} B^2 h} \sim 1\times 10^2\,
\left({h\over 3\,GM_{\rm BH}/c^2}\right)^{-1}
\ee
where $h$
is the vertical scale height of the dissipating layer.  This shows
that bulk Compton drag is likely to be less effective than
3-wave couplings at damping MHD turbulence at the base of a 
black hole corona.  We emphasize that the 
net effect is still to deposit the wave energy directly in
the electrons, but by the mechanism of electrostatic heating.

\subsection{Compton Cooling}\label{compcool}

We showed in the preceding section that electrostatic heating of the
light charges in a turbulent magnetoplasma can occur gradually, through
many small impulses.   Heating can, therefore,
be compensated by an incoherent cooling process.
In this section, we compare the relative importance of inverse-Compton
and synchrotron cooling.

Compton drag will damp the motion of the charges on a timescale
\be
t_{\rm IC} = {3m_ec\over 4\sigma_{\rm T} \gamma_e U_\gamma} =
    \left({3\over 4\gamma_e\ell_\gamma}\right) {r\over \Gamma_c c}
\ee
where $U_\gamma$ is the ambient radiation energy density
and 
\be\label{lbulk}
\ell_\gamma = {\sigma_{\rm T} U_\gamma\over m_ec^2} {r\over \Gamma_c}
\ee
is the corresponding compactness in the bulk frame.\footnote{
Of photons with rest frame energies $E_\gamma \la m_ec^2/\gamma_e$.}
The energy equation
\be
{\partial\gamma_e\over\partial t} = 
   \left(\varepsilon_{\rm cas} \sigma_\pm -
       {4\gamma_e^2\beta_e^2\over 3}\ell_\gamma\right)\,{\Gamma_c c\over r}.
\ee
has the equilibrium solution
\be\label{gamslow}
\langle\gamma_e^2\beta_e^2\rangle^{1/2} = 
\left({3\varepsilon_{\rm cas} \sigma_\pm\over 4\ell_\gamma}\right)^{1/2}
= \left({3\varepsilon_{\rm cas}\over 4\tau_\pm}
{B^2\over 8\pi U_\gamma}\right)^{1/2}.
\ee
The corresponding Compton parameter can be expressed in terms
of the Thomson depth $\tau_\pm = n_\pm \sigma_{\rm T} r/\Gamma_c$ as
\be
y_{\rm C} = {4\over 3}\langle\gamma_e^2\beta_e^2\rangle \tau_\pm = 
    \varepsilon_{\rm cas} {B^2\over 8\pi U_\gamma}.
\ee
This parameter can be large early on in the process of turbulent
damping, but must quickly saturate at a value $y_{\rm C} \la 1$,
which a significant fraction of the available magnetic energy
density has been transferred to the radiation field (Thompson 1994).
The heated particles can sustain relativistic energies only 
if $\tau_\pm \la 1$.

Modulations in the volumetric heating rate will occur over a range
of lengthscales and timescales in the fluid.  
The above estimate of the equilibrium Lorentz factor
does not apply when the cascade timescale becomes shorter than
the inverse Compton time-scale.  We examine this distinct 
regime of `flash' heating in \S \ref{flasheat}.

Let us now estimate the relative importance of synchrotron cooling.
When $\tau_\pm \ga 1$, the electrons and positrons are maintained
at a sub-relativistic temperature.  Direct cyclo-synchrotron emission
is suppressed by the large optical depth at the cyclotron resonance,
\be
\tau_{\rm cyc} \sim {\pi^2en_e(r/\Gamma_c)\over B} \sim 
               {\tau_\pm\over \alpha_{\rm em}(B/B_{\rm QED})}.
\ee

Even when the torsional wave spectrum is cut off at a frequency
much lower than the electron gyrofrequency $eB/m_ec\gamma_e$,
the light charges will still
acquire momentum perpendicular to the magnetic
field by upscattering ambient photons.  An electron moving relativistically
along the magnetic field receives a transverse momentum
\be
p_\perp = \sin\theta' {E_\gamma'\over c},
\ee
when it scatters a photon into a direction $\theta'$ measured with
respect to the field.   Here $E_\gamma' \simeq
\gamma_e E_\gamma (1 - \cos\theta) \la m_ec^2$ 
is the photon energy in the electron
rest frame, and $\theta$ its angle of propagation in the fluid frame.
The perpendicular momentum gained per scattering 
is obtained by averaging over $\theta$ and
$\theta'$, using the differential cross section $d\sigma_{\rm T}/d\Omega' =
(3\sigma_{\rm T}/16\pi)$$(1~+~\cos^2\theta')$ and assuming an isotropic 
photon distribution in the fluid frame.  One finds
\be
\langle p_\perp^2\rangle = 
    {4\over 5}\left(\gamma_e{E_\gamma\over c}\right)^2
\ee
at fixed energy $E_\gamma$.
The effects of successive scatterings accumulate
as a random walk, giving
\be\label{pperpav}
\langle p_\perp^2\rangle = 
    {4\over 5}\sigma_{\rm T} \left({r\over \Gamma_c}\right)
\;\int dE_\gamma {dn_\gamma\over dE_\gamma} 
   \left({\gamma_e E_\gamma\over c}\right)^2  = 
     {4\gamma_e^2\ell_\gamma\over 5}
 \,\left({\langle E_\gamma^2\rangle m_e\over\langle E_\gamma \rangle}\right)
\ee
over a timescale $r/\Gamma_c c$.

The synchrotron cooling rate can now be calculated as
\be
\dot E_{\rm synch} = 2\left({p_\perp\over m_ec}\right)^2
\sigma_{\rm T} {B^2\over 8\pi} c.
\ee
Comparing with the inverse-Compton power gives
\be\label{lratio}
{\dot E_{\rm synch}\over \dot E_{\rm IC}} =
       {3\over 2}\left({p_\perp\over \gamma_e m_e c}\right)^2\,
        {B^2\over 8\pi U_\gamma}
      \sim \left({\langle E_\gamma^2\rangle \over
              \langle E_\gamma \rangle m_ec^2}\right)\,\ell_B.
\ee
In this expression the radiative compactness $\ell_\gamma$ (eq. \ref{lbulk})
has been rescaled to the total magnetic energy density,
$\ell_B = \ell_\gamma (B^2/8\pi U_\gamma)$.  
The inverse-Compton emission from an ensemble of charges of
energy $\gamma_e$ is beamed into an angle
\be\label{dtheta}
\delta\theta \sim {\langle p_\perp^2\rangle^{1/2}\over \gamma_e m_ec}.
\ee

The anisotropy of the electron distribution, and the relative
output in synchrotron and inverse-Compton radiation, 
depends on the spectral distribution of the ambient radiation field.
When the radiation field has a Wien distribution with a bulk-frame
temperature $T_r$, then
\be
{\langle E_\gamma^2\rangle\over \langle E_\gamma\rangle m_ec^2} =
{4k_{\rm B}T_r\over m_ec^2} = {3\over\Gamma_c}\cdot
{k_{\rm B}T_{r,{\rm obs}}\over m_ec^2}.
\ee
If the observed temperature $T_{r,{\rm obs}}$ is typical of the spectral 
peak of gamma-ray bursts, $k_{\rm B}T_{r,{\rm obs}} 
= {4\over 3}\Gamma_c k_{\rm B}T_r \la m_ec^2$, then the bulk-frame 
temperature will be far below $m_ec^2/k_{\rm B}$.  In this case, the 
equilibrium pitch angle of the radiating electrons is small,
\be
{\langle p_\perp^2\rangle\over (\gamma_e m_ec)^2}
\sim {2\,\ell_\gamma\over\Gamma_c}\cdot{k_{\rm B}T_{r,{\rm obs}}\over m_ec^2}.
\ee
The synchrotron power is also small relative to the inverse-Compton power
as long as $\ell_B \la \Gamma_c$ (which is indeed the case when
the deceleration is tied to the end of pre-acceleration; eq. [\ref{compfin}]).

Similar conclusions are reached when the
target radiation field has a high-energy tail.
The scattering is in the Thomson regime as long
as $E_\gamma \la m_ec^2/\gamma_e$.  At higher energies the
cross section is suppressed by a factor $\sim {3\over 4}
(m_ec^2/E_\gamma')\ln(E_\gamma'/m_ec^2)$.
When the spectrum above the observed peak energy
$E_{\rm peak}$ is a power law
with energy index $\beta = -1$ up to a maximum energy $E_{\rm max}
> \Gamma_c(m_ec^2/\gamma_e)$, the average in eq. (\ref{pperpav}) becomes
\be
{\langle E_\gamma^2\rangle\over \langle E_\gamma\rangle}
\rightarrow {m_ec^2\over \gamma_e}\,
{1 + (3/8)\ln^2(\gamma_e E_{\rm max}/\Gamma_c m_ec^2)\over
 \ln({\Gamma_c m_ec^2/\gamma_e E_{\rm peak}})} \sim 
{m_ec^2\over\gamma_e\ln\Gamma_c},
\ee
where the last equality holds for $\gamma_e \sim 1$.
The pitch angle of the electrons saturates at
\be\label{eqpitch}
{\langle p_\perp^2\rangle\over (\gamma_e m_ec)^2} \sim 
                   {\ell_\gamma\over \gamma_e\ln\Gamma_c}
   \;\;\;\;\;\;\;\;
   \;\;\;\;\;\;(\beta = -1),
\ee
and the ratio of synchrotron power to inverse-Compton power at
\be\label{lratiob}
{\dot E_{\rm synch}\over \dot E_{\rm IC}} \sim
     {\ell_B\over \gamma_e\ln\Gamma_c}
   \;\;\;\;\;\;\;\;
   \;\;\;\;\;\;(\beta = -1).
\ee

The above expressions for the r.m.s. pitch angle apply to a charge
that is heated continously over the expansion time
$\sim r/\Gamma_c c$.  The pitch angle
is smaller if a charge is accelerated suddenly along the
magnetic field, and then cools passively on a timescale
$\sim (\ell_\gamma\gamma_e)^{-1}\,r/\Gamma_c c$.  Since
$\langle p_\perp^2\rangle$ grows linearly with time, one has
\be\label{eqpitchd}
{\dot E_{\rm synch}\over \dot E_{\rm IC}} \;\sim\;
     {\ell_B\over \ell_\gamma}\,
{\langle p_\perp^2\rangle\over (\gamma_e m_ec)^2} \;\sim\;
                   {\ell_B\over \ell_\gamma\gamma_e^2\ln\Gamma_c}
   \;\;\;\;\;\;\;\;
   \;\;\;\;\;\;(\beta = -1).
\ee
The r.m.s. pitch angle of the cooling particle never grows
larger than $1/\gamma_e$, and the emission cone of an ensemble
of particles is comparable in width to that of a single particle.

These expressions are easily generalized to softer gamma-ray spectra,
with an energy index $-2 \la \beta \la -1$.  One finds
\be\label{eqpitchc}
{\dot E_{\rm synch}\over \dot E_{\rm IC}} \;\sim\;
     {\ell_B\over \ell_\gamma}\,
{\langle p_\perp^2\rangle\over (\gamma_e m_ec)^2} \;\sim\;
                   {\ell_B\over \gamma_e}\,
      \left|{1+\beta\over 2+\beta}\right|\,
\left({\Gamma_c m_ec^2\over \gamma_e E_{\rm peak}}\right)^{1+\beta}
\ee
in the case of continuous heating; and
\be\label{eqpitche}
{\dot E_{\rm synch}\over \dot E_{\rm IC}} \;\sim\;
                   {\ell_B\over\ell_\gamma \gamma_e^2}\,
      \left|{1+\beta\over 2+\beta}\right|\,
\left({\Gamma_c m_ec^2\over \gamma_e E_{\rm peak}}\right)^{1+\beta}
\ee
in the case of sudden heating followed by passive Compton cooling.

\subsection{Flash Electron/Positron Heating}\label{flasheat}

We have argued that the relativistic outflow is likely to 
contain a variable magnetic field, which switches sign on
a lengthscale that is small compared with $c\Delta t$
(but still much larger than the size of the neutronized
torus that feeds the central black hole).  As a result,
it is plausible that stochastic bursts of dissipation
(driven by magnetic reconnection) occur on timescales
that are small compared with $\Delta t$ (that is, on
lengthscales that are small compared with $r/\Gamma_c$ in the
frame of the contact).  This means that a fraction
$(\delta B_0/B)^2$ of the energy of the shock relativistic
outflow can be transferred {\it locally} to the electrons and positrons
quickly enough that they reach the equilibrium energy
\be\label{gamfin}
\langle\gamma_e\rangle \;=\; \sigma_\pm \left({\delta B_0\over B}\right)^2
\ee
before they are able to cool.  This sets an upper bound on the
size $L_0$ of the heated region.  One requires that $L_0/c 
\la t_{\rm cool}(\langle\gamma_e\rangle) =
 3m_ec/4\langle\gamma_e\rangle\sigma_{\rm T} U_\gamma$, or equivalently that
\be\label{coolbound}
{L_0\over r/\Gamma_c} \;\la\; {1\over \langle\gamma_e\rangle \ell_\gamma}
  \;=\; {\tau_\pm\over\ell_B\ell_\gamma}
       \left({\delta B_0\over B}\right)^{-2}.
\ee
Here we have related the magnetization parameter of the pairs
to the scattering optical depth via $\sigma_\pm = \ell_B/\tau_\pm$.

\section{Beaming of the Inverse-Compton Radiation in 
a Turbulent Magnetofluid}\label{gbeam}

We have found that a relativistic shell of ejecta begins to dissipate
strongly when the compactness of the radiation streaming across the
forward shock has dropped below a characteristic value (\S \ref{shockdec}).
This results in a characteristic radiative compactness 
$\ell'$ within the moving
shell (eq. [\ref{compfin}]), and a characteristic optical depth to 
scattering between the reverse and forward shocks (the sum of eqs. 
[\ref{taurdecel}] and [\ref{taushocked}]).  Because the dissipation
is concentrated in a narrow range of radius, one obtains
an explanation for the weak evolution of the temporal
power spectrum that is typically seen within a gamma-ray burst.

If the gamma-ray emission is triggered by the interaction with the
external medium, then the observation of many well-separated subpulses
in some gamma-ray bursts forces a significant constraint on the
the emission mechanism:  the gamma-rays must be beamed in 
the bulk frame (in the frame
of the contact discontinuity).  By contrast,
high-frequency variability from internal shocks is smoothed out 
due to the curvature of the ejecta shell, unless the internal shocks
are occurring well inside the radius where the reverse shock passes
through the ejecta shell (Sari \& Piran 1997).  Emission on a timescale
$\delta t \ll r/2c\Gamma_c^2$ must be localized to a small fraction $\sim
(2\Gamma_c^2c\delta t/ r)^2$ of the surface area of the shell that is visible
to the observer.  The emission occurs at a radius
$r \sim 2\Gamma_c^2c\Delta t$ when it is triggered by the interaction 
with the external medium.  This means that $\sim (\Delta t/\delta t)^2$ 
independent regions of the shell contribute to each time interval 
$\delta t$, and only a fraction of them must be visible to the observer
in a highly variable burst.

What is the cause of this beaming?
Reconnection in a high-$\sigma$ magnetofluid
can, in principle, create bulk motions as fast as
\be\label{gambulk}
\gamma_{\rm bulk} \la {1\over (1-V_A^2/c^2)^{1/2}} = (1 + 2\sigma)^{1/2}
\ee
where $V_A^2 = (B')^2/4\pi\rho'$ (Blackman \& Field 1994) and
$\sigma^{-1} = \sigma_p^{-1} + \sigma_\pm^{-1}$.  So one could
ascribe the gamma-ray variability to stochastic variations in the direction
of bulk motion (Lyutikov \& Blandford 2003).  It is not clear, however,
that significant elements of the fluid will be able to 
accelerate to relativistic speeds, as they evidently do in the 
outer magnetosphere of a Soft Gamma Repeater during a giant flare
(e.g. Thompson \& Duncan 2001).  In contrast with an isolated neutron star,
whose magnetic dipole field pressure drops off rapidly with radius
(as $\sim r^{-6}$), one expects the pressure to 
equilibrate rapidly after the relativistic outflow has passed through
the reverse shock.   An element of magnetofluid may therefore
feel a strong drag force off the ambient fluid as it attempts to move
relativistically.  

A quite different beaming mechanism is afforded by our model
of decaying Alfv\'enic turbulence (\S \ref{aldamp}).  The energy
of the waves is extracted by electrostatically accelerating
electrons and positrons along the magnetic field.  As a result, 
the radiation of these charges is collimated along the local 
direction of the field lines  (Fig. 4).

The beaming pattern depends in a subtle way on how the charges are
heated.  The charges start out with small perpendicular motion
in the case of flash heating (where energy is transferred
from the torsional waves on a timescale much shorter than the 
cooling time; \S \ref{flasheat}).  Their radiation is therefore
beamed into a solid angle $\sim \pi/\gamma_e^2$.
Repeated inverse-Compton scatterings impart some gyrational motion.
By the time a charge has lost half of its initial kinetic
energy, the pitch angle has increased to $\langle p_\perp^2\rangle^{1/2}
/\gamma_em_ec
\sim 1/\gamma_e(\ln\Gamma_c)^{1/2}$  (in the case of
hard high-energy photon spectrum; eq. [\ref{eqpitch}]).   
Here $\gamma_e \sim \sigma_\pm (\delta B_0'/B')^2$ depends on the
fraction of the magnetic energy that is transferred quickly to the charges
(eq. [\ref{gamfin}]).   The beaming pattern of a large collection
of charges does not change significantly from that of a single charge.

\vskip .2in
\centerline{{
\vbox{\epsfxsize=7.5cm\epsfbox{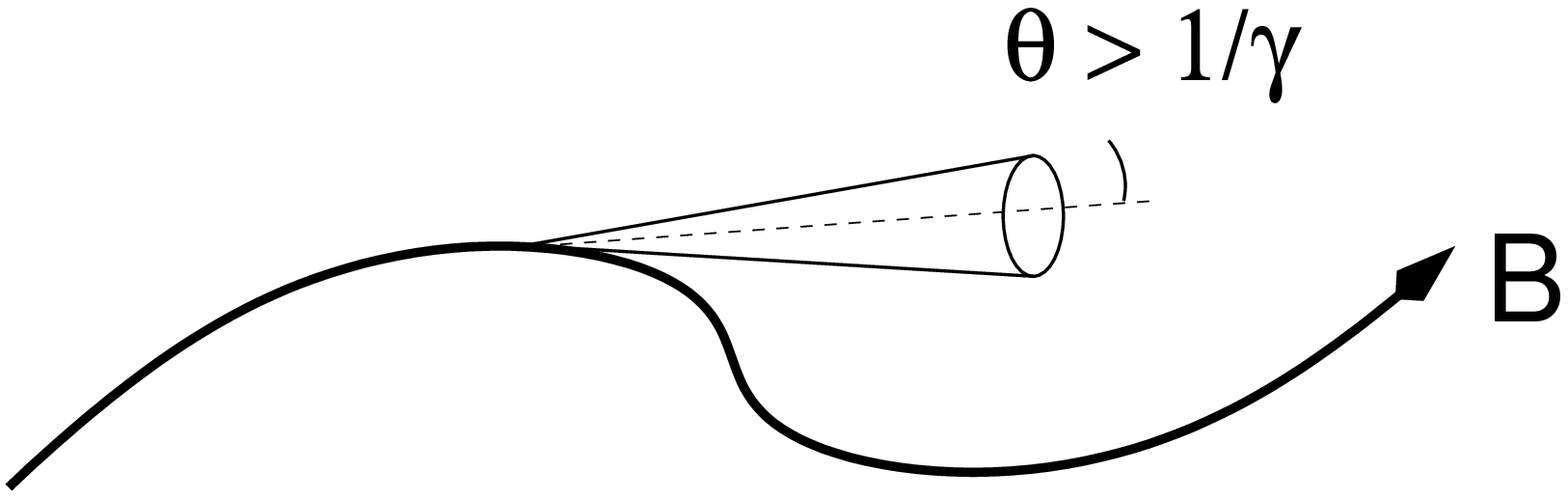}}}}
%\begin{figure}[ht]
%\epsscale{.75}
%\plotone{fig4.eps}
\figcaption{Torsional MHD waves in a magnetofluid 
transfer energy to the light charges
by forming a high-frequency spectrum of current fluctuations,
and an accelerating electric field.  The charges cool
primarily by inverse-Compton scattering ambient photons along
the local direction of the magnetic field.  The observed duration of
a gamma-ray spike decreases with energy (eqs. [\ref{etcorb}],
[\ref{etcorc}]).  The scaling with energy
depends on whether cooling is faster than the shift in beaming
direction that is
caused by the long-wavelength undulations of the magnetic field.}
\label{figbeam}
%\end{figure}
\vskip .2in

The heating may, alternatively, be slow
enough to be balanced by inverse-Compton drag. The electrons and
positrons will then maintain a pitch angle (\ref{eqpitch}) --
but at a somewhat smaller Lorentz factor,
$\langle \gamma_e^2\rangle^{1/2} \sim (3\sigma_\pm/4\ell_\gamma')^{1/2}$
(eq. [\ref{gamslow}]).  
The radiation of many charges
is now beamed into a larger solid angle $\pi{\theta'}^2 \sim 
\pi\langle p_\perp^2\rangle/(\gamma_e m_ec)^2 \sim \pi\ell_\gamma'/
\gamma_e\ln\Gamma_c$.  
One sees that beamed radiation is a direct consequence of the damping of 
Alfv\'enic turbulence.

The beaming pattern of the gamma-rays, as seen by the observer, is
sensitive to the presence of slow undulations of the contact, through
an angle larger than the beaming angle $\delta\theta'$.
The inverse-Compton radiation
is beamed in the directions
parallel and anti-parallel to the magnetic field.  An observer situated
in the rest frame of the central engine will
see the seed thermal photons emitted within an angle $\sim 1/\Gamma_c$ 
of the radial direction.  The inverse-Compton photons
will, by contrast, be visible only in two spots of angle diameter
$\sim \delta\theta'/\Gamma_c$ that are displaced by an angle 
$\sim 1/2\Gamma_c$ from the center of the visible patch of the
ejecta shell.

\subsection{Correlation between Variability Timescale and \\ Photon 
Energy}\label{evar}

We now consider the temporal pattern of the beamed radiation that
is emitted by the heated pairs.  A torsional wave
wave of an amplitude $\xi$ and a
period $P' = 2\pi/(k_\parallel c)$ imparts a tilt
$\delta B'/B' = k_\parallel\,\xi$ to the magnetic field.   This tilt varies
with time, and so the wave motion forces the radiation beam to 
to sweep past an observer on the timescale
\be\label{tfluc}
\delta t'_B \sim {\delta\theta'\over \partial(k_\parallel\xi)/\partial t'}
 = {\delta\theta'\,P'\over 2\pi(\delta B'/B')} \propto k_\parallel^{-1/2}
\ee
(in the bulk frame).
The fluctuation timescale is shorter in the observer's frame,
$\delta t_B = \delta t'_B/\Gamma_c$.

This expression for $\delta t'_B$ 
holds when $\delta B'/B' \ga \delta\theta'$, which
is possible only over a certain range of wavenumbers
in a critically balanced Alfv\'enic cascade.  The conservation of energy flux
(eq. [\ref{casflux}]) implies that $\delta B'/B' \propto k_\parallel^{-1/2}$.  
The minimum fluctuation timescale is
\be
{\delta t_B\over\Delta t} \sim
{\varepsilon_0\over 2\pi}\,\left({\delta\theta'\over \delta B_0'/B'}\right)^2,
\ee
given a wave amplitude $\delta B_0'$ at an outer 
scale $L_0 = \varepsilon_0\,(\Gamma_c c\Delta t)$ in the bulk frame.

Two other timescales for variability are relevant here when the
particles undergo flash heating.  
The motion of the charges along the magnetic field is damped on the timescale
\be
{\delta t_{\rm cool} \over \Delta t} \sim {1\over\gamma_e \ell_\gamma'}
\ee
(as seen by an observer who is at rest with respect to the central engine).
Finally, the beaming angle $\delta\theta'$ broadens
in response to repeated Compton scatterings.  In particular, the r.m.s.
pitch angle becomes comparable to $\sim 1/\gamma_e$ on a timescale 
\be
{\delta t_\theta\over \Delta t} \sim 
{\ln\Gamma_c\over \gamma_e \ell_\gamma'}.
\ee
(This expression assumes a hard photon spectrum,
$\beta = -1$, and is easily generalized to softer spectra using
eq. [\ref{eqpitchc}].)

All of these 
timescales are potentially resolvable in a gamma-ray burst
light curve.  By contrast, if the emission mechanism were synchrotron
radiation,  then the cooling time would be 
far too short to agree with the observed width of the
gamma-ray sub-pulses.
One requires a Lorentz factor $\gamma_e \ga 10^4$ to generate
photons with a observed energy of 100 keV or greater (eq. [\ref{magval}]).
The corresponding synchrotron 
cooling time is $\Delta t/\ell'_\gamma \gamma_e 
\la 10^{-4}\,\Delta t$.  

Each of these variability 
timescales become shorter as the energy of the inverse-Compton photon 
increases.  When photons from the advected thermal peak are the principal
seeds, one has $E_\gamma \sim \gamma_e^2\,E_{\rm peak}$ and
\be\label{etcorc}
\delta t_{\rm cool},\;\delta t_\theta \;\propto\; E_\gamma^{-1/2};
\ee
\ba\label{etcorb}
\delta t_B &\propto& \delta\theta' \;\propto\; E_\gamma^{-1/4}
\;\;\;\;\;\;\;\;\;\;({\rm continuous~heating});\nn
\delta t_B &\propto& \delta\theta' \;\propto\; E_\gamma^{-1/2}
\;\;\;\;\;\;\;\;\;\;({\rm flash~heating}).
\ea
Which of these scalings is most applicable depends on the
speed of Compton cooling relative to the change in beaming direction
that is driven by long-wavelength undulations of the magnetic field.
It is interesting to recall, in this regard, that
some gamma-ray bursts are composed of spikes which show noticeable
asymmetry between rise and decay - the so-called FRED behavior
(e.g. Fenimore et al. 1996).  Other bursts are highly variable but
composed of spikes with no clear asymmetry between rise and decay.
We associate this second type of variability with fluctuations in
beaming direction on a timescale that is shorter than the cooling
time.  FRED pulses demonstrate the scaling between width and
frequency that is characteristic of cooling, $\delta t \propto
E_\gamma^{-1/2}$ (eq. [\ref{etcorb}]; e.g. Fenimore et al. 1995; Norris
et al. 1996).  An analysis which separates out the two types of
variability has not yet been done, but would be illuminating.

\subsection{Pair Creation:  Effects of Beamed Emission 
and \\ Inertia of the Breakout Shell}\label{gamgam}

The ejecta are photon rich during the
deceleration process.  The number of photons emitted per unit area
inside radius $r$ is
\be
N_\gamma = {L_\gamma\over\langle E_\gamma\rangle}\,
\left({r\over 2\Gamma_c^2 c}\right).
\ee
This can be normalized to the electron column by\footnote{We
ignore spectral evolution for the purpose of this comparison.}
\be
{\sigma_{\rm T} N_\gamma\over \tau_\pm}
= {\ell_+\over \varepsilon_+\tau_\pm}\,
\left({\langle E_\gamma\rangle\over m_ec^2}\right)^{-1}.
\ee
Substituting eqs. (\ref{compact}), (\ref{elldec}), and (\ref{gamcr})
gives
\ba
{\sigma_{\rm T} N_\gamma\over \tau_\pm} &=&
{\ell_{\rm crit}\widetilde\Gamma_{\rm eq}^{1/2}\over 
  2^{3/2}\varepsilon_+\tau_\pm}\,
\left({\langle E_\gamma\rangle\over m_ec^2}\right)^{-1}\,
\left({r\over R_{\rm decel-}}\right)^{-1/2}\nn
&\simeq& 10^3\,\tau_\pm^{-1}\,
\left({\langle E_\gamma\rangle\over m_ec^2}\right)^{-1}\,
\left({r\over R_{\rm decel-}}\right)^{-1/2}.
\ea

The shocked outflow will, itself, have a modest optical
depth to scattering (eq. [\ref{taurdecel}]).    The larger contribution
to the optical depth $\tau_\pm$ comes from gamma-ray collisions
within the shell.  An estimate can be obtained
directly from the observed gamma-ray spectrum, under the assumption
that the high-energy power-law extends up to the pair-creation
threshold $\sim m_ec^2$ in the emitting frame (e.g. Baring \& Harding
1997).  The standard estimate assumes
{\it isotropic emission} of the gamma-rays in the bulk frame.

Given an observed peak energy $E_{\rm peak}$, the density of
photons of energy $m_ec^2$ in the bulk frame is 
\be\label{ngam}
n_\gamma'(m_ec^2) \sim {L_{\gamma}\over 4\pi r^2
\Gamma_cE_{\rm peak}c\ln(m_ec^2/E_{\rm peak})}
\,\left({\Gamma_c m_ec^2\over E_{\rm peak}}\right)^{-1}.
\ee
We specialize here to a hard power-law with a photon energy index $-1$ 
extending from energy $E_{\rm peak}$ 
up to a bulk frame energy $\sim m_ec^2$.   
The density (\ref{ngam}) can be re-expressed as
\be
\sigma_{\rm T} n_\gamma'(m_ec^2)\,{r\over \Gamma_c} \simeq
   {\ell_\gamma'\over \ln(m_ec^2/E_{\rm peak})},
\ee
where $\ell_\gamma'$ is the bulk frame compactness of the photons,
defined analogously to eq.(\ref{ellbulk}),
\be
\ell_\gamma' = {\sigma_{\rm T} L_{\rm rel}\over 4\pi\Gamma_c^3 m_ec^3 r}.
\ee
The density of
created pairs is $n_\pm' \simeq 2\varepsilon_{\gamma\gamma}\,
\sigma_{\rm T}\,
[n_\gamma'(m_ec^2)]^2\, (r/\Gamma_c)$, where $\varepsilon_{\gamma\gamma}$
normalizes the frequency-averaged photon collision cross section to
$\sigma_{\rm T}$, and is $\varepsilon_{\gamma\gamma} \simeq 0.1$ for a 
$\beta = -1$ photon energy index (Svensson 1987).
The scattering depth in the region of shocked relativistic material
behind the contact is then
\be\label{taushocked}
\tau_\pm(r < R_c) = \varepsilon_{\gamma\gamma}
{2(\ell_\gamma')^2\over \ln^2(\Gamma_c m_ec^2/E_{\rm peak})} \simeq
0.9\,L_{\rm rel\,51}^{-1/2}\,\dot M_{w\,-5}^{5/6}\,V_{w\,8}^{-5/6}
\Delta t_1^{-1/3}.
\ee
Here we have substituted eq. (\ref{lrest}) using $\varepsilon_+ = 
0.25$ at the radius where half the ejecta are shocked.  

There is, however, considerable reason to doubt this assumption of
isotropic emission.  Some bursts show strong variability, which in
the framework advanced in this paper demands that this emission
be beamed in the bulk frame (\S \ref{gbeam}).  
In such a situation, the production rate of pairs rises considerably.
Suppose, for example, that the emission is collimated within cones
of angular width $2\delta\theta' \sim 2/\gamma_e$ that run anti-parallel 
(in the bulk frame).  We fix the net emissivity per unit volume.  
The intensity of gamma-ray photons in the oncoming beam increases
by a factor $2\pi/\pi(\delta\theta')^2 \sim 2\gamma_e^2$.  In addition,
the threshold condition for pair creation  by the two colliding photons,
\be
E_{\gamma 1}E_{\gamma 2}(1-\cos\theta_{12}) > 2(m_ec^2)^2,
\ee
is more easily satisfied.  If the photons are moving anti-parallel,
$\theta_{12} \simeq \pi$, then the threshold energy for pair creation 
is reduced by a factor $\sim 1/\sqrt{2}$ compared with the r.m.s. value in
an isotropic photon gas.  Assuming the same spectrum as in eq. (\ref{ngam}),
the pair creation rate therefore increases by a factor $\sim 2$.

The breakout shell also has a strong effect on the compactness
of the shocked fluid.  Inward from the point where $\Gamma_c$
reaches its peak value, the compactness rises rapidly toward small $r$,
$\ell'(r) \propto r^{-2}$ (eq. [\ref{lscale}]).  The proportion of
the relativistic ejecta which has been shocked decreases only slowly 
inward, as $\Delta r \propto r^{1/3}$.  
The optical depth in pairs is proportional to $(\ell')^2$.  Combining
these scalings, one sees that $\tau_\pm$ is
an extremely strong function of the thickness of the shocked relativistic
shell,
\be
\tau_\pm \propto (\Delta r)^{-12}.
\ee
In addition, the relativistic jet material can develop a mildly
relativistic motion with respect to the fragmenting breakout shell.
This reduces the energy threshold for pair creation by allowing some
of the gamma-ray photons to side-scatter off the denser shell material.

We conclude that a substantial fraction of the relativistic
outflow will dissipate at optical depths $\tau_\pm \ga 1$,  when
the effects of beaming and the breakout shell are taken into account.

\section{Implications for the Spectra of \\ Gamma-Ray Bursts}\label{summar}

We have explored in some detail the physics of
a relativistic outflow, composed of a gas of seed thermal photons and
a stochastically reversing magnetic field, which is launched into
the dense wind of a Wolf-Rayet star.  In this section, we outline
the implications of our results for the non-thermal spectra of
gamma-ray bursts.  We first address the origin of the high-energy
non-thermal continuum that is a defining feature of GRBs.  Second,
we show how the observations of gamma-rays at energies
$\ga 10^2$ times the peak energy provide a valuable diagnostic
of the physics of pair-loading, the density of the ambient medium, and
the compactness of the zone in which the ejecta decelerate.
Third, we address the spectral transition 
that occurs between the end of the prompt GRB phase and the beginning
of the afterglow emission (a good example being GRB 980923: Giblin
et al. 1999).  We then explain how prompt optical synchrotron emission
may be suppressed (without self-absorption) in a turbulent magnetoplasma.
Finally, we connect our model with a previous suggestion (Thompson 1994)
that thermal radiation generated close to the central engine is
the dominant source of inverse-Compton seeds for the non-thermal
radiation of GRBs.  The physical mechanisms that we have examined in this
paper can also be applied to spectrally harder classes of bursts -- 
the short GRBs 
(Kouveliotou et al. 1993) and the giant flares of the SGRs
(e.g. Hurley et al. 2005).  In these types of bursts,
one requires that the outflow 
go into (nearly) free expansion much closer to the engine than it does
in the Wolf-Rayet/jet model.

Two general lessons emerge here.  First, a close correspondence between
the peak of the seed thermal spectrum and the peak of the final GRB 
spectrum arises much more easily if the heating of the pairs
is spatially and temporally distributed, instead of being localized
at a strong shock wave (or a few such shock waves within each causal patch
of the outflow).  Even in a situation where the magnetic energy density
greatly exceeds the rest energy density of the radiating particles,
rapid Compton cooling allows the mean particle Lorentz factor to remain
close to $m_ec^2$ in the fluid rest frame.  

Second, single-box models of
the radiation spectrum can provide some subtle insights into the
interplay between different radiative processes (e.g. Pe'er \&
Waxman 2004b), but ultimately the GRB emission problem is 
one of radiative transport.  The prompt GRB emission continues only as long
as the thermal seed photons continue to overlap with the dissipating
magnetized shell.  The prompt 100 MeV component of the GRB spectrum
is a byproduct of the dissipative processes occurring two distinct
regions:  the inverse-Compton emission behind the contact discontinuity;
and the side-scattering of these gamma-rays in the 
Wolf-Rayet wind.   The fluid that passes through the reverse
shock builds up pairs by collisions between gamma-rays, but
this pair creation is extended in time.  As a result,
the mean energy per particle has a strong negative gradient
away from the reverse shock.  A high-energy spectral tail
is the byproduct of this inhomogeneous distribution of particle energies.

\subsection{Power-Law Spectral Tails above the Peak Frequency}\label{ic}

The spectra of gamma-ray bursts are characteristically
non-thermal, and in many cases appear to have a power-law shape
above the peak energy $E_{\rm peak}$.
BATSE burst data do not characteristically cover a wide
energy range above the peak energy; but some Ginga bursts
with lower peak energies have high-energy
power-law tails that cover more than two orders of magnitude
in frequency (Strohmayer et al. 1998). 

Compton upscattering of the seed thermal photons will
preserve the relation between peak energy and isotropic
luminosity if i) the optical depth through the dissipating shell is
$\tau_\pm \ga 1$; and ii) the particle heating is distributed
broadly through the fluid.  We have seen that
the thermal photons follow the Amati et al. (2002) relation 
-- both in slope and normalization -- if they are generated in
the turbulent core of the jet near the photosphere of the Wolf-Rayet star
(eq. [\ref{amati}]).   And the scattering depth can attain values
near unity in an outflow of isotropic
luminosity $10^{51}\,L_{\rm rel\,51}$ ergs s$^{-1}$ and
duration $10\,\Delta t_1$ s, moving in a Wolf-Rayet wind with
mass-loss rate $10^{-5}\,\dot M_{w\,-5}\,M_\odot$ yr$^{-1}$
(eq. [\ref{taushocked}]).  This requires a hard
photon spectrum (energy index $\beta = -1$), which is consistent
with a significant majority of BATSE bursts (Band et al. 1993) and
essentially all of the sample of Ginga bursts analyzed by
Strohmayer et al. (1998) (excepting those with high peak energies
in which the high-energy tail was not fully sampled).  
Even when $\dot M_w < 10^{-5}\,M_\odot$ yr$^{-1}$, the optical depth
can be strongly enhanced by the residual
inertia of the breakout shell, and
by angular inhomogeneities in the outflow that allow side-scattering
of photons off slower, denser material (\S \ref{gamgam}).

A net optical depth $\tau_\pm \ga 1$ in the dissipation zone
is required for several reasons.  First, the energy radiated by
each charge is inversely proportional to $\tau_\pm$.
The energy of the inverse-Compton photons scales as
$\tau_\pm^{-2}$ at small optical depths, and so the peak of the
inverse-Compton spectrum lies far above the peak of the thermal seed
photon spectrum.  
Second, a hard inverse-Compton tail $dL_\gamma/dE_\gamma 
\propto E_\gamma^{-1/2}$ is generated below the peak energy
if the base of the non-thermal electron spectrum lies at
an energy $\gamma_e \gg 1$.  Third, a significant fraction
of seed blackbody photons must be scattered, so as to avoid a
strong and localized thermal bump in the transmitted spectrum .

A lower bound to the luminosity of the seed photons is obtained
by demanding that the first-order inverse-Compton (IC) photons carry
a significant fraction of the bolometric output of the burst.
As the heated particles cool off, and the
energy density of the IC photons may begin to exceed
that of the blackbody seeds, in which case the first-order IC photons become
the dominant coolant for the heated pairs.
Nonetheless, because $\ell' > 1$, it is possible for the particles
to be energized repeatedly  -- followed by cooling
off the ambient radiation field -- so that the 
instantaneous energy remains close to $\gamma_e \sim 1$.
Starting with a compactness $\ell_{\rm bb}'$ in seed photons, it
is possible to radiate a total energy $\ell' \sim (\ell_{\rm bb}')^2$
this way.  The seed photons must carry a minimal fraction
of the outflow energy 
\be
{\ell'_{\rm bb}\over \ell'} \ga (\ell')^{-1/2}
\ee
to avoid the appearance of a 
prominent $E_\gamma^{1/2}$ tail in the soft gamma-ray spectrum.

The existence of some bursts (or subcomponents of bursts)
with very soft high-energy
spectra (Pendleton et al. 1997) is, of course, fully consistent
with this model.  The thermal seed photons will not be significantly
reprocessed if the Wolf-Rayet wind has a mass-loss rate much
less than $\sim 10^{-5}\,M_\odot$ yr$^{-1}$.  At the same time,
one can expect that the optical depth will continue to drop as
the ejecta expand, so that the later parts of a burst will
have a tendency to be spectrally softer.  So there is a general
expectation that NHE {\it sub}-components of bursts will be concentrated
toward the end of the burst.  A good example is
GRB 920622B (Fig. 2a in Pendleton et al. 1997).  

There is, nonetheless, a basic tension in this model between
maintaining the $E_{\rm peak}-L_{\rm iso}$ relation, and allowing
rapid variability in a burst light curve.  It should be emphasized
that beaming of the inverse-Compton photons will be maintained
when $\tau_\pm \ga 1$, because the motion of the electrons and
positrons is collimated along the direction of the magnetic field.
Nonetheless, strong beaming cannot be maintained below the 
thermal peak energy at $\tau_\pm \ga 1$, because the Compton
parameter $y_{\rm C} = {4\over 3}\tau_\pm \langle \beta_e^2\rangle$
is limited to a value $y_{\rm C} \la 1$.  
The relative absence
of bursts with smooth light curves at the highest measured
fluxes (Norris et al. 2005) may be connected to the modest
decrease in $\tau_\pm$ that is expected at higher luminosities,
$\tau_\pm \propto L_{\rm rel}^{-1/2}$.  The hypothesis that
greater variability corresponds to lower optical depths is
testable:  more variable bursts should be spectrally harder, on
the average, than the Amati et al. (2002) relation would imply.

A hard gamma-ray spectrum depends on the
presence of inhomogeneities
in the outflow.  We now consider three possibilities that are suggested
by this model.

\subsubsection{First-order Fermi Acceleration of Pairs at a \\
Mildly Relativistic
Reverse Shock, followed by \\ Inverse-Compton Cooling}

A non-thermal positron tail extending upward from a minimum energy
$\gamma_e \sim 1$, and having a significant optical depth $\tau_\pm \ga 1$,
would generate a hard, high-energy tail to the gamma-ray spectrum
by cooling off the seed thermal photons (Thompson 1997).  
The asymptotic Lorentz factor that the jet material attains after
escaping the Wolf-Rayet star (eq. [\ref{gambcrit}])
is barely a factor 2 larger than
the equilibrium Lorentz factor of the contact during the deceleration
phase (eq. [\ref{gamrdecel}]).  This means that the reverse shock
wave is mildly relativistic, and the mean energy of the thermal
pairs behind the shock is indeed $(1-2)m_ec^2$.  However, our
deceleration model suggests that the optical depth of the pairs
crossing the reverse shock, and the energy they carry, 
are too small to produce the desired effect.

Collisions between gamma-rays that flow
upstream across the reverse shock will generate a somewhat
smaller depth, $\tau_\pm = 0.1\,\tau_{\pm\,-1}$, than
they do in the downstream fluid .  (This optical depth
is sensitive to the amount of cooling and compression that occurs downstream
of the shock.  The bulk motion 
of the fluid and the beaming of the emission downstream of the 
reverse shock both have the effect of suppressing
the pair creation rate upstream.)
This means that the pairs carry only a fraction
\be
\varepsilon_\pm \sim {\tau_\pm\over \ell'} \sim 10^{-2}\tau_{\pm\,-1}
\ee 
of the outflow energy when they hit the reverse shock.

The proportion of the outflow energy
carried by the ions, $\varepsilon_b = \Gamma_{b\,\rm crit}/\Gamma_b$,
can easily be larger than $\varepsilon_\pm$, 
even though the number of positrons exceeds the
number of ions by a factor $\sim (m_p/m_e)\varepsilon_\pm/\varepsilon_b$.
Particle-in-cell simulations of pair-dominated shocks show that
the shocked positrons can acquire a non-thermal tail by absorbing
the cyclotron waves emitted by the ions (Hoshino et al. 1992).
However, the ions are only mildly relativistic behind the reverse shock,
and so the frequency of the ion ring is too low to resonate
with the gyromotion of the positrons.  
(While Fermi-acceleration of the ions is possible,
such a non-thermal ion tail would presumably be a much weaker emitter
of ion cyclotron waves.)   

Two other problems with this mechanism
present themselves.  First, rapid variability in the gamma-ray
emission is not possible at a reverse shock that is triggered by
the deceleration off the ambient medium (Sari \& Piran 1997).
And, second, non-thermal positrons of an energy
$\gamma_e \sim 10^2\varepsilon_B'^{-1/2}$ will emit synchrotron
photons in the optical band (\S \ref{opsync}).  The observed upper bound of
$\sim 10^{-4}-10^{-3}$ to the fraction of the bolometric output that
is carried in the optical-IR band
(Akerlof et al. 1999; Vestrand et al. 2005) then requires that
$\varepsilon_B' \la 10^{-4}$ upstream of the
reverse shock. 

For all of these reasons, we turn to other mechanisms for energizing
pairs behind the reverse shock, that are special to strongly
magnetized plasmas.

\subsubsection{Gradient in Electron Energy due to \\ Gradual 
Pair Loading}\label{gradheat}

The scattering depth builds up after the relativistic outflow passes through
the reverse shock.  The minimal optical depth of the entrained 
electrons and ions is less than $(m_e/m_p)\ell' \la 10^{-2}$;
the optical depth in pairs is roughly an order of magnitude larger.

There is, therefore, a {\it distribution} of magnetization parameters
$\sigma_\pm$ in the fluid between the reverse shock and the contact.
As pairs are created by photon collisions, the shocked magnetofluid will also
be heated continuously (by reconnection of the reversing magnetic 
field; \S \ref{flasheat}).  This leads to a steady decrease in 
the mean energy (eqs. [\ref{gamslow}], [\ref{gamfin}]) of the
heated electrons and positrons (Fig. 5).

The energy of the inverse-Compton photons decreases in parallel with
(\ref{gamfin}).  To illustrate 
the effect, we take the pair density to incease as a power of time, 
$\sigma_\pm \propto B^2/n_\pm \propto t^{-\alpha}$, and the wave energy density to decrease
as $\delta B_0^2 \propto t^{-\beta}$.  One then has
$\gamma_e \propto t^{-(\alpha+\beta)}$ during repeated episodes of
flash heating, followed by Compton cooling.
The scaling is slightly different, $\gamma_e \propto t^{-(\alpha+\beta+1)/2}$,
if the heating is slow enough to be balanced continuously
 by Compton cooling.
The energy of the Comptonized thermal photons decreases with
distance behind the reverse shock, as
$E_\gamma \propto t^{-2(\alpha+\beta)}$ (flash heating) and
$E_\gamma \propto t^{-(\alpha+\beta+1)}$ (continuous heating).
When the pair-creating photons are supplied by the emission at smaller
radius, one has $\alpha \simeq 1$.  
The fluence $F_{\rm IC}$ 
in high energy photons is proportional to $\delta B_0^2 t$, and so 
\be\label{fluspec}
{dF_{\rm IC}\over d\ln E_\gamma} \propto t^{-\beta+1} \propto
E_\gamma^{-\gamma},
\ee
where
$\gamma = (1-\beta)/2(1+\beta)$ (flash heating) and
$\gamma = (1-\beta)/(2+\beta)$ (continuous heating).  The 
same spectrum is obtained, independent of the mode of heating, when
the tubulent intensity is independent of distance behind
the reverse shock ($\beta = 0$).  In this case, the integrated
spectrum is hard, $dF_{\rm IC}/d\ln E_\gamma 
\propto E_\gamma^{-1/2}$.  One also recovers the 
observed scaling (Fenimore et al. 1995) 
between peak duration and frequency, $t \propto E_\gamma^{-1/2}$, 
independent of the mode of heating.

\vskip .2in
\centerline{{
\vbox{\epsfxsize=7.5cm\epsfbox{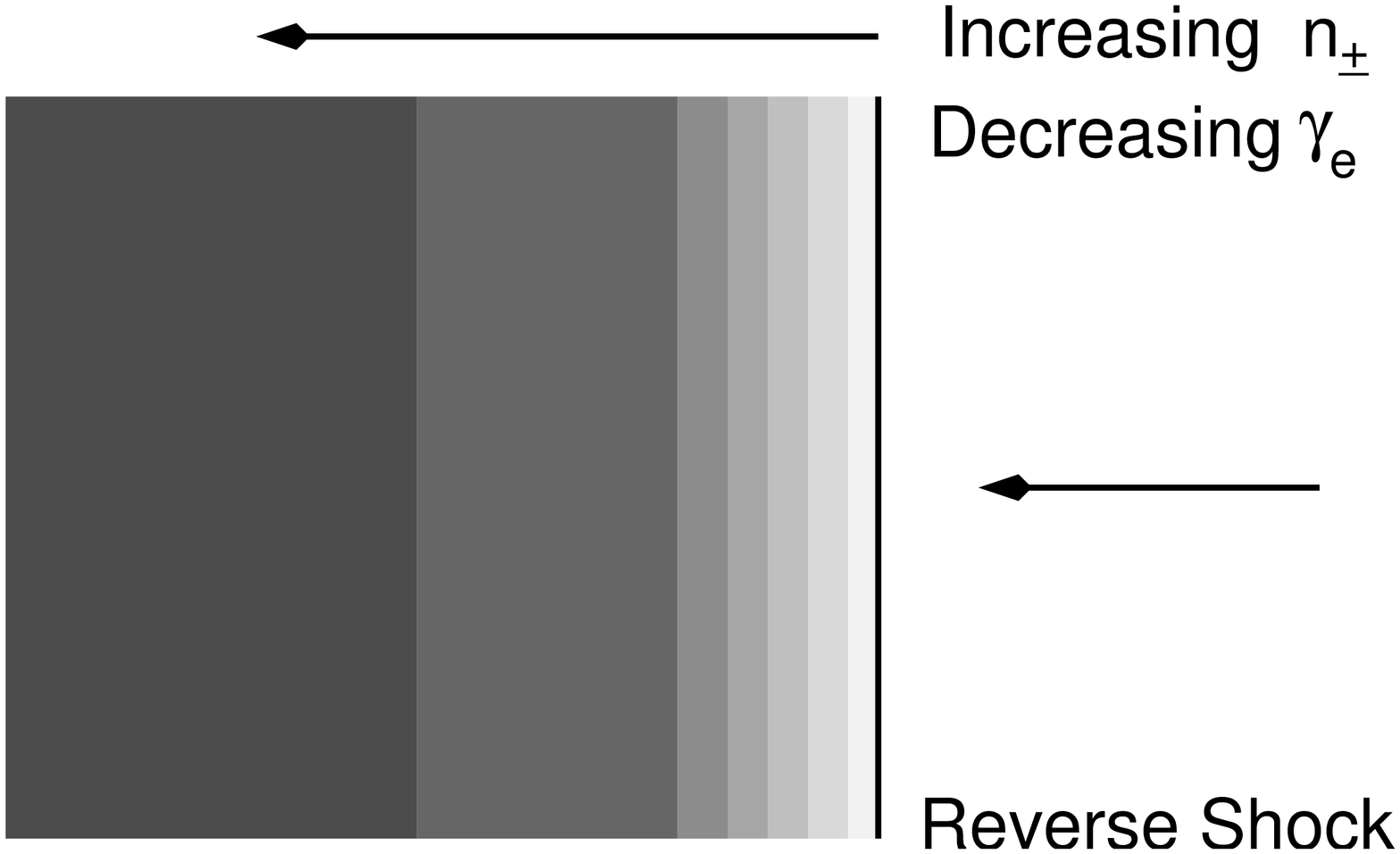}}}}
%\begin{figure}[ht]
%\epsscale{.9}    
%\plotone{fig5.eps}
\figcaption{The pair density builds up gradually after the magnetofluid
passes through the reverse shock wave, due to a finite rate
of photon collisions.  This leads to a gradual
decrease in the mean energy of the charges that are heated
electrostatically by decaying, high-frequency MHD turbulence.  A
high energy gamma-ray spectral tail emerges from the Compton
cooling of this extended particle energy spectrum (eq. [\ref{fluspec}]).}
\label{figgrad}
%\end{figure}
\vskip .2in

An important feature of this mechanism is that the emission of the 
soft photons lags behind that of the hard photons:  the softer photons
are emitted at later times on average as the pair density builds up.
The presence of such a hard-soft lag is a well established feature
of a subclass of gamma-ray bursts with smooth and easily
separable sub-pulses (Norris et al. 2005, and references therein).
Its existence therefore provides interesting circumstantial evidence
for this type of feedback between pair creation and distributed heating.

\subsubsection{Turbulent Cells Driven by Reconnection}

The turbulence in the outflow is expected to be strongly intermittent
if the magnetic field reverses sign in a stochastic manner.  
The size $L_0 = \varepsilon_0(r/\Gamma_c)$ of the
`cells' can be related to the reversal scale of the field
through eq. (\ref{louterb}).  They are typically smaller
than the causal scale $r/\Gamma_c$, so that the expansion of the outflow has
a small effect on their internal dynamics.

A high-energy gamma-ray tail can be generated in
cells with optical depths $\Delta \tau_\pm \ga 0.1$, when the photons are 
multiply scattered by the bulk turbulent motions (Thompson 1994, 1997).
While this mechanism could operate in bursts with smooth light curves,
it is disfavored in highly variable bursts.  Strong beaming of the
inverse-Compton photons requires relativistic bulk motion in this case,
so that the various orders of Compton scattering should be 
well separated in frequency.  A smooth power-law spectrum is not
naturally produced unless the dissipation is spread over a wider
range of radius.

\subsection{Compton Cooling of Pairs behind the Forward Shock}\label{hetail}

A pair-rich medium sweeps past
the forward shock during the prompt deceleration phase. 
Although the number of positrons
in this flow consistently exceeds the number of ions, the proportion of the
kinetic energy carried by pairs gradually declines from a value
${\cal M}_\pm-1 \sim 1$ at $r \sim R_{\rm decel-}$ to $\sim
(R_{\rm decel+}/R_{\rm decel-})^{-1/2} \sim 0.2$ when the reverse
shock has finished passing through the ejecta shell
(eq. [\ref{mdecel}]).  The total energy dissipated
at radius $r$ scales as $\sim (r/R_{\rm decel+})^{3/2}$; a fraction 
$\sim {1\over 2}$ of this being dissipated behind the forward shock.

The kinetic energy of the pairs is readily radiated away behind
the forward shock:  the cooling time is a small fraction
$\sim 1/\ell'\Gamma_\pm$ of the flow time behind the shock.  
We have argued that the magnetic field is likely to
carry a smaller fraction (\ref{varbfor}) of the energy behind the forward
shock than do the gamma-ray photons.  
The shocked pairs therefore lose energy by inverse-Compton
scattering the gamma-ray photons (Thompson 1997; Beloborodov 2005;
Fan, Zhang, \& Wei 2005).  

The spectral distribution of these inverse-Compton photons 
provides a direct probe the dynamics of the ejecta shell,
and of the coupling between the ions and the lighter charges. 
First let us consider the case where the ions decouple entirely.
The shocked pairs have a mean energy
$\Gamma_\pm = \Gamma_c/2\Gamma_{\rm amb}$, and the
characteristic energy of the inverse-Compton photons is
\be
E_{\rm IC} \sim {4\over 3}\Gamma_\pm^2\,E_{\rm peak} \sim 
    {4\over 3}\,\left({r\over R_{\rm decel-}}\right)^{3/2}\,E_{\rm peak}.
\ee
At the end of the prompt emission phase ($r = R_{\rm decel+}$), one finds
\be\label{eicmax}
{E_{\rm IC}(R_{\rm decel+})\over E_{\rm peak}}
     \sim {4\over 3}\left({R_{\rm decel+}\over 
    R_{\rm decel-}}\right)^{3/2} = 1\times 10^2 \,\varepsilon_+^{-1}\,
      L_{\rm rel\,51}^{1/2}\,\dot M_{w\,-5}^{-3/2}\,V_{w\,8}^{3/2}\, 
      \Delta t_1,
\ee
which is $E_{\rm IC}/E_{\rm break} \sim 400\,L_{\rm rel\,51}^{1/2}
\dot M_{w\,-5}^{-3/2}$ for $\varepsilon_+ = {1\over 4}$.  
This means that the inverse-Compton
photons have an energy less than $\sim m_ec^2$ in the frame of the contact
only if $E_{\rm peak} \la 50$ keV.  
It will be recalled that $\Gamma_\pm \propto \Gamma_c^{-3}$, and so
\be
E_{\rm IC} \propto \Gamma_c^{-6}.
\ee
The energy of the inverse-Compton photons is a strong
function of the Lorentz factor of the contact, and a modest increase
or decrease in the external inertia will push $E_{\rm IC}$ to higher
or lower energies.

The total power in Comptonized photons 
at energies $E_\gamma \leq E_{\rm IC}$ is 
\be\label{icspec}
{dL_{\rm IC}\over d\ln E_\gamma} \simeq
L_{\rm rel}\,({\cal M}_\pm -1)\left({R_{\rm decel+}\over 
R_{\rm decel-}}\right)^{-1/2}\left({E_\gamma\over E_{\rm IC}}\right)^{1/2}.
\ee
Unless the input gamma-ray spectrum is soft at high energies, this
inverse-Compton tail will not be visible, due to the combination of
small factors $({\cal M}_\pm-1)(R_{\rm decel+}/R_{\rm decel-})^{-1/2}$.  

A rapid coupling between the gyromotion of the protons and positrons
could cool the protons.  The PIC simulations of 
Hoshino et al. (1992) suggest that the positrons will develop a 
hard, non-thermal tail (number index $d\ln N_{e^+}/d\ln\gamma \ga -2$)
when $m_p n_p \ga m_e n_{e^+}$.  However, resonant excitation of
the gyromotion of positrons of energy $\sim \gamma_{e^+}$ requires
that the proton ring emit a significant power in cyclo-synchrotron
waves at a frequency $n(eB/\Gamma_\pm m_p c) = 
(m_p/m_e)(eB/\gamma_{e^+} m_p c)$.  Thermal positrons ($\gamma_{e^+} = 
\Gamma_\pm$) behind the shock will be heated only if
$0.3\Gamma_\pm^2 \ga (m_p/m_e)\Gamma_\pm^{-1}$, i.e., only if 
$\Gamma_\pm \ga 20$ and $n \ga 6\times 10^3$.  In the 
case where the proton ring has a small thermal spread, which limits
the harmonic to a value $n \leq n_{\rm max} \ll m_p/m_e$, one requires
a secondary mechanism for creating non-thermal positrons with 
an energy $\gamma_{e^+} \ga n_{\rm max}^{-1}(m_p/m_e)\Gamma_\pm$.
Unless the non-thermal component of the spectrum is very hard, its
energy density is a modest fraction of the thermal positron energy
density behind the shock (which itself is $\la 10\,\%$ of the energy 
density of the protons). 
For this reason, it is plausible that the protons will retain
a significant fraction of their kinetic energy after thermalizing
downstream of the shock.
Rapid proton cooling requires a collective process that persists 
longer than the synchro/Compton cooling time of the absorbing particles.

It is also useful to contrast this calculation with previous estimates,
which gives some sense of the subtleties involved.  
The cooling of synchrotron-emitting particles is dominated by
inverse-Compton scattering of the synchrotron photons if the
energy density of the cooling particles exceeds the energy density of
the magnetic field and the synchrotron radiation is
sufficiently soft (e.g. Sari \& Esin 2001).  It has been suggested
that the prompt gamma-ray photons will be inverse-Compton scattered
at the forward shock, thereby triggering a pair cascade (Fan et al. 2005).
This calculation however neglected pair creation in the external
medium and assumed rapid energy equilibration between ions and electrons, 
thereby yielding $\langle\gamma_e\rangle \sim 10^3$ rather than 
$\langle\gamma_e\rangle \sim 10$ behind the forward shock.   
Pair creation in the Wolf-Rayet wind guarantees at least a $\sim 10$\%
conversion efficiency of bulk kinetic energy to inverse-Compton
photons at the forward shock.  
A high-energy particle tail is, of course,
probably present at the forward shock;  but the lower-energy
part of the inverse-Compton spectrum will be dominated by cooling 
particles that start with an energy
$\sim \langle\gamma_e\rangle$.  

It has also been noted that
electrons emitting optical synchrotron photons could upscatter
the gamma-ray peak photons to GeV-TeV energies (Beloborodov 2005).
However, Compton cooling of these particles cannot,
by itself, explain the low optical efficiency that is observed in 
GRBs 990123 and 041219a (Akerlof et al. 1999; Vestrand et al. 2005)
if the magnetic field is strong in the gamma-ray emitting region
(\S \ref{opsync}).  One infers, instead, that the particles must
be restricted to a more limited range of energy and pitch angle
than one would expect from standard shock-acceleration.

Compton GRO detected a very bright burst, GRB 941017, whose bolometric
output was dominated by a high-energy component with an energy
index $\beta \simeq 0$ above $1-10$ MeV (Gonz\'alez et al. 2003).
This high-energy emission probably involved the inverse-Compton scattering
of the prompt burst photons by energetic electrons and positrons.  This
spectrum
was harder than eq. (\ref{icspec}), and is more consistent with 
continuously heated particles.  For this reason, and because of the
relatively low normalization of (\ref{icspec}), we favor emission from
pairs that are heated
by a reconnecting magnetic field behind the reverse shock.
A more detailed discussion of this event is given in \S \ref{941017}.

\subsection{Harder Classes of Gamma-Ray Bursts}

Even when an extended Wolf-Rayet envelope is
absent, one still obtains a characteristic temperature for thermalization
at the base of the outflow.  The free expansion phase begins at 
a smaller radius.  In a rapidly rotating source, this
thermalization radius can be normalized
to the light cylinder, 
$R_{\rm bb} \sim cP/2\pi = 5\times 10^6\,(P/{\rm msec})$ cm;
whereas in a Soft Gamma Repeater flare it can be normalized
to the radius of the star.
The blackbody temperature is then higher than we deduced
for a thermalization radius of $R_\star = 2\times 10^{10}$ cm
(eq. [\ref{amati}]);  it works out to
\be
k T'_{\rm bb} = 0.5\,{(\varepsilon_{\rm bb}L_{50})^{1/4}\over
\Gamma(R_{\rm bb})^{1/2}\,P_{\rm msec}^{1/2}}\,
\left({\Delta\Omega\over 4\pi}\right)^{1/4}\;\;\;\;{\rm MeV}.
\ee
The normalization here is well above that 
relation between $E_{\rm peak}$ and $L_{\rm iso}$ for long bursts
with measured redshifts.

The same conclusion holds for two recent short gamma-ray bursts
with tentative redshift measurements:  GRB 050509b
($z \simeq 0.2$;  Bloom et al. 2005), and GRB 050709 ($z \simeq 0.16$;
Price, Roth, \& Fox 2005).  In both cases, the spectrum is therefore
consistent a binary neutron star merger as the source of the burst; and,
in the case of GRB 050509b, the energetics is marginally consistent
with an extragalactic SGR flare.  (The accretion-induced 
collapse of a white dwarf following a binary white dwarf merger
could give rise to the birth of a young, active 
magnetar in a galaxy with a low star formation rate:
King, Pringle, \& Wickramasinghe 2001; Thompson \& Duncan 1995.)

\subsection{Suppression of Optical-IR Synchrotron Emission}\label{opsync}

The presence of the Wolf-Rayet wind forces the relativistic ejecta
to decelerate at a radius where electrons of an energy $\gamma_e 
\sim 10^2$ will radiate in the optical band behind the
reverse shock (eq. [\ref{esync}]). 
Observations of direct optical-IR emission from two gamma-ray bursts
(990123: Akerlof et al. 1999; and 041219a: Vestrand et al. 2005; 
Blake et al. 2005), and upper bounds on other bursts (Akerlof et al. 2000)
therefore allow stringent constraints to be placed
on the particle distribution in the outflow.   The fraction of the bolometric
power released in the optical band in these two bursts was $\sim 10^{-4}
 - 3\times 10^{-3}$ in different parts of GRB 990123, and $\sim 1-3 \times
10^{-4}$ in GRB 041219a.   Electrons/positrons of an energy 
$\gamma_e \sim 10^2$ must, therefore, contribute only modestly to the 
energy density in the emitting region;  or have small pitch angles.

Taking the simplest case
of mono-energetic electrons with a fixed pitch angle
$\kappa$, the absorption coefficient at frequency 
$\nu' \la 0.3\gamma_e^2(\sin\kappa)\nu_{Be}' = 
0.3\gamma_e^2(\sin\kappa)(eB'/2\pi m_ec)$ is
\be
\alpha_\nu' = {8\pi^2en_\pm'\over 3^{4/3}\,\Gamma(1/3)\,(\sin\kappa) B'}
\left[\gamma_e{\nu'\over (\sin\kappa)\nu_{Be}'}\right]^{-5/3}
\ee
in the bulk frame (e.g. Rybicki \& Lightman 1979).
The optical depth $\tau_\nu = \alpha_\nu' (r/\Gamma_c) 
\propto \gamma_e^{-5/3}$,
and so there is a critical energy $\gamma_e(\tau_\nu=1)$ above which the medium
is optically thin at any given frequency.  We normalize this 
frequency\footnote{We are most interested in the case of anisotropic emission,
with a small pitch angle $\kappa \ll 1$, along a magnetic field with a
non-radial orientation in the bulk frame.  
The mean frequency of a synchrotron photon
therefore transforms as $\nu = \Gamma_c\nu'$ from the bulk frame to
the rest frame of the engine.}
to $\nu' = \Gamma_c^{-1}({\rm 1~eV}/h) 
   \simeq 5\times 10^{12}\,(\Gamma_c/50)^{-1}$ Hz in the bulk frame.
Relating the pair density $n_\pm'$ to the Thomson optical depth $\tau_\pm'$,
one finds
\be
\gamma_e(\tau_\nu=1) = 0.9\,{\tau_\pm^{3/5}(\sin\kappa)^{2/5}
  \over \alpha_{\rm em}^{3/5}\,(B'/B_{\rm QED})^{3/5}}\,
  \left({\nu'\over \nu_{Be}'}\right)^{-1}.
\ee
Setting $B' = B/\Gamma_c = 2\times 10^3\,B_5(\Gamma_c/50)^{-1}$ G 
in the gamma-ray emission zone
(we expect the magnetic field to have
a weak dependence on $L_{\rm rel}$ and $\dot M_w$: eq. [\ref{magval}])
gives
\be
\gamma_e(\tau_\nu=1) = 3\times 10^4
       \,B_5^{2/5}\tau_\pm^{3/5}\,(\sin\kappa)^{2/5}\,
\left({\Gamma_c\over 50}\right)^{3/5}.
\ee
In practice, the optical depth in such energetic particles is limited
to $\tau_\pm \sim 1/\gamma_e^2$, and so we have
\be
\gamma_e(\tau_\nu=1) = 1\times 10^2
       \,B_5^{2/11}\,(\sin\kappa)^{2/11}\,
\left({\Gamma_c\over 50}\right)^{3/11}.
\ee

Synchrotron radiation in the optical band requires the electrons to
be more energetic than
$0.3\gamma_e^2(\sin\kappa)h(\Gamma_c\nu_{Be}') > 1~{\rm eV}$ or, equivalently,
\be
\gamma_e \ga \gamma_{e\,\rm min} = 50 B_5^{-1/2}\,(\sin\kappa)^{-1/2}.
\ee
The optical synchrotron emission is self-absorbed if
$\gamma_{e\,\rm min} < \gamma_e < \gamma_e(\tau_\nu=1)$, and is
determined only by the temperature $k_{\rm B}T_e' \simeq {1\over 3}
\gamma_e m_ec^2$ of the electrons.  The
energy density of optical photons in the bulk frame is
\be
(\nu U^O_\nu)' = {k_{\rm B}T_e'\over \pi^2}\,\left({2\pi\nu'\over c}\right)^3.
\ee
The isotropic luminosity of the optical photons ($h\nu' = 1~{\rm eV}/h$) is
\ba
{\nu L_\nu^O\over L_{\rm rel}}\biggr|_{R_{\rm decel+}}
    &=& 4\pi r^2 c\,\Gamma_c^2\, (\nu U^O_\nu)' / L_{\rm rel} \nn
 &=& 1\times 10^{-2}\,\left({\gamma_e\over 10^2}\right)\,
            L_{\rm rel\,51}^{-1/4}\,\dot M_{w\,-5}^{-1/4}\,V_{w\,8}^{1/4} \,
              \Delta t_1^{3/2}.\nn
\ea
Here we have substituted expressions (\ref{fdecelb}) and (\ref{gamrdecel}) for
the deceleration radius $R_{\rm decel+}$ and the Lorentz factor at this radius.

One possible way of avoiding excess optical/IR emission 
is to inject electrons with random Lorentz factors
$\gamma_e \sim 10^4-10^5$, so that their synchrotron radiation
is in the gamma-ray band (e.g. Rees \& M\'esz\'aros 1994).  
The synchrotron output in the optical/IR band is then suppressed by
a factor $(\hbar\omega/E_{\rm peak})^{1/2} \sim 10^{-2}-10^{-2.5}$.
However, such a large mean energy per particle
is inconsistent with the level of pair creation that must necessarily
occur if decelerating agent is a dense Wolf-Rayet wind or breakout shell
(\S \ref{gamgam}).  In such a compact environment,
the mean energy per electron/positron is reduced to
$\sigma_\pm \sim \ell_B'/\tau_\pm \sim 3-10$.  The cooling spectrum emitted
by the particles ($dL/d\ln E_\gamma \propto E_\gamma^{1/2}$) is
also  inconsistent with the observed low-energy spectra of
GRBs (Ghisellini \& Celotti 1999).  

The optical synchrotron emission is also suppressed if the
high-energy portion of the gamma-ray spectrum
arises from inverse-Compton scattering of thermal seed photons,
and if shock acceleration is {\it not} the principal energization mechanism.
The inverse-Compton
spectrum extends up to an energy $\sim 4\gamma_e^2 k_{\rm B}T_{\rm bb}$,
which means that a distribution of 
particle energies up to $\gamma_e \la 10-30$ is sufficient to create
an extended gamma-ray spectral tail.
The maximum energy to which electrons and
positrons may be flash heated by decaying MHD turbulence
is $\gamma_e \sim \sigma_\pm$, which is
typically less than $\sim 10^2$ in the prompt deceleration phase.  
Their synchrotron emission is also suppressed by a factor
$\sim 1/\gamma_e^2\ln\Gamma_c$ compared with the
inverse-Compton emission (in the case of a flat 
particle energy spectrum; eq. [\ref{eqpitchd}]).  

By contrast, the particle spectrum that is generated by shock
acceleration is cut off by synchrotron cooling at a much higher energy,
where the synchrotron energy is $\sim \alpha_{\rm em}^{-1}\,m_ec^2$.
First-order fermi acceleration is therefore 
disfavored as the mechanism for heating particles behind the reverse shock,
given that a GRB radiates only $\la 10^{-3}$ of its bolometric
output in the optical-IR band.
The same conclusion applies to other acceleration mechanisms which
operate on a short timescale, including the resonant absorption of 
ion cyclotron waves by positrons (Hoshino et al. 1992).

\subsection{Transition of Prompt Gamma-Ray Burst to 
    Afterglow}\label{spectrans}

In the model advanced in this paper,
the prompt gamma-ray emission continues as long as the
shell of seed thermal photons overlaps the shell of turbulent magnetofluid.
We now examine the transition from the prompt burst emission to the
afterglow.  A
good example is GRB 980923 (Giblin et al. 1999).  At such a transition,
the prompt emission is often much more variable than the afterglow.

Several factors point to the
region between the reverse shock and the contact as the source
of the prompt gamma-ray emission.  First,
strong variability in the burst emission requires beaming of the
radiation in the bulk frame (\S \ref{gbeam}).
 Second, the radiative efficiency is 
larger when the pairs contribute a larger fraction of the particle
inertia, which is the case behind the contact (compare eqs.
[\ref{taurdecel}] and [\ref{taushocked}])  And, third, the magnetic 
field is expected to be stronger between the reverse shock and the contact
(\S \ref{shockwr}).

The Wolf-Rayet wind remains pair-loaded and mildly relativistic
beyond the point at which the reverse shock has completed
its passage through the ejecta shell.  We have found that it is still
being pushed to $\Gamma_{\rm amb} \sim 2\,\dot M_{w\,-5}^{2/3}\,
\Delta t_1^{-2/3}$
at the radius $R_{\rm recel+}$ (eqs. [\ref{fdecelb}] and [\ref{gamactrans}]),
independent of $L_{\rm rel}$.  Beyond this point,
$\Gamma_{\rm amb}$ decreases inversely with radius.
One infers that the ambient medium becomes fully quiescent only
after the seed thermal photons have decoupled from the forward shock.
The time since the burst scales as $r/\Gamma_c^2 \propto r^{3/2}$,
and so one expects an observable transition in the brightness of the X-ray
synchrotron emission from the forward shock around a time
$t \sim 30\,(1+z)\,\dot M_{w\,-5}$ s following a burst.  
Beyond this point, a much larger fraction of the energy density behind
the forward shock is in slowly cooling ions than during the prompt burst
phase.  It should be emphasized that there is no clear connection between
the amplitude of the prompt gamma-ray emission and the immediate
post-flare synchrotron emission in this model.  

A rapid decline in the X-ray flux has been observed following 
several Swift bursts, the most plausible explanation for which is
off-axis emission from the prompt emission phase (Tagliaferri et al. 2005).
Since the overlap of the prompt MeV photons with the forward shock will
not cut of sharply (we expect $\tau_{\rm T} \sim 1$ at the end of the 
prompt emission phase), this would require strong dissipation of the 
non-radial magnetic field behind the reverse shock, if the ejecta shell
were spherical. 

The  contact discontinuity is, however, expected to be strongly nonspherical
(\S \ref{gamgam}).  We have argued that Rayleigh-Taylor modes are excited 
in the cooling breakout shell over a wide range of angular scales. 
Relativistic ejecta penetrating through holes in the breakout shell will
form `subjets' that are boosted with respect to the shell material. 
The size of the holes decreases with the radius at which they are created,
and the final generation of holes is created on an 
angular scale $\la \Gamma_c^{-1}$.  One may therefore expect a few 
subjets to be present within each causal patch of
area $(r/\Gamma_c)^2$ (cf. Kumar and Piran 2000).

We note that
some bursts show very smooth pulses which decay exponentially.
These events appear to be dimmer on average, and are rare amongst
the brightest bursts (Norris et al. 2005).  Some possibly involve a different
emission geometry, such as the breakout of a mildly relativistic
shock from the surface of the Wolf-Rayet star (Tan et al. 2001) that is
powered by a buried jet.   It should also be noted that the optical
depth in pairs grows toward lower values of the isotropic luminosity
$L_{\rm rel}$ and higher Wolf-Rayet mass loss rates $\dot M_w$,
because the relativistic outflow is stopped more rapidly:
$\tau_\pm \propto L_{\rm rel}^{-1/2}\,\dot M_w^{5/6}$ (eq. [\ref{taushocked}]).
Variability in the burst light curve would be washed out when
$\tau_\pm \gg 1$.

There are suggestions from the analysis of broadband
afterglow emission that the mass density in the ambient wind
material is as low as $\dot M_{w\,-5}/V_{w\,8} \sim 10^{-3}-10^{-2}$ in
a few bursts (Kumar \& Panaitescu 2003; Chevalier et al. 2004).
In this case, one can still define a radius at which the 
the photon shell will decouple from the relativistic ejecta
(in the absence of a breakout shell).
Taking the Lorentz factor of the ejecta to be
$\Gamma_{\rm rel} = \Gamma_{b\,\rm crit}$ (eq. [\ref{gambcrit}])
gives the decoupling radius
\ba
R_{\rm overlap} &\;=\;& 2\Gamma_{\rm rel}^2 c\Delta t \nn
  &\;=\;& 3\times 10^{15}\,
 {\varepsilon_{\rm bb}^{1/2} L_{\rm rel\,51}^{3/4}\Delta t_1^{5/4}\over
    (1-\varepsilon_{\rm bb})^{3/4}}\,
\left({R_\star\over 2\times 10^{10}~{\rm cm}}\right)^{-1/2}\;\;\;\;{\rm cm}.\nn
\ea
In this expression, we have made use of the relation $L_{\rm iso} = 
L_{\rm rel}/(1-\varepsilon_{\rm bb})$.  The compactness in the bulk frame is
\ba\label{ellover}
\ell' &\simeq& {\sigma_{\rm T}L_{\rm rel}\Delta t\over 4\pi 
        \Gamma_{\rm rel} R_{\rm overlap}^2 m_ec^2}\nn
 &=& 1.0\,{(1-\varepsilon_{\rm bb})^{15/8}\over\varepsilon_{\rm bb}^{5/4}
            L_{\rm rel\,51}^{7/8}\,\Delta t_1^{13/8}}\,
 \left({R_\star\over 2\times 10^{10}~{\rm cm}}\right)^{5/4}.\nn
\ea
Note that a much lower compactness would be obtained if the ejecta began
to expand freely from a smaller radius $\sim 10^7-10^8$ cm (rather
than the photosphere $R_\star$ of the Wolf-Rayet star).  
It should also be emphasized that $\ell'$ is sensitive to the asymptotic
Lorentz factor of the ejecta shell, $\ell' \propto \Gamma_{\rm rel}^{-5}$.
This means that, if the reverse shock is sub-relativistic, then the 
photon shell will decouple from the ejecta shell before the reverse shock has
completed its passage through the ejecta shell.  However, a sub-relativistic
reverse shock implies that
$\Gamma_{\rm rel} \la 50\,L_{\rm rel\,51}^{1/4}\Delta t_1^{-1/6}$ 
(eq. [\ref{gamrdecel}]), 
in which case $\ell' \ga  6\,L_{\rm rel\,51}^{-1/4}\,\Delta t_1^{-1/6}$
at the radius of decoupling.  The portion of the magnetized shell
which dissipates in the presence of the thermal photon bath will, therefore,
do so at a large compactness.

\subsection{The Peculiar (?) Burst 941017}\label{941017}

The hard, high-energy spectral component that
was observed in GRB 941017 (Gonz\'alez et al. 2003) has interesting
implications for the mechanism of particle heating.   The 
energy index was observed to flatten from $\beta \simeq -{1\over 2}$
to $\beta \simeq 0$ at high energies.  The high-energy component
decayed more slowly, and so the transition energy between the
two power-law components also decreased with time, from $\sim 10$ MeV
down to $\sim 300$ keV.  The burst had a very long duration,
$T_{90} \simeq 10^2$ s, and its fluence  was one of the dozen
highest measured by BATSE.  The source was, therefore, probably located
at a cosmological redshift $z < 1$.
It is not clear how atypical this high-energy emission is, given the
brightness of GRB 941017 and the relative difficulty of detecting
it in lower-fluence bursts.  The presence of strong high-energy emission
in bursts with relatively soft 1-10 MeV spectra would significantly boost
the pair creation rate and the optical depth in the emitting plasma.

Inverse-Compton
scattering of the $\sim 0.5$ MeV peak photons is the most obvious
source of the high-energy component.  If the cooling particles are
isotropically distributed, then the $\beta = 0$ index is inconsistent
with passive Compton cooling ($\beta = -{1\over 2}$) and so the
particles must be continuously heated (Stern \& Poutanen 2004).
The slope of the rising spectrum would then mirror that of the
spectrum below $E_{\rm peak}$;  indeed $\beta \simeq 0$ is the most
commonly measured low-energy spectral slope in GRBs.  

This line of reasoning allows one to place some significant constraints
on the mechanism of particle heating.    In particular, one
can show that the heated particles must occupy a small fraction of
the volume of the ejecta, during the peak of the burst emission.
 Given that the dominant contribution to the
bolometric output of GRB 941017 is at high energies, each electron
or positron must, on average, radiate a total energy
$\sim \sigma_\pm m_ec^2$ in inverse-Compton
photons. Note that the value of the magnetization parameter
$\sigma_\pm$ in the high-energy emission zone may be significantly
larger than the average value within the ejecta (\S \ref{gradheat}).
The cooling time
at a random Lorentz factor $\gamma_e \ga 30$ is much shorter than the burst
duration, $t_{\rm cool}/\Delta t \sim 1/\ell'\gamma_e$ (as measured
by the observer). 
This means that each particle must perform its emission duties
over a very short timescale, $t_{\rm emission}/\Delta t 
\sim (\sigma_\pm/\gamma_e)t_{\rm cool} \sim \sigma_\pm/\ell'\gamma_e^2
\la 10^{-4}\sigma_\pm$.
Energy must, therefore, be transferred to the particles over a short, 
but not microscopic timescale.  For example, shock acceleration does
not satisfy this constraint, because the residency time of a particle
of energy $\gamma_e \sim 30$ near the shock is orders of magnitude
shorter than $t_{\rm cool}$.   One infers that the outflow must
have small-scale structure, on a scale $\la 10^{-4}\sigma_\pm\,c\Delta t
\sim (0.01~{\rm s})\,\sigma_\pm\,c$, which allows dissipation to take place so
rapidly.  Indeed, reconnection of a reversing magnetic field in the outflow
will have the desired effect (\S \ref{shockrel}; \ref{gradheat}).

Could the $\beta = 0$ spectrum be a signature of slow Compton cooling
of the emitting particles?   This requires
the compactness to be tuned to a value somewhat smaller 
than $\sim \gamma_e^{-1} \la 0.03$.
During the peak of the burst emission,
we deduce $\ell' \simeq 5\,L_{\rm rel\,51}^{-1/4}
\dot M_{w\,-5}^{5/12}\,V_{w8}^{-5/12}$ 
from  eq. (\ref{compfin}), which indicates that the external wind inertia
must be exceedingly small for this explanation to hold.
It is, nonetheless, clear 
that the compactness of the $\sim 1$ MeV photons must drop rapidly toward
the end of the burst.  The increasing dominance of the high-energy
component is therefore consistent with a drop in $\tau_\pm$ and
an increase in the cooling time (at fixed $\gamma_e$) at the
reverse shock.   

Synchrotron self-Compton emission by $\gamma_e \sim 3\times 10^4$ electrons
at the forward shock
is another possible high-energy emission mechanism
(Granot \& Guetta 2003; Pe'er \& Waxman 2004a), 
which was motivated by the delayed onset of the high-energy component
in GRB 941017 and the apparent need for
slow cooling to explain the hard $\beta = 0$ spectrum.  We  note that for
the parameters $\varepsilon_B \la 10^{-4}$ and $\Gamma_c \sim 300$ 
adopted in this model, Compton cooling by the MeV photons will, in fact,
dominate\footnote{Including the Klein-Nishina suppression of the 
scattering cross section.} cooling by optical synchrotron photons
as long as the gamma-ray photons overlap the forward shock (Beloborodov 2005). 

\section{Conclusions}\label{conc}

We have examined, in some detail, the hypothesis that
long gamma-ray bursts are emitted by relativistic material that
breaks out from the photosphere of a Wolf-Rayet star, and then
propagates through the surrounding dense wind.  We have started
with a key simplifying assumption, that there are only two major sites of
dissipation in the outflow:  where the relativistic jet breaks
out of the Wolf-Rayet star (at $R_\star \sim 2\times 10^{10}$ cm);  
and the much larger zone where the reverse shock makes significant
backward motion through the ejecta shell (at a distance $10^{14}-
10^{15}$ cm).  The key components of the relativistic outflow are
a non-radial magnetic field and a thermal bath of hard X-ray photons.  The
magnetic field changes sign many times,
due to the stochasticity of the dynamo operating in the central engine.
The photons are created in the turbulent jet core as it is
heated by Kelvin-Helmholtz instabilities during the passage of
the jet through the envelope of the Wolf-Rayet star.  The kinetic
energy of the entrained baryons can be comparable to the energy carried
by these other two components, but does not have to be.

Our analysis has covered the deceleration of the ejecta shell, as
influenced by pair-creation in the ambient wind material;
the entrainment of a thin shell of stellar material by the jet head;
and the structure and composition of the fluid on either side of the
contact discontinuity.
We have described a mechanism by which
turbulent motions in an expanding,
relativistic magnetofluid will transfer
energy to electrons and positrons, and then (by Compton scattering)
to an ambient radiation field.  This inverse-Compton radiation is {\it beamed}
along the magnetic field in the bulk frame, thereby providing an
explanation for the rapid variability of many gamma-ray bursts, without
resort to an ill-defined source of internal variability in the outflow.
The ultimate source of seed photons is the relativistic jet itself
(and not in its sub-relativistic cocoon)
as it suffers Kelvin-Helmholtz instabilities during its emergence
from the Wolf-Rayet star.  

The normalization and slope of
the Amati et al. (2002) relation between
peak energy and isotropic energy of long gamma-ray bursts is reproduced,
essentially without free parameters.   This relation is {\it predicted}
to flatten from $E_{\rm peak} \sim 
L_{\rm iso}^{1/2}$ to $E_{\rm peak} \sim L_{\rm iso}^{1/4}$ below
a luminosity $L_{\rm iso} \sim 3\times 10^{50}$ ergs s$^{-1}$.  
This transition corresponds to a wide jet opening angle,
$\theta \ga 1/\sqrt{3}$, at the surface of the Wolf-Rayet star.

One also obtains an explanation
for why the short gamma-ray burst GRB 050509b (Bloom et al. 2005)
did not follow this relation:  the relatively hard spectrum of this burst
reflected a smaller radius at which the ejecta went into free expansion,
and at which the temperature of the seed black body photons was frozen.
The bright initial spikes of the SGR flares are another example of
energetic bursts which deviate from the Amati et al. relation, and are
consistent with a smaller thermalization radius of
10-100 km (Hurley et al. 2005; Palmer et al. 2005).

We close with a summary list of some addition conclusions.

\subsection{Additional Conclusions}
\begin{enumerate}

\item{} The passage of the reverse shock wave through the ejecta
shell occurs when the compactness of the radiation field near 
the contact discontinuity has dropped to a characteristic value
$\ell' \sim 10$.  At this
point, the gamma-ray flux across the foward shock is no longer sufficient
to push the ambient medium up to a speed comparable to that 
of the contact discontinuity.  There is no
detachment of the pair-loaded wind material from the ejecta shell.

\item{} The prompt gamma-ray emission takes place in a medium with
an optical depth $\sim 1$ in electron-positron pairs.  Pairs
generated by side-scattering of gamma-ray photons in the Wolf-Rayet wind
are the dominant source of scattering particles in between the forward
shock and the contact.  Pairs are created at a somewhat higher rate
by collisions between photons in the turbulent region between the
reverse shock and the contact.   This shocked material becomes optically
thin only outside $\sim 10^{15}$ cm from the central engine.  

\item{}  The shocked Wolf-Rayet wind and the shocked relativistic outflow
are themselves subject only to a weak Rayleigh-Taylor instability at
their mutual contact discontinuity.  However, the shell of matter that
is entrained by the jet head during the final breakout from the star
is much denser.  This breakout shell will gradually accelerate
as it intercepts relativistic ejecta from behind, and
will suffer from a strong corrugation instability as it cools.
The radiation that is trapped by the breakout shell has a temperature of
$\sim 2$ keV when the shell becomes optically thin.

\item{} The interaction of the relativistic ejecta with the 
fragmenting breakout
shell offers a hybrid mechanism for triggering dissipation and
gamma-ray emission.  In this case, the breakout shell does not
form a truely `external' medium, since its structure depends on its
interaction with the relativistic wind material at a smaller radius.
It is especially important when the mass-loss
rate in the Wolf-Rayet wind preceding the burst is much less than
$10^{-5} M_\odot$ yr$^{-1}$.   The role of
the corrugation instablity in generating long null regions in gamma-ray
bursts should be investigated.  

\item{} Electrons and positrons in the outflow are electrostatically
heated in regions of strong MHD turbulence.  The Kolmogorov energy
flux in torsional waves is damped at a smaller frequency than the
ion gyrofrequency.  We find that  
similar conclusion applies to the magnetic corona of a black hole,
and to any magnetofluid in which $B^2/8\pi$ is larger than the rest
energy density in electrons (pairs).

\item{} The radiation emitted by charges in a turbulent magnetofluid
is beamed along the background magnetic field.  The damping
of MHD turbulence in a medium with $(B')^2/8\pi \gg n_\pm' m_ec^2$
leads to electrostatic acceleration of charges along the magnetic field.
Indeed, when the gamma-ray emission
is triggered by the deceleration of the relativistic ejecta, one
requires beaming in the bulk frame to explain the
fast variability observed that is observed in many bursts.

\item{} The outflow is photon rich ($n_\gamma/n_\pm \sim 10^2-10^3$)
and the photons are the dominant
coolant of the electrons and positrons 
-- even if the magnetic field has a higher
energy density.  Numerical 
simulations (e.g. Pe'er, M\'esz\'aros, \& Rees 2005)
which do not account for the anisotropy or spatial
inhomogeneity of the 
electron distribution will lead to qualitatively different results.  

\item{}
The observed correlation $\delta t \propto E_\gamma^{-1/2}$ between photon
energy and the width of gamma-ray pulses arises naturally from inverse-Compton
cooling of pairs that have been flash-heated up to
a limiting Lorentz factor $\langle\gamma_e\rangle \sim \ell_B' (\delta B'/B')^2
/\tau_\pm \ga 1$.  Some substructure in GRBs does not show clear
asymmetries between rise and decay;  we ascribe it to variations in
the direction of beaming associated with wave excitations of the 
background magnetic field.

\item{}
The electrons/positrons that are heated by decaying MHD turbulence 
behind the reverse shock have a negative gradient in energy away from
the shock.  There is, correspondingly, a gradient in the energy
of the inverse-Compton photons.  The harder spectra that are often
observed at the beginning of GRB sub-pulses could be explained by
this effect.

\item{} Synchrotron emission in the optical range is suppressed because
the heated electrons and positrons have small pitch angles.  
The inverse-Compton emission of MeV-GeV gamma-rays only requires
particle energies up to $\gamma_e \sim 30-10^2$, which is smaller
than the energy that is needed for optical synchrotron emission
($\gamma_e \sim 50(\sin\kappa)^{-1/2}$) 
when the small pitch angle is taken into account.

\item{}  The magnetization is very different in the fluids that
have passed through the foward and reverse
shocks:  the magnetic field probably contributes a much smaller
fraction of the pressure in front of the contact.  The Weibel instability
generates a very small-scale magnetic field.  This field has a tendency
to smooth out and increase its scale downstream of the shock
(Medvedev et al. 2005).  However, if the rate of smoothing is at least
$\sim 10^{-4}$ of the maximal rate allowed by causality, then
the r.m.s. smoothed field will be weaker than the field which is
swept up from the Wolf-Rayet wind and linearly compressed behind
the forward shock.

\item{} The thermal photons advected out from the 
sub-photospheric region of the Wolf-Rayet star have
a finite duration, and a transition from prompt burst to
afterglow can be explained by a drop in the thermal photon density.
A modest 
anisotropy in the gamma-ray emission on a angular scale $\sim \Gamma_c^{-1}$,
due to corrugation instabilities of the breakout shell, could also
influence the shape of the transition to the afterglow regime.

\item{}  The gamma-rays themselves are the dominant coolant
at the foward shock.  As pre-acceleration turns off, the medium
ahead of the shock develops relativistic motion with respect to the
contact discontinuity (a differential Lorentz factor $\sim 20$).  
The pairs which pass across the forward shock emit inverse-Compton
gamma-rays with an energy ${4\over 3}\Gamma_\pm^2 \sim 500$ times the 
peak energy of the burst.    This emission is, however, too
spectrally soft and too weak to explain the rising high-energy component of GRB
941017.  We argue that the high-energy emission in GRB 941017 requires
rapid (but not too rapid) heating of the particles behind the reverse
shock, most probably triggered by magnetic reconnection.

\item{} The global MHD instabilities identified by Lyutikov and Blandford
(2003) are effective at tangling the magnetic field
while the jet is still working its way through
the envelope of the Wolf-Rayet star.   They are probably less important
in the prompt emission zone at
$\sim 10^{14}-10^{15}$ cm.  The Lorentz factor of the
contact is still large at this stage, $\Gamma_c \sim 50$, and much
larger than the inverse of the jet opening angle.  Smaller angular
structures in the outflow will be created by the corrugation instability
of the breakout shell; but the growing mode will have a fairly
large angular scale close to the star, $\theta \sim \Gamma_c^{-1}
\propto (r/R_\star)^{-1/3}$, where the Lorentz factor of the contact
is still modest.  The angular scale decreases as the breakout shell
is accelerated outward, but at the end of the prompt deceleration 
phase the contact is moving nearly as fast as the ejecta.

\end{enumerate}

We close this paper with a question:  What is the main outstanding
theoretical problem relating to the prompt emission of gamma-ray bursts? 
This question is far from being rhetorical:  the proper design of
numerical experiments to study the 3-dimensional behavior of plasmas --
especially in the extreme regimes that are encountered in GRB outflows --
requires a basic understanding of the dominant physical effects
that are likely to be encountered.

Most recent research would seem to point to the following answer:
one must understand the structure and intermittency of the outflow from
a stellar-mass black hole that is surrounded by a neutronized torus.  
However, this problem is vast, as it encompasses the
mass flow in the torus; the Blandford-Znajek process
operating in the black hole magnetosphere;  the acceleration and
collimation of the outflow near the engine;  the interaction of this
jet with the envelope of the Wolf-Rayet star; nuclear processes occurring
over a wide range of radii; and the various
instabilities that occur in the jet outside the star.  In this
picture, the emission problem {\it cannot} simply be reduced
to the problem of understanding particle acceleration and radiation
at a mildly relativistic shock wave, since the
details of the microphysics depend strongly on where the shock
occurs in the outflow, the strength of the magnetic field, and
the feedback of radiation processes on the particle composition
and energy distribution.  

We are suggesting here that, in fact, the theoretical
problem is a bit more tractable.  Because the 
non-thermal emission is triggered
by the interaction with the external wind and the material of the
breakout shell, one can focus down on a
small part of the outflow, and in particular on the problem of how
long-wavelength magnetohydrodynamic turbulence is damped in a fluid
satisfying the following conditions:
approximate equipartition between thermal radiation
and magnetic field; $B^2/8\pi \ga n_\pm m_ec^2$; and
compactness $\ell \sim 10$.  
Because the dissipation zone has been localized to a well-defined region,
one can now begin to examine the radial structure of the outflow
in a constrained way, and to understand the origin of 
distinct emission components such as was observed in 
GRB 941017 (Gonz\'alez et al. 2003).  

From this perspective, the classical gamma-ray bursts
have a greater affinity with the giant flares of the Soft Gamma Repeaters
that has been generally assumed in the recent literature.  Much of the early
confusion about whether gamma-ray bursts originate in the magnetospheres
of Galactic neutron stars, or in relativistic fireballs at cosmological
distances, turns out to be connected to a similarity in 
the mechanism underlying these spectacular releases of radiant energy.

\acknowledgments
The author would like to thank Maxim Lyutikov for several discussions of
particle heating during the early stages of this work, Yoram Lithwick for
discussions of MHD turbulence,
and Chris Matzner for discussions of shock breakout.  This work was
supported by the NSERC of Canada.

\end{document}